\documentclass[twocolumn,prd,nofootinbib,showpacs,superscriptaddress,amsmath,amssymb,preprintnumbers]{revtex4-1}

\usepackage{tikz}
\usepackage{axodraw4j}
\usepackage{dsfont}
\usepackage{picture}
\usepackage{ebezier}
\usepackage{pstricks}

\usepackage{color}
\usepackage{graphicx}
\usepackage{dcolumn}
\usepackage{bm}
\usepackage{multirow}
\usepackage{hyperref}
\usepackage{natbib}
\usepackage{soul}

\DeclareMathOperator{\Tr}{Tr}

\DeclareMathOperator{\sgn}{sgn}
\let\Re\relax
\let\Im\relax
\DeclareMathOperator{\Re}{Re}
\DeclareMathOperator{\Im}{Im}
\DeclareMathOperator{\PV}{P}

\DeclareMathOperator{\vol}{Vol}

\newcommand{\dbar}[1]{\bar{\bar{#1}}}

\makeatletter
\newcommand{\subalign}[1]{
  \vcenter{
    \Let@ \restore@math@cr \default@tag
    \baselineskip\fontdimen10 \scriptfont\tw@
    \advance\baselineskip\fontdimen12 \scriptfont\tw@
    \lineskip\thr@@\fontdimen8 \scriptfont\thr@@
    \lineskiplimit\lineskip
    \ialign{\hfil$\m@th\scriptstyle##$&$\m@th\scriptstyle{}##$\crcr
      #1\crcr
    }
  }
}

\makeatother
  
\begin{document}

\preprint{RIKEN-QHP-181, RIKEN-STAMP-6}

\title{From quantum to classical dynamics: The relativistic $O(N)$ model in the framework of the real-time functional renormalization group}

\author{D.~Mesterh\'azy}
\email{mesterh@itp.unibe.ch}
\affiliation{
  \mbox{Albert Einstein Center for Fundamental Physics, Universit\"at Bern, Sidlerstrasse 5, 3012 Bern, Switzerland}
}
\altaffiliation[On leave from:]{
  \textit{Department of Physics, University of Illinois at Chicago, 845 West Taylor Street, Chicago, IL 60607, USA}
}
\author{J. H.~Stockemer}
\email{j.stockemer@thphys.uni-heidelberg.de}
\affiliation{
  Institut f\"ur Theoretische Physik, Universit\"at Heidelberg, Philosophenweg 16, 69120 Heidelberg, Germany
}
\author{Y.~Tanizaki}
\email{yuya.tanizaki@riken.jp}
\affiliation{
  \mbox{Department of Physics, The University of Tokyo, 7-3-1 Hongo, Bunkyo-ku, Tokyo 113-0033, Japan}}
\affiliation{
  Theoretical Research Division, Nishina Center, RIKEN, Wako, Saitama 351-0198, Japan
}

\date{\today}

\begin{abstract}
We investigate the transition from unitary to dissipative dynamics in the relativistic $O(N)$ vector model with the $\lambda\! \left(\varphi^{2}\right)^{2}\!$ interaction using the nonperturbative functional renormalization group in the real-time formalism. In thermal equilibrium, the theory is characterized by two scales, the interaction range for coherent scattering of particles and the mean free path determined by the rate of incoherent collisions with excitations in the thermal medium. Their competition determines the renormalization group flow and the effective dynamics of the model. Here we quantify the dynamic properties of the model in terms of the scale-dependent dynamic critical exponent $z$ in the limit of large temperatures and in $2 \leq d \leq 4$ spatial dimensions. We contrast our results to the behavior expected at vanishing temperature and address the question of the appropriate dynamic universality class for the given microscopic theory.
\end{abstract}
 

\maketitle

\section{Introduction}
\label{Sec:Introduction}

Efforts to understand the properties of extreme states of matter in QCD \cite{Wilczek:1999ym, Rajagopal:2000wf} rely crucially on the identification and characterization of universal fluctuations. Understanding their properties is essential to settle the presence and nature of the critical end point \cite{Stephanov:2004wx}, marking the end of the first order transition between the hadronic and quark-gluon plasma (QGP) phase. To establish the hydrodynamics of the QGP near the QCD critical point, it is in particular their dynamic properties that are of considerable interest \cite{Son:2004iv}. Near the critical point the characteristic fluctuations display the phenomenon of critical slowing down which has important ramifications: It controls the maximal achievable correlation length in the process of the rapid cooling of the quark-gluon plasma through the critical region \cite{Berdnikov:1999ph,*Rajagopal:2000yt,*Bettencourt:2001xd} and thereby sets the typical strength of event-by-event fluctuations in heavy ion collisions probed at RHIC or the LHC \cite{Stephanov:1998dy,*Stephanov:1999zu}. On the other hand, it constrains important nonequilibrium effects as, e.g., the production of topological defects \cite{Balachandran:2001qn,Rajantie:2001ps}. It is a remarkable fact that these phenomena are determined only by a few quantities characteristic of the universality class of the critical point. While the identification of the static universality class is dictated by principles of symmetry and the dimensionality of the problem \cite{Pisarski:1983ms,Rajagopal:1992qz}, to identify the proper dynamic universality class requires an understanding of relevant modes at the phase transition (i.e., the conservation laws respected by the dynamics). Unfortunately, there are no general principles that allow to distinguish whether a particular mode is relevant at the critical point, or if its effect is irrelevant for the dynamic critical behavior. To answer this question it is necessary to reconcile well-established effective models for critical dynamics used in the vicinity of classical phase transitions \cite{Hohenberg:1977ym,Tauber:2014} with our understanding of the underlying coherently propagating excitations. In the past, there have been several attempts in this direction using perturbative methods as, e.g., the dynamic renormalization group \cite{Boyanovsky:2000nt, *Boyanovsky:2001pa, *Boyanovsky:2003ui, Nakano:2011re}, as well as nonperturbative methods, e.g., AdS/CFT duality \cite{Maeda:2008hn, Buchel:2010gd, DeWolfe:2011ts}, classical-statistical lattice field theory simulations \cite{Berges:2009jz}, or $n$-particle irreducible (nPI) effective actions \cite{Morimatsu:2014fna}. We propose a new attempt to make a solid connection between dynamic critical phenomena and microscopic quantum physics using the nonperturbative functional renormalization group (RG) \cite{Wetterich:1989xg, *Wetterich:1992yh}. This formalism has the potential to shed new light on the smooth quantum-to-classical crossover and enables us to address the question of the appropriate dynamic universality class for a given microscopic theory. In particular, we provide a continuous connection between the microscopic relativistic and nonrelativistic dissipative dynamics at nonvanishing temperature -- going beyond conventional approaches to dynamic critical behavior based on effective models. Our work constitutes a first important step towards the goal to fully unravel the dynamic properties in both the quantum and classical domain in the framework of the real-time functional renormalization group.

\begin{figure}[!t]
\centering
\includegraphics[width=0.45\textwidth]{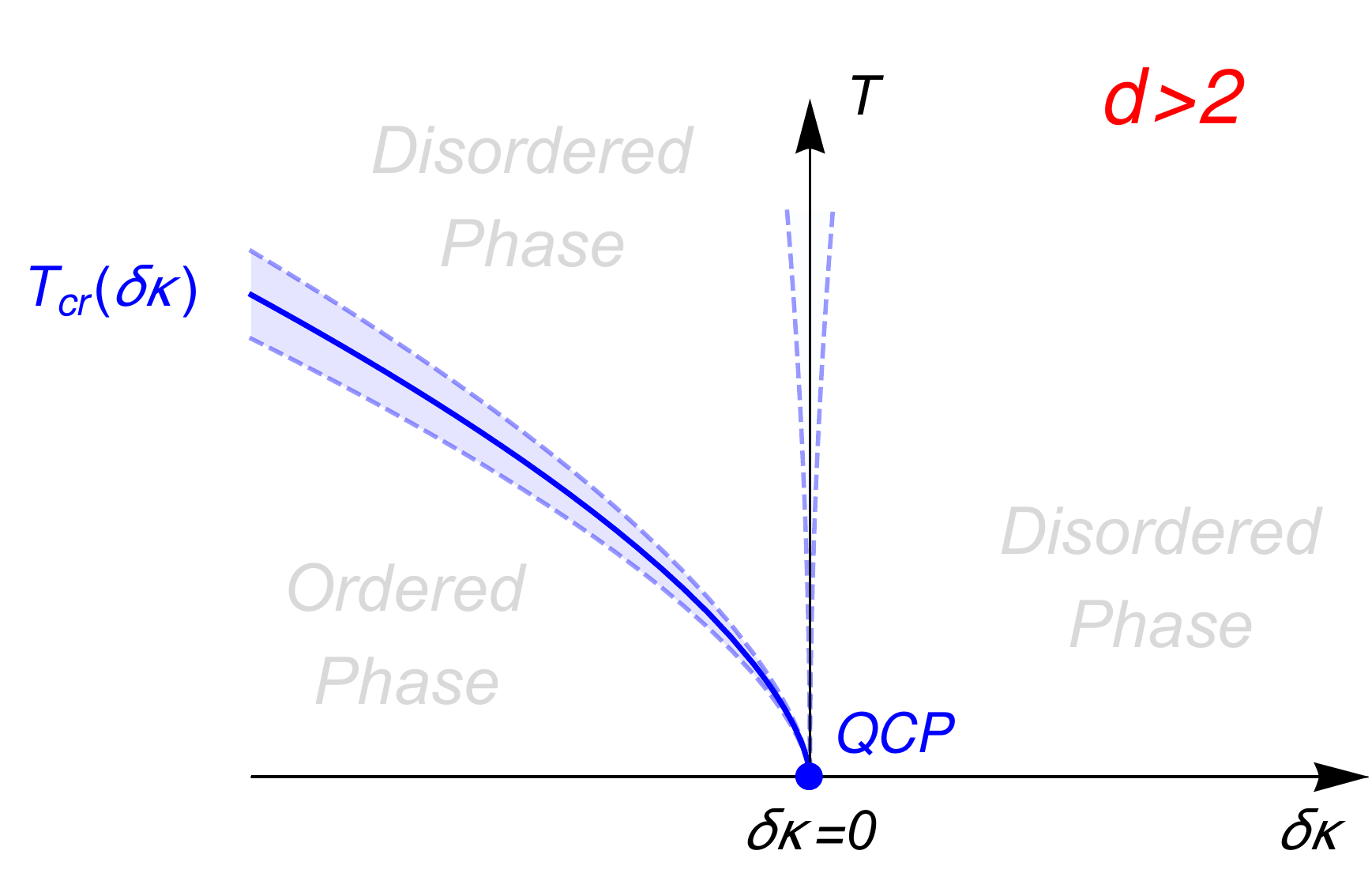} 
\caption{\label{Fig:PhaseDiagramSchematic}Illustration of the expected phase diagram for the $O(N)$ vector model in $d > 2$ spatial dimensions. It features a line of second-order phase transitions $T_{cr} = T_{cr}(\delta \kappa)$ (blue, continuous line) that separates an ordered from a disordered phase, and terminates at a quantum critical point ($T = \delta \kappa = 0$). As this critical line is approached, the correlation length $\xi = \xi(T, \delta \kappa)$ diverges. Shaded areas highlight those regions in the phase diagram where critical fluctuations become important. The narrow region around $\delta \kappa$ is controlled primarily by the QCP and becomes smaller as the spatial dimension $d$ is increased. It vanishes completely above $d = 3$, in which case the effect of fluctuations can be neglected. The critical region around the thermal transition $T_{cr}(\delta \kappa)$ similarly decreases in size until fluctuations cease to be important above its upper critical dimension $d = 4$.}
\end{figure}

In this paper, we consider the relativistic $O(N)$ vector model with the $\lambda\! \left(\varphi^{2}\right)^{2}\!$ interaction and investigate the transition from the microscopic dynamics to low-energy effective dynamics in the vicinity of a continuous phase transition. Let us briefly explain qualitative properties of this model. It can be seen to describe an interacting gas of relativistic bosonic particles in thermal equilibrium where the interaction is short-ranged and determined by the $s$-wave scattering length \cite{Leggett:2001zz}. Ultraviolet renormalization tells us that a finite number of parameters, which determines the low-energy scattering of particles in the vacuum, totally characterizes the microscopic dynamics. On the other hand, the screening of soft momenta at nonvanishing temperature is achieved by infrared renormalization. Thus, there are two competing processes that play a role in the system: The coherent scattering of particles leads to a renormalization of the coupling and mass parameters of the model, whereas incoherent collisions with thermal excitations in the medium endow the particles with a finite lifetime. The importance of these two processes for the critical behavior of the theory strongly depend on the spatial dimension as well as the temperature of the system.

At zero temperature ($T = 0$) and in spatial dimension $d \geq 2$, the relativistic $O(N)$ model features a second-order phase transition which separates a disordered (symmetric) phase from an ordered phase with a nonvanishing field expectation value. In the vicinity of such a quantum critical point (QCP) \cite{Sachdev:2011} the system is characterized by a diverging correlation length $\xi \sim |\delta \kappa|^{-\nu}$, where $\delta \kappa = \kappa-\kappa_{\mathrm{c}}$ is a microscopic parameter that controls the deviation from the critical point and $\nu$ is the correlation-length exponent. The diverging correlation length can be related to the vanishing of a relevant spectral gap $\Delta \sim \xi^{-z}$, where $z$ defines the associated dynamic critical exponent.\footnote{We only indicate the leading divergence in $\xi$ and $\Delta$ and thus the provided scaling holds up to an appropriate dimensionful constant (and additional subleading scaling corrections). Note, that both the correlation length and the spectral gap depend on the temperature $T$, the parameter $\delta \kappa$, and possibly other external fields, in general.} There is no characteristic scale at the QCP -- the system fluctuates at all scales and thus some physical quantities, such as scaling exponents, amplitude ratios, etc., can be universally classified without knowing the details of the microscopic physics.

Temperature is a relevant perturbation at the QCP. It yields a different universality class for the thermal transition or may even render the transition unstable. While for $d > 2$ the zero-temperature transition lies in the $(d+1)$-dimensional universality class of the classical $O(N)$ model, at nonvanishing temperature the system experiences an effective dimensional reduction and the appropriate characterization of universal fluctuations is given by the $d$-dimensional theory. For spatial dimension $d = 2$ the QCP is in the three-dimensional $O(N)$ universality class, however for $T\neq 0$ one expects that the field expectation value always vanishes, in accordance with the Mermin-Wagner theorem \cite{Mermin:1966fe}.\footnote{There is an important exception to the theorem and that is the theory with a discrete order parameter ($N = 1$). At $T \neq 0$ and in spatial dimensions $d > 1$, this model features a continuous phase transition in the $d$-dimensional Ising universality class, while the QCP lies in the $(d+1)$-dimensional Ising universality class.} Indeed, for vector models with $N \geq 3$ this is the case and there are no indications for an ordered low-temperature phase. Nevertheless, for $N = 2$ the system experiences a Kosterlitz-Thouless transition at some critical temperature, which separates a high-temperature disordered phase from a low-temperature phase with algebraic order. This observation does not contradict the Mermin-Wagner theorem, which cannot exclude a low-temperature phase with algebraically decaying correlations (see, e.g., Ref.\ \cite{Pelissetto:2002} and references therein). In the presence of a nonvanishing temperature we may identify different regimes for the dynamic correlations and response of the system: In the classical regime $\Delta \ll T$, all modes are highly occupied up to the scale set by the temperature. On the other hand, when the temperature is sufficiently small, the thermal occupation of modes is exponentially suppressed, i.e., $e^{-\omega(\boldsymbol{p}) / T} \ll 1$, for $\omega(\boldsymbol{p}) \gtrsim \Delta$. This defines the quantum regime, where occupation numbers are small and the commutator of field operators cannot be neglected. Close to the QCP, in the quantum regime, the dynamic critical exponent $z = 1$ in the relativistic $O(N)$ model which is a consequence of Lorentz symmetry. As the temperature is increased and the system crosses over to the classical regime, similar to the change in static universality class, the dynamic critical exponent changes ($z > 1$). It is this interplay of temperature and dynamics that we address in the following. These arguments are summarized in Fig.\ \ref{Fig:PhaseDiagramSchematic} which shows an illustration of the expected phase diagram at zero and nonvanishing temperatures.

We provide a picture of the processes that lead to the demise of coherent quasiparticle excitations and demonstrate how an effective classical description of many-body systems emerges at nonvanishing temperature. For that purpose we employ the nonperturbative functional RG \cite{Wetterich:1989xg, *Wetterich:1992yh} that allows us to study the impact of temperature and dimensionality independent of the assumption of small coupling or small deviations from the upper critical dimension. Similar work along these lines has been pursued, where the properties of classical phase transitions and its static critical exponents have been examined in the real-time formalism, e.g., at the example of the real scalar theory \cite{D'Attanasio:1996zt, Litim:1998yn} and gauge theories \cite{D'Attanasio:1996fy}. Dynamic scaling properties have been addressed in the framework of the functional RG in the context of effective models for critical dynamics \cite{Tauber:2014}. This includes in particular, Model A \cite{Canet:2006xu} and \mbox{Model C} \cite{Mesterhazy:2013naa} in the Halperin-Hohenberg classification of dynamic universality classes \cite{Hohenberg:1977ym}. Further applications include stochastically driven systems out of equilibrium \cite{Canet:2011wf,Sieberer:2013,*Sieberer:2013lwa,Mathey:2014yxa}, e.g., reaction-diffusion and percolation problems \cite{Canet:2003yu}, the Kardar-Parisi-Zhang (KPZ) equation \cite{Canet:2009vz, *Kloss:2012ms, *Kloss:2013xva}, hydrodynamic turbulence \cite{Canet:2014dta,Mathey:2014xxa}, and nonequilibrium steady states in closed systems \cite{Berges:2008sr,*Berges:2012ty,Gasenzer:2008zz,*Gasenzer:2010rq}. Close to a quantum phase transition (QPT) the functional renormalization group has been applied to a number of problems in the imaginary-time formalism. These include nonrelativistic systems where a crossover was observed to a quasi-relativistic action with second (imaginary) time derivatives \cite{Wetterich:2008} and the two-dimensional $O(N)$ model \cite{Rancon:2014cfa,*Rose:2015bma} for which analytic continuation was used to derive spectral functions in the quantum critical regime. The relative importance of quantum and thermal fluctuations in the vicinity of a QPT was addressed for the two-dimensional quantum Ising model \cite{Strack:2009}, as well systems with fermions \cite{Jakubczyk:2008cv}.

The outline of this paper is as follows: In \mbox{Sec.\ \ref{Sec:Real-time n-point functions and generating functionals}} we review real-time formalisms to determine $n$-point functions -- a necessary prerequisite to introduce the nonperturbative real-time functional RG. In \mbox{Sec.\ \ref{Sec:Nonperturbative functional renormalization group in the real-time formalism}} we discuss the properties of the functional flow and its application to calculate real-time response in equilibrium systems. We define the relativistic $O(N)$ vector model and associated scale-dependent effective action in \mbox{Sec.\ \ref{Sec:Relativistic O(N) model}} and discuss the role of different interaction vertices as well as consistent truncation schemes for the RG flow. The main part of this work consists of \mbox{Secs.\ \ref{Sec:Quantum regime} -- \ref{Sec:Limiting behavior of renormalization group equations in the classical regime}} where we investigate different possible regimes for the dynamics, the properties of scaling solutions, and how our results compare quantitatively to known results from the $\epsilon = 4-d$ and large-$N$ expansion. We close in Sec.\ \ref{Sec:Quantum-classical transition} with our perspective on the quantum-to-classical transition in the low-temperature regime.

\section{Real-time $n$-point functions and generating functionals}
\label{Sec:Real-time n-point functions and generating functionals}

The correlations and response of a system to external perturbations are fully captured by the time-ordered $n$-point functions:
\begin{eqnarray}
&& \hspace{-20pt} \big\langle \hat{T} \lbrack \hat{\varphi}_{a_{1}}(x_{1}) \cdots \hat{\varphi}_{a_{n}}(x_{n}) \rbrack \big\rangle \nonumber\\
&& \equiv \left. (-i)^{n} \frac{1}{Z[J]} \frac{\delta^{n} Z[J]}{\delta J^{a_{1}}(x_{1}) \cdots \delta J^{a_{n}} (x_{n})} \right\vert_{J = 0} ,
\label{Eq:TimeOrderedNPtFunction}
\end{eqnarray}
where the Heisenberg-field operators $\hat{\varphi}^{a}(x) = e^{i \hat{H} t} \hat{\varphi}^{a}(0, \boldsymbol{x}) e^{-i \hat{H} t}$ are given in the $N$-component vector representation of the $O(N)$ symmetry group and the functional derivatives with respect to the real-valued external sources $J^{a}(x)$ are evaluated at spacetime points $x = (t,\boldsymbol{x})$. For a system that is prepared in the groundstate or in thermal equilibrium, the generating functional in Eq.\ \eqref{Eq:TimeOrderedNPtFunction} can be defined as
\begin{equation}
Z[J] = \Tr \left\{ e^{-\beta \hat{H}} \, \hat{T} \left\lbrack e^{i\int_x J_a(x)\hat{\varphi}^a(x)} \right\rbrack \right\} . 
\label{Eq:TimeOrderedGeneratingFunctional}
\end{equation}
Here, $\beta = 1/T$ defines the inverse temperature, $\hat{H}$ is the (interacting) Hamiltonian, and the operator $\hat{T}$ orders the fields on the real line. However, systematic perturbation theory based on \eqref{Eq:TimeOrderedGeneratingFunctional} is ill-defined (see, e.g., Refs.\ \cite{Niemi:1983ea,*Niemi:1983nf,Landsman:1986uw}). To circumvent this point, we introduce the closed time contour $\mathcal{C}$ \cite{Keldysh:1964ud,Schwinger:1960qe}, so that the generating functional can be written as
\begin{equation}
Z[J] = \Tr \left\{ e^{-\beta \hat{H}} \, \hat{T}_{\mathcal{C}} \left\lbrack e^{i \int_{x}^{\mathcal{C}} J_{a}(x) \hat{\varphi}^{a}(x)} \right\rbrack \right\} .
\label{Eq:GeneratingFunctional}
\end{equation}
The external sources are taken to have nonvanishing support on parts of $\mathcal{C}$ in the complex time plane, upon which the operator $\hat{T}_{\mathcal{C}}$ defines the time ordering of the fields. The contour integration in Eq.\ \eqref{Eq:GeneratingFunctional} can be expressed via the parametrization $\mathcal{C}: [0,1] \ni s \mapsto t(s) \in \mathbb{C}$, $\int_{x}^{\mathcal{C}} \,\cdots \equiv \int^{[0,1]} ds\, t'(s) \int d^{d}x \,\cdots$, with $t(0) = t(1) = t_{i}$, and $t_{i}$ real. While the purpose of this section is to illustrate how to determine the real-time dynamic correlation functions and response, for now, we will impose no restrictions on the time path $\mathcal{C}$. In the following, we examine different possible choices and highlight their importance and shortcomings.

\begin{figure*}[!t]
\begin{minipage}[b]{0.47\textwidth}
\centering
\includegraphics[width=0.7\textwidth]{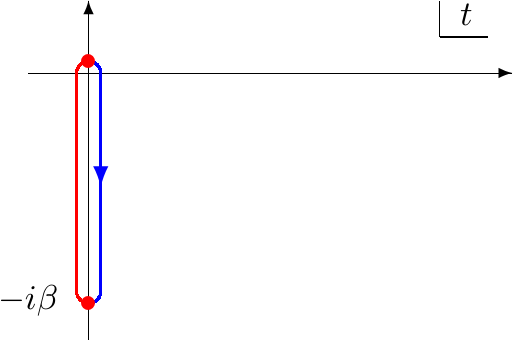}
\vskip 0pt
\caption{\label{Fig:ComplexContourImaginaryTime}Imaginary-time contour $\mathcal{C}_{-i\beta}$ in the complex time plane. The right branch defines the segment of the contour on which the sources are nonvanishing and is equipped with a imaginary-time ordering (indicated by the arrow). The left branch closes the contour and defines a periodicity constraint for the fields that live on the right contour (at initial and final imaginary time). In the corresponding Euclidean action this is expressed by the boundary condition $\varphi^{a}(t = t_{i}) = \varphi^{a}(t = t_{i} -i \beta)$. The particular choice on which branch the fields should propagate is arbitrary.}
\end{minipage} \qquad
\begin{minipage}[b]{0.47\textwidth}
  \centering
  \includegraphics[width=0.66\textwidth]{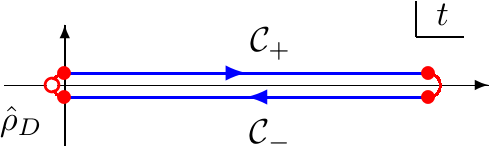}
\vskip 44pt
\caption{\label{Fig:ComplexContourRealTime}Real-time contour $\mathcal{C} = \mathcal{C}_{+} \oplus\, \mathcal{C}_{-}$ in the complex time plane. The upper and lower part of the contour define those segments on which the external sources are nonvanishing. The statistical operator $\hat{\rho}_{D} = Z^{-1} e^{-\beta \hat{H}}$ is inserted at initial time $t_{i}$, while the contour winds around at final time $t_{f}$. The endpoints of the contour are completely arbitrary and the generating functional does not depend on the specific choice of $t_{i}$ or $t_{f}$. However, initial and final segments define constraints for the fields that live on the upper and lower branches of the contour, e.g., in the real-time action on the closed time path, we impose the constraint \mbox{$\varphi_{+}^{a}(t = t_{f}) = \varphi_{-}^{a}(t = t_{f})$} at final time.}
\end{minipage}
\end{figure*}

The imaginary-time formalism (ITF) \cite{Matsubara:1955ws,Dolan:1973qd,Pisarski:1987wc} provides a convenient means to calculate imaginary time-ordered correlation functions in the groundstate or in thermal equilibrium. In this framework, the fields live on an imaginary-time contour $\mathcal{C} = \mathcal{C}_{-i\beta}$, where $\mathcal{C}_{-i \beta} = \{ {t\in \mathbb{C}} \,|\, \Re t = t_{i},\, \Im t \in [-\beta, 0] \}$ (see Fig.\ \ref{Fig:ComplexContourImaginaryTime}). That is, in the ITF the $(d+1)$-dimensional Euclidean theory appears compactified in the time direction with perimeter $\beta = 1/T$ and the imaginary-time Green functions satisfy a periodicity constraint, the Kubo-Martin-Schwinger (KMS) condition \cite{Kubo:1957mj,Martin:1959jp}. Based on the generating functional \eqref{Eq:GeneratingFunctional}, this formalism allows us to determine thermodynamic properties of interacting many-body systems, either by perturbative or nonperturbative means. The analytic continuation of imaginary-time correlation functions furthermore allows for the calculation of dynamic properties, e.g., the response to external perturbations and associated transport properties. However, this analytic continuation, in practice, is an ill-posed numerical problem. In the absence of a method that produces exact results at imaginary time, additional assumptions on the analyticity properties and ultraviolet (UV) behavior of correlation functions are required (see, e.g., Refs.\ \cite{Gubernatis:1991zz,Jarrell:1996}). The validity of such assumptions are especially questionable in the vicinity of a thermodynamic singularity, e.g., at a second-order phase transition where the system displays nonanalytic behavior. Either way, the final result will be sensitive to the values of some \emph{ad hoc} parameters that might not have an immediate physical significance. To determine the dynamic correlations or response, it is therefore desirable to employ a framework that allows for a direct calculation in real time without the need for analytic continuation.

Real-time formalisms (RTF) \cite{Zhou:1980ky, Niemi:1983ea, *Niemi:1983nf, Calzetta:1986cq, *Calzetta:1986ey, Landsman:1986uw} employ a different choice for the time contour $\mathcal{C}$ that appears in the definition of $n$-point functions \eqref{Eq:TimeOrderedNPtFunction} and their generating functional \eqref{Eq:GeneratingFunctional}. While in principle there is an infinite set of possible time paths, we consider a time-ordered contour consisting of two branches only, $\mathcal{C} = \mathcal{C}_{+} \oplus\, \mathcal{C}_{-}$, with $t(0) = t(1) = t_{i}$ and $t(s)$ real for all $s \in\lbrack 0, 1 \rbrack$ \cite{Keldysh:1964ud,Schwinger:1960qe}. This contour, illustrated in \mbox{Fig.\ \ref{Fig:ComplexContourRealTime}}, is defined with forward and backward time ordering on the branches $\mathcal{C}_{+}$ and $\mathcal{C}_{-}$, respectively. The closed time path (CTP) generating functional \eqref{Eq:GeneratingFunctional} becomes
\begin{eqnarray}
Z[J_{+}, J_{-}] &=& \Tr \bigg\{ \hat{T} \left\lbrack e^{i \int_{x} J_{+,a}(x) \hat{\varphi}^{a}(x)} \right\rbrack \nonumber\\
&& \times\:  e^{-\beta \hat{H}} \, \left( \hat{T} \left\lbrack e^{i \int_{x} J_{-,a}(x) \hat{\varphi}^{a}(x)} \right\rbrack \right)^{\!\dagger} \bigg\} ,
\label{Eq:ContourGeneratingFunctional}
\end{eqnarray} 
where we have used the cyclicity of the trace and introduced external sources $J_{+}$ and $J_{-}$ on the forward and backward time-ordered segments of the contour. We assume in the following that the sources have nonvanishing support at times $t_{i} < t < t_{f}$ and fall off sufficiently fast at spatial infinity.\footnote{The spacetime integration is defined as: $\int_{x} \cdots \equiv \int^{[t_{i},t_{f}]} dt \int d^{d}x \cdots$ and it is understood that the limits $t_{i} \rightarrow -\infty$ and $t_{f}\rightarrow \infty$ are taken at the end.} This doubling of the degrees of freedom allows us to employ the usual definition of the time ordering operator $\hat{T}$ for real times. Furthermore, writing the generating functional in the form \eqref{Eq:ContourGeneratingFunctional} makes the conjugation symmetry \cite{Niemi:1983ea, *Niemi:1983nf} clear:
\begin{equation}
Z[J_{+},J_{-}]^{\ast} = Z[J_{-},J_{+}] .
\label{Eq:ConjugationSymmetryGeneratingFunctional}
\end{equation}
Of course, in the absence of sources the physical content of \eqref{Eq:ContourGeneratingFunctional} is no different than that provided by evaluating the generating functional on the imaginary contour $\mathcal{C}_{-i\beta}$, and we have $Z^{\textrm{ITF}} = Z^{\textrm{RTF}}$.

In the following, we aim at a nonperturbative evaluation of the generating functional using the functional RG in the RTF. Its construction relies on the path integral representation of the generating functional on the CTP, the derivation of which \cite{Niemi:1983ea, *Niemi:1983nf, Chou:1984es, Calzetta:1986cq, *Calzetta:1986ey, Landsman:1986uw} is based on the insertion of a complete set of states associated to the Heisenberg-field operator $\hat{\varphi}(x)$. We distinguish the fields on the two branches of the CTP and introduce two distinct degrees of freedom, $\varphi_{+}$ and $\mathcal{\varphi}_{-}$, on the forward and backward time-ordered segments, respectively:
\begin{equation}
\hat{\varphi}(t,\boldsymbol{x}) | \varphi_{\alpha}(\boldsymbol{x}) ; t \in \mathcal{C}_{\alpha} \rangle = \varphi_{\alpha}(t,\boldsymbol{x}) | \varphi_{\alpha}(\boldsymbol{x}) ; t \in \mathcal{C}_{\alpha} \rangle ,
\end{equation}
where $\alpha \in \{ + , - \}$. At large time $t_{f}$, at which we identify the two branches of the contour (cf. Fig.\ \ref{Fig:ComplexContourRealTime}) the fields satisfy the property $\varphi_{+}(t_{f}) = \varphi_{-}(t_{f})$ and we will choose the value of $t_{f}$ in the following such that $t_{f}\rightarrow \infty$. Proceeding along these lines we first decompose the generating functional \eqref{Eq:ContourGeneratingFunctional} as a product of transition matrix elements between states that live on the CTP, i.e.
\begin{widetext}
\begin{equation}
Z[J_{+},J_{-}] = \int\!\! \prod_{\alpha \in \{+, -\}} [d\varphi_{\alpha}(t_{i})] [d\varphi_{\alpha}(t_{f})]\,  \left\langle \varphi_{+} ; t_{i} \right\vert e^{-\beta \hat{H}} \left\vert \varphi_{-} ; t_{i} \right\rangle \left\langle \varphi_{-} ; t_{i} | \varphi_{-} ; t_{f} \right\rangle_{J_{-}} \left\langle \varphi_{-} ; t_{f} | \varphi_{+} ; t_{f} \right\rangle \left\langle \varphi_{+} ; t_{f} | \varphi_{+} ; t_{i} \right\rangle_{J_{+}} , 
\label{Eq:GeneratingFunctionalProductForm}
\end{equation}
\end{widetext}
where the subscripts $J_{+}$ and $J_{-}$ indicate that the corresponding matrix elements should be evaluated in the presence of external sources. The functional measure is defined as $\int [d\varphi_{\alpha}(t)]\, \equiv \mathcal{N} \int \prod_{\boldsymbol{x}\in \mathbb{R}^{d}} \, d\varphi_{\alpha}(t,\boldsymbol{x})$ with some appropriate normalization constant $\mathcal{N}$ and a proper regularization of the infinite product. The matrix elements for which the external sources are nonvanishing can be expressed in terms of a functional integral over a phase factor which is fully determined by the dynamics and the coupling of the fields to external sources. On the $\mathcal{C}_{\alpha}$, $\alpha \in \{ + , -\}$, segments they take the following form:
\begin{equation}
\left\langle \varphi_{\alpha} ; t_{f} | \varphi_{\alpha} ; t_{i} \right\rangle_{J_{\alpha}} = \int [d\varphi_{\alpha}'] \, e^{i \{  S[\varphi_{\alpha}] + \int_{x} J_{\alpha,a}(x) \varphi_{\alpha}^{a}(x) \}} .
\end{equation}
Here, the functional integration runs over all field configurations with fixed values of the field both at initial $\varphi_{\alpha} = \varphi_{\alpha}(t_{i})$ and final time $\varphi_{\alpha} = \varphi_{\alpha}(t_{f})$. Evaluating the product of all matrix elements in \eqref{Eq:GeneratingFunctionalProductForm} we obtain the path integral representation of the full generating functional:
\begin{widetext}
\begin{equation}
Z[J_{+},J_{-}] = \int\!\! \prod_{\alpha \in \{+, -\}} [d\varphi_{\alpha}] \, e^{i \{  S[\varphi_{+},\varphi_{-}] + \int_{x} J_{+,a}(x) \varphi_{+}^{a}(x) - \int_{x} J_{-,a}(x) \varphi_{-}^{a}(x) \}} ,
\label{Eq:ContourGeneratingFunctionalPI}
\end{equation}
\end{widetext}
where we have introduced the CTP action
\begin{equation}
S[\varphi_{+},\varphi_{-}] = S[\varphi_{+}] - S^{\ast}[\varphi_{-}] + \mathcal{F}[\varphi_{+},\varphi_{-}] ,
\label{Eq:CTPAction}
\end{equation}
and the full functional measure 
\begin{equation}
\int\!\! \prod_{\alpha \in \{ +,- \}} [d\varphi_{\alpha}]\, \equiv \int\!\! \prod_{\alpha \in \{ +,- \}} \prod_{t_{i} \leq t \leq t_{f}} [d\varphi_{\alpha}(t)] .
\end{equation}
The functional integral goes over all possible field configurations. Any constraints on the fields have been absorbed into the CTP action \eqref{Eq:CTPAction}, in the form of the constraint-fluctuation (CF) functional $\mathcal{F}[\varphi_{+},\varphi_{-}]$. We provide the following formal definition in terms of the matrix elements at initial and final time
\begin{eqnarray}
\mathcal{F}[\varphi_{+},\varphi_{-}] &=& -i \ln Z  -\: i \ln \left\langle \varphi_{+}; t_{i} \right\vert \hat{\rho}_{D} \left\vert \varphi_{-} ; t_{i} \right\rangle \nonumber \\ && -i \ln \left\langle \varphi_{-}; t_{f} \vert \varphi_{+} ; t_{f} \right\rangle ,
\label{Eq:ConstraintFunctional}
\end{eqnarray}
where $\hat{\rho}_{D} = Z^{-1} e^{- \beta \hat{H}}$, so that $\Tr \hat{\rho}_{D} = 1$. Of course, to give any meaning to this object, we need to consider regularizations of the contributions on the RHS. This is immediately clear if we examine, e.g., the final time constraint which can be expressed in the following way: $\left\langle \varphi_{-}; t_{f} \vert \varphi_{+} ; t_{f} \right\rangle = \delta \lbrack \varphi_{-}(t_{f}) - \varphi_{+}(t_{f}) \rbrack$. Independent of the regularization we take note of the following properties: The CF-functional references only the field values at the boundaries $t_{i}$ and $t_{f}$ and, in equilibrium, it is homogeneous in time. Also, while the first term on the RHS of Eq.\ \eqref{Eq:ConstraintFunctional} is field-independent and therefore does not contribute in the calculation of correlation functions the remaining terms do. They eventually lead to nonvanishing mixing contributions and additional couplings between the $\varphi_{+}$ and $\varphi_{-}$ fields in the CTP effective action. It is these terms that affect the correlations and response of the system and have to be accounted for to resolve the real-time properties of the theory in equilibrium.

From Eq.\ \eqref{Eq:ContourGeneratingFunctionalPI} we obtain the generating functional for connected correlation functions in the path integral formalism. It is defined as
\begin{equation}
W[J_{+},J_{-}] = -i \ln Z[J_{+},J_{-}] .
\end{equation} 
Field expectation values are obtained by functional differentiation
\begin{equation}
\phi_{\alpha}(x) = \frac{\delta W[J_{+},J_{-}]}{\delta J_{\alpha}(x)} ,
\label{Eq:AverageFieldsRTF}
\end{equation}
as are the connected correlation functions
\begin{equation}
G_{\alpha \beta}(x,y) = \frac{\delta^{2} W[J_{+},J_{-}]}{\delta J_{\alpha}(x) \delta J_{\beta}(y)} , 
\end{equation}
where Greek indices $\alpha, \beta \in \{ +, - \}$ denote fields and external sources on the distinct branches of the CTP. We construct the effective action in the RTF via the Legendre transform
\begin{eqnarray}
&& \hspace{-20pt} \Gamma[\phi_{+} , \phi_{-}] = W[J_{+}, J_{-}] \nonumber\\ 
&& \hspace{35pt} -\: \int_{x} \left\{ J_{+,a}(x) \phi_{+}^{a}(x) - J_{-,a}(x) \phi_{-}^{a}(x) \right\} ,
\end{eqnarray}
for which the sources are considered to be functionals of the fields, i.e., $J_{+} = J_{+}[\phi_{+},\phi_{-}]$ and $J_{-} = J_{-}[\phi_{+},\phi_{-}]$, as obtained from inverting the relations \eqref{Eq:AverageFieldsRTF}. From the effective action we derive the equations of motion in the presence of nonvanishing sources
\begin{equation}
\frac{\delta \Gamma[\phi_{+} , \phi_{-}]}{\delta \phi_{\alpha}(x)} = - \sgn(\alpha) J_{\alpha}(x) .
\end{equation}
When the external sources are set to zero, in the groundstate or in thermal equilibrium, the field configurations that extremize the effective action are homogeneous both in space and time. While in the construction of the effective action we employed the $\phi_{+}$ and $\phi_{-}$ fields, a much more convenient parametrization of the physical degrees of freedom is instead given by the average and difference fields (retarded/advanced basis)
\begin{equation}
\phi = \frac{1}{2} ( \phi_{+} + \phi_{-} ) , \qquad \tilde{\phi} = \phi_{+} - \phi_{-} ,
\label{Eq:RetardedAdvancedBasis}
\end{equation}
and the corresponding external sources 
\begin{equation}
J = \frac{1}{2} (J_{+} + J_{-}), \qquad \tilde{J} = J_{+} - J_{-} .
\end{equation}
In this parametrization, the equilibrium field configurations are: $\phi\,\vert_{J = \tilde{J} = 0} = \phi_{+}\,\vert_{J = \tilde{J} = 0} = \phi_{-}\,\vert_{J = \tilde{J} = 0} = \langle \hat{\varphi} \rangle$, $\tilde{\phi}\,\vert_{J = \tilde{J} = 0} = 0$, and we define the retarded propagator in the following way
\begin{equation}
G^{\textrm R}(x,y) = \frac{\delta^{2} W[J,\tilde{J}]}{\delta \tilde{J}(x) \delta J(y)} ,
\label{Eq:RetardedCommutator}
\end{equation}
which (in the absence of sources) is equivalent to the expectation value of the anti-symmetric field commutator, i.e., $\left. G^{\textrm R}(x,y) \right\vert_{J = \tilde{J} = 0} = i\theta(x^0-y^0) \langle [\hat{\varphi}(x),\hat{\varphi}(y)]\rangle$. The statistical correlation function is given by
\begin{equation}
F(x,y) = -i \frac{\delta^{2} W[J,\tilde{J}]}{\delta \tilde{J}(x) \delta \tilde{J}(y)} ,
\label{Eq:StatisticalCorrelator}
\end{equation}
which, also in the absence of sources, can be written as the (connected) average anti-commutator $\left. F(x,y)\right\vert_{J = \tilde{J} = 0} = \frac{1}{2}\langle \{ \hat{\varphi}(x), \hat{\varphi}(y) \} \rangle$. While the derivation of the effective action relies on the Legendre-transform with respect to the external sources, in the end we are interested only in the case when they are set to zero. To avoid a proliferation of notation in the following, it will therefore be understood that all quantities are specified in the absence of sources unless stated otherwise.

Equation \eqref{Eq:GeneratingFunctionalProductForm} essentially contains two different functional averages: The conditional average over the real-time evolution of the fields at times $t > t_{i}$, as well as an averaging over the groundstate or statistical operator $\hat{\rho}_{D}$ at initial time $t_{i}$. These two contributions are not distinguished within the ITF, where the functional average is performed over all (imaginary-time) fluctuations at once. Thus, in principle, the RTF allows us to study separately the effect of the initial conditions and the dynamics. Here, however, we ask specifically about the scale dependence of different dynamical processes in equilibrium as a function of a given set of microscopic parameters in the action. That is, we do not show how thermal correlations are built up in real time starting from an arbitrary initial state. This is an interesting problem in its own right and there are other frameworks that address the issue of thermalization \cite{Berges:2000ur,*Berges:2001fi,Calzetta:2002ub,Juchem:2003bi,Sengupta:2003db,Balasubramanian:2011ur,*Balasubramanian:2010ce}. We assume that the averaging over the initial statistical operator has been taken into account to the effect of imposing relations between different $n$-point functions dictated by the properties of equilibrium fluctuations. Typically, these relations are expressed in frequency space and this implies that the statistical operator $\hat{\rho}_{D}$ has been inserted in the infinite past $t_{i} \rightarrow -\infty$, so that the system satisfies time translation invariance.

The derivation of the CTP action \eqref{Eq:CTPAction} makes it clear, that the calculation of correlation functions will not depend on the field-independent contribution $- i \ln Z$ to the CF-functional. In fact, it can be ignored for the most part when dealing with a system in equilibrium and this is what we will do in the following. However, this procedure leads to some subtle issues in the calculation of the thermodynamic free energy, which we address here: The free energy density in the RTF is defined as
\begin{equation}
F = \frac{i}{\beta} V - \frac{1}{\beta \vol_{d+1}} \ln Z ,
\label{Eq:FreeEnergy}
\end{equation} 
where $\vol_{d+1} \equiv \int^{[t_{i},t_{f}]} dt \int d^{d}x$ and 
\begin{equation}
V = - \frac{\Gamma[\phi = \langle \hat{\varphi} \rangle,\tilde{\phi} = 0] }{\vol_{d+1}} ,
\end{equation} 
is the CTP effective potential. It depends on the value of the homogeneous field expectation value $\langle \hat{\varphi} \rangle$ and we have pulled out the field-independent contribution $\sim \ln Z$ from the CTP effective action to make it explicit. In equilibrium, one finds the effective potential $V$ to be zero in a calculation where all zero momentum 1PI diagrams are summed over \cite{Niemi:1983ea,*Niemi:1983nf,Evans:1986ws,*Evans:1988ub}. Of course, this is simply a consequence of the conjugation symmetry \eqref{Eq:ConjugationSymmetryGeneratingFunctional}, which constrains the free energy to be real-valued. However, it is important to realize that derivatives of the CTP effective action might be nonzero even in the absence of sources. A careful analysis shows that they match the contributions from the real part $- \frac{1}{\beta \vol_{d+1}} \ln Z$ \cite{Evans:1986ws,*Evans:1988ub}. Thus, the derivatives of the effective action carry the same information as $\ln Z$ and the free energy can be reproduced also within the CTP formalism when the field-independent contribution to the effective action is not taken into account explicitly.

The presence of  conjugation symmetry \eqref{Eq:ConjugationSymmetryGeneratingFunctional} constrains the form of possible parameters and couplings of the theory. For the CTP effective action it takes the form
\begin{equation}
\Gamma[\phi,\tilde{\phi}]^{\ast} = - \Gamma[\phi,-\tilde{\phi}] .
\label{Eq:ConjugationSymmetryEffectiveAction}
\end{equation}
If we expand $\Gamma[\phi,\tilde{\phi}]$ in the fields $\phi$ and $\tilde{\phi}$
\begin{widetext}
\begin{eqnarray}
&& \hspace{-20pt} \Gamma[\phi,\tilde{\phi}] = \sum_{m,n} \int_{x_{1}, \ldots , x_{m}; y_{1} , \ldots , y_{n}} \Gamma^{(m,n)}(x_{1} , \ldots, x_{m} ; y_{1} , \ldots , y_{n}) \phi(x_{1}) \cdots \phi (x_{m}) \tilde{\phi}(y_{1}) \cdots \tilde{\phi}(y_{n}) , 
\label{Eq:VertexExpansion} \\
&& \hspace{-20pt} \Gamma^{(m,n)}(x_{1} , \ldots, x_{m} ; y_{1} , \ldots , y_{n}) \equiv \frac{\delta^{m+n} \Gamma[\phi,\tilde{\phi}]}{\delta \phi(x_{1}) \cdots \delta \phi(x_{m}) \delta \tilde{\phi}(y_{1}) \cdots \delta \tilde{\phi}(y_{n})} , 
\end{eqnarray}
\end{widetext}
where the summation over internal indices is implied, the symmetry \eqref{Eq:ConjugationSymmetryEffectiveAction} divides the contributions to the effective action into two distinct classes, characterized by either real or imaginary (amputated) 1PI correlation functions $\Gamma^{(m,n)}(x_{1} , \ldots, x_{m} ; y_{1} , \ldots , y_{n})$ (which are related to the odd and even powers of $\tilde{\phi}$ in \eqref{Eq:VertexExpansion}, respectively). In fact, already from the form of the CF-functional \eqref{Eq:ConstraintFunctional}, it is clear that imaginary couplings and parameters to the CTP action are necessarily present when initial fluctuations and boundary conditions are taken into account. They play an important role over a wide range of scales and ensure the consistency of the thermal flow by providing relations between real and imaginary vertices of the Wilsonian effective action. The general aim of this paper is to demonstrate how these vertices might be generated and to show how an effective classical description emerges in the RTF when all fluctuations have been accounted for.

\section{Nonperturbative functional renormalization group in the real-time formalism}
\label{Sec:Nonperturbative functional renormalization group in the real-time formalism}

The nonperturbative RG employed in this work is based on the flow of generating functionals for correlation functions. In particular, for systems with spatial and time translation invariance, it is useful to consider the functional flow of the scale-dependent 1PI effective action \cite{Wetterich:1989xg, *Wetterich:1992yh}. There exists a large number of reviews in the literature and we will not provide all the details here, see, e.g., Refs.\ \cite{Berges:2000ew,Morris:1993qb,Bagnuls:2000ae,Polonyi:2001se,Pawlowski:2005xe,Gies:2006wv}. However, we will highlight some of the properties that appear in the physical (real-time) representation.

The construction of the functional flow equation relies on a modification of the original CTP action, i.e., $S[\varphi,\tilde{\varphi}] \rightarrow S[\varphi, \tilde{\varphi}] + \Delta_{k} S[\varphi,\tilde{\varphi}]$, where 
\begin{equation}
\Delta_{k} S[\varphi,\tilde{\varphi}] = - \int_{x,y} \tilde{\varphi}^{a}(x) R_{k, a b}(x,y) \varphi^{b}(y) .
\label{Eq:BilinearSourceTerm}
\end{equation} 
The function $R_{k}$ is taken to be nonnegative and its properties are chosen so that \eqref{Eq:BilinearSourceTerm} defines a momentum-dependent mass term, i.e.
\begin{equation}
R_{k,a b} (x^{0},y^{0}; \boldsymbol{p}) = R_{k} (\boldsymbol{p}) \delta(x^{0} - y^{0}) \delta_{a b} ,
\end{equation}
where $t_{i} \leq x^{0}, y^{0} \leq t_{f}$. It depends on an additional scale parameter $k$ which controls the effective mass of different modes. Thus, by an appropriate choice of $R_{k}$, we may regulate the infrared (IR) divergences associated to the massless modes in the vicinity of a second order (or weakly first order) phase transition (see, e.g., \mbox{Ref.\ \cite{Parisi:1993sp}}). Further regulating contributions quadratic in the difference field $\tilde{\varphi}$ might be necessary in the presence of additional symmetries \cite{Canet:2011wf}, but this is not the case here. For our purposes, it is sufficient to consider a regulator coupling the $\varphi$ and $\tilde{\varphi}$ fields only. Such a regulator is compatible with the symmetries in the presence of a fixed reference frame -- the rest frame in which we define the temperature of the system.

The modification of the action by the insertion \eqref{Eq:BilinearSourceTerm} endows the generating functional for connected correlation functions $W[J,\tilde{J}]$ with a scale dependence, which we highlight by an additional index $k$:
\begin{equation}
W_{k}[J,\tilde{J}] = -i \ln Z_{k}[J,\tilde{J}] .
\label{Eq:SchwingerFunctional}
\end{equation}
Any $n$-point correlation function derived from $W_{k}$ will inherit this scale dependence and by functional differentiation with respect to the external sources, we obtain the field expectation values
\begin{equation}
\phi_{k}(x) = \frac{\delta W_{k}[J,\tilde{J}]}{\delta \tilde{J}(x)} , \qquad 
\tilde{\phi}_{k}(x) = \frac{\delta W_{k}[J,\tilde{J}]}{\delta J(x)} ,
\label{Eq:FieldExpectationValues}
\end{equation}
in the presence of nonvanishing sources $J, \tilde{J}$, and $R_{k}$. Similarly, the (connected) retarded propagator and statistical correlation function read
\begin{eqnarray}
G_{k}^{\textrm R}(x,y) &=& \frac{\delta^{2} W_{k}[J,\tilde{J}]}{\delta \tilde{J}(x) \delta J(y)} , \\
F_{k}(x,y) &=& -i \frac{\delta^{2} W_{k}[J,\tilde{J}]}{\delta \tilde{J}(x) \delta \tilde{J}(y)} .
\end{eqnarray}
The advanced propagator is not independent and given by the relation $G^{\textrm A}_{k} (x,y) = \left\lbrack G^{\textrm R}_{k} (y,x) \right\rbrack^{T}$, while the anom\-a\-lous propagator $\tilde{F}_{k}(x,y) = -i \frac{\delta^{2} W_{k}[J,\tilde{J}]}{\delta J(x) \delta J(y)}$ vanishes in the absence of sources and therefore does not yield any contribution to the RG flow.

The scale-dependent generating functional of 1PI correlation functions $\Gamma_k[\phi,\tilde{\phi}]$ is defined by the (partial) Le\-gen\-dre transform of Eq.\ \eqref{Eq:SchwingerFunctional}, with respect to the sources $J$ and $\tilde{J}$, for fixed (i.e., $k$-independent) values of the fields $\phi$ and $\tilde{\phi}$:
\begin{eqnarray}
\Gamma_{k}[\phi,\tilde{\phi}] &=& W_{k}[J,\tilde{J}] - \Delta_{k} S[\phi,\tilde{\phi}] \nonumber\\
&& -\: \int_{x} \left\{ \tilde{J}_{a}(x) \phi^{a}(x) + J_{a}(x) \tilde{\phi}^{a}(x) \right\} .
\label{Eq:EffectiveAction}
\end{eqnarray}
Here, $J = J[\phi,\tilde{\phi}]$ and $\tilde{J} = \tilde{J}[\phi,\tilde{\phi}]$ are in fact $k$-dependent and are determined by inverting relations \eqref{Eq:FieldExpectationValues} for fixed $\phi$ and $\tilde{\phi}$. From Eq.\ \eqref{Eq:EffectiveAction} it is straightforward to derive the exact RG flow equation: Taking the scale derivative with respect to the logarithmic scale parameter $s = \ln ( k/\Lambda )$ (where $\Lambda$ refers to some fixed reference scale, which will typically be identified with the UV cutoff), we obtain:\footnote{By writing the flow equation in the frequency-momentum representation, we assume that the limits $t_{i} \rightarrow -\infty$ and $t_{f}\rightarrow \infty$ have been taken, thus restoring time translation invariance. Here and in the following we define the frequency-momentum integration as: $\int_{p} \cdots \equiv \frac{1}{(2\pi)^{d+1}} \int d\omega \, d^d |\boldsymbol{p}| \cdots$.}
\begin{equation}
\frac{\partial}{\partial s} \Gamma_{k} = i \int_{p} \, \Tr \left\{ \frac{\partial}{\partial s} R_{k}(\boldsymbol{p}) \Re G^{\textrm R}_{k}(\omega, \boldsymbol{p}) \right\} ,
\label{Eq:FlowEquation}
\end{equation}
where the integration runs over real frequencies and momenta, $p = (\omega, \boldsymbol{p})$, and the trace $\Tr \,\{ \cdots \}$ is evaluated over internal field indices. Both the scale derivative of the CTP effective action, on the LHS of Eq.\ \eqref{Eq:FlowEquation}, as well as the retarded propagator on the RHS are defined in the presence of arbitrary sources $J$ and $\tilde{J}$ and therefore depend on the fields. At this level the functional flow equation Eq.\ \eqref{Eq:FlowEquation} corresponds to an infinite hierarchy of coupled differential equations. Taking functional derivatives with respect to the fields, we may derive the flow equations for $n$-point functions to arbitrary order. To solve this hierarchy however, we need to close the infinite set of coupled equations which defines a particular truncation of the scale-dependent CTP effective action $\Gamma_{k}$. The so-obtained flow equations are then evaluated at the global minimum of the effective action, determined by \eqref{Eq:FieldExpectationValues} in the absence of sources, $J = \tilde{J} = 0$. Of course, any truncation of \eqref{Eq:FlowEquation} amounts to an approximation of the full theory and restricts both the possible degrees of freedom and interactions that might become relevant in the low-energy effective theory. While sophisticated truncations have been developed to solve these flow equations (see, e.g., Refs.\ \cite{Morris:1994ie, Litim:2002cf, Canet:2003qd, Blaizot:2005wd, *Blaizot:2005xy, Benitez:2009xg, Litim:2010tt, Rosten:2010vm}) typically the quality of a particular truncation can be improved considerably if one has some understanding of the relevant operators in the low-energy regime. In the following section we will consider possible truncations for the $O(N)$ model that allow us to follow the flow of the scale-dependent effective potential in the vicinity of the second order phase transition (at vanishing or nonvanishing temperature). This allows us to address the impact of the characteristic scales of the system determined by temperature and microscopic interactions and to resolve the dynamic scaling behavior.

Let us finally comment on the form of the regulator function $R_{k}(\boldsymbol{p})$. So far it seems, that we might choose any function as long as it is nonnegative. However, if we are to make sure that Eq.\ \eqref{Eq:FlowEquation} defines a functional flow between the microscopic action and CTP effective action, we need to impose additional constraints on the regulator function $R_{k}$, that enforce compatibility with these boundary conditions. In particular, the following properties should hold: $\lim_{k\rightarrow \Lambda} \Gamma_{k}[\phi,\tilde{\phi}] = S[\varphi,\tilde{\varphi}]$ and $\lim_{k\rightarrow 0} \Gamma_{k}[\phi,\tilde{\phi}] = \Gamma[\phi , \tilde{\phi}]$. Of course, when $J = \tilde{J} = 0$, the effective action should depend only on $\phi = \langle \hat{\varphi} \rangle$ while $\tilde{\phi} = 0$. These conditions can be implemented by the following requirements: $\lim_{k \rightarrow \Lambda} R_{k} \sim \Lambda^{2} \rightarrow \infty$, and \mbox{$\lim_{k \rightarrow 0} R_{k} = 0$}, respectively. Specifically, in the main part of this work, concerned with the critical dynamics at nonvanishing temperature (cf.\ Sec.\ \ref{Sec:Classical regime}), we employ the following frequency-independent regulator function
\begin{equation}
R_{k} (\boldsymbol{p}) = Z_{k}^{\bot} (k^{2} - \boldsymbol{p}^{2}) \theta( k^{2} - \boldsymbol{p}^{2} ) ,
\label{Eq:LitimRegulator}
\end{equation}
while at zero temperature we choose to work with a similar $(d+1)$-dimensional Euclidean regulator (cf.\ \mbox{Sec.\ \ref{Sec:Quantum regime}}). $Z_{k}^{\bot}$ is a scale-dependent factor which appears in the definition of our truncation for the effective action $\Gamma_{k}$ (see Sec.\ \ref{SubSec:Low-energy effective theory}). \mbox{Eq.\ \eqref{Eq:LitimRegulator}} is also known as the Litim regulator and satisfies an optimization criterion that reduces the spurious scheme dependence from the truncated flow equations \cite{Litim:2000ci, *Litim:2001up}. Compared to other regulator functions it has the clear advantage that it allows us to derive analytic expressions for the nonperturbative RG functions. However, a word of caution is in order at this point: Different choices of regulators might affect the quality of the final results. To understand the possible effects that such a choice might entail, we compare the performance of different regulators in Sec.\ \ref{Sec:Limiting behavior of renormalization group equations in the classical regime}. That is, we apply an expansion around the upper critical dimension $d_{cr} = 4$ for the classical phase transition and compare to known results from the $\epsilon$-expansion. A similar analysis is provided for the large-$N$ expansion. This will give an indication to what extent results obtained within the functional renormalization group are regulator dependent for a given truncation of the scale-dependent effective action.

Before we proceed, let us remark that we will drop the index $k$ from now on to avoid a proliferation of notation, unless it is necessary to highlight the scale dependence of parameters or couplings of the model.

\section{Relativistic $O(N)$ model}
\label{Sec:Relativistic O(N) model}

In this section, we discuss the general structure of the CTP effective action and its truncations with the objective to smoothly connect the coherent and incoherent dynamics in the UV and IR, respectively. In Sec.\ \ref{SubSec:Microscopic Theory (T = 0)} the CTP action at $T = 0$ is discussed in detail, whose structure governs the microscopic coherent dynamics at arbitrary $T$. In Sec.\ \ref{SubSec:Low-energy effective theory}, our \emph{ansatz} on the two-point propagators is explicitly constructed so that the fluctuation theorem holds exactly at the linear level. This fixes the scaling dimensions of the fields, which we provide in Sec.\ \ref{SubSec:Scaling dimensions and dynamic scaling relations}. Finally, in Sec.\ \ref{SubSec:Local interaction approximation}, we discuss consistent truncations of the vertex expansion of the CTP effective action. These considerations allow us to address the effectiveness as well as the limitations of the assumptions implicit in our \emph{ansatz}.

\subsection{Microscopic theory ($T = 0$)}
\label{SubSec:Microscopic Theory (T = 0)}

Here, we provide the CTP action for the relativistic $O(N)$ theory at zero temperature. We employ the $(\varphi_{+}, \varphi_{-})$ basis on the $\mathcal{C}_{+}\oplus\, \mathcal{C}_{-}$ contour (see Sec.\ \ref{Sec:Real-time n-point functions and generating functionals}) and start with the noninteracting theory $S = S_{0}$, for which we derive the propagators and discuss the implications of the CF-functional (cf.\ Eqs.\ \eqref{Eq:CTPAction} and \eqref{Eq:ConstraintFunctional}). This section serves essentially as an introduction to the structure of the $O(N)$ theory in the real-time formalism (see also Ref.\ \cite{Kamenev:2011} which provides an extensive introduction to real-time methods in field theory). The propagators are defined and their properties outlined. While most of them generalize also beyond perturbation theory, care is taken to clearly highlight these properties and distinguish them from the assumptions valid only in the perturbative regime. At the example of the free theory, we argue that the effect of the boundary constraints and initial fluctuations can be fully captured by introducing an effective CTP action. The free action can be written in terms of contributions from the two distinct branches of the contour plus an additional contribution that takes the form of a mixing term $\sim \varphi_{+} \varphi_{-}$:
\begin{widetext}
\begin{equation}
S_{0}[\varphi_{+},\varphi_{-}] = \frac{1}{2} \int_{p} \left( \varphi_{+}(-p) , \, \varphi_{-}(-p) \right) \begin{pmatrix} \omega^{2} + i 0^{+} - \omega^{2}(\boldsymbol{p}) & - i 0^{+} (\sgn(\omega)  + 1) \\ i 0^{+} (\sgn(\omega) - 1) & -\left( \omega^{2} - i 0^{+} - \omega^{2}(\boldsymbol{p}) \right) \end{pmatrix} \begin{pmatrix} \varphi_{+}(p) \\ \varphi_{-}(p) \end{pmatrix} , \label{Eq:FreeActionPMStateBasis}
\end{equation}
\end{widetext}
where $\omega(\boldsymbol{p}) = \pm \sqrt{\boldsymbol{p}^{2} + m^{2}}$ defines the characteristic frequency of the modes. Note, that the spectrum of characteristic frequencies and propagators of \eqref{Eq:FreeActionPMStateBasis} are invariant under $O(N)$ symmetry (internal indices are not explicitly written out, unless necessary). This is by no means obvious: In fact, a naive doubling of the degrees of freedom on the CTP would yield a theory which is invariant under independent $O(N)$-rotations of the $\varphi_{+}$ and $\varphi_{-}$ fields. However, such a symmetry is unphysical -- the fields on the upper and lower branch of the CTP are not independent due to the presence of the boundary conditions that enter in the construction of the path integral, cf.\ Sec.\ \ref{Sec:Real-time n-point functions and generating functionals}. We include the nondiagonal mixing contributions $\sim \varphi_{+,a} \varphi_{-}^{a}$ in Eq.\ \eqref{Eq:FreeActionPMStateBasis} precisely to account for the nonvanishing correlation of the fields and to constrain the spectrum of the theory.\footnote{This situation is very much reminiscent of the situation that one encounters in the replica approach to spin glasses \cite{Mezard:1987}. As an example consider two identical copies of a classical lattice Ising-spin model $\{ s_{i}^{\alpha} \}$, where the Greek index $\alpha$ labels the different replicas, while the Latin indices $i,j$ label the sites of the discrete spatial lattice. The Hamiltonian, $H = \sum_{\alpha = 1,2} \sum_{i,j} J_{i j} s_{i}^{\alpha} s_{j}^{\alpha} - \varepsilon \sum_{i} s_{i}^{1} s_{i}^{2}$, is defined via the exchange couplings $J_{i j}$, which are not necessarily nearest-neighbor and randomly alternate in sign. It includes an additional mixing term which introduces nonvanishing correlations between replicas. Depending on the specific choice for the probability distribution of the couplings, one might observe a large degeneracy of equilibrium states for such a spin system at sufficiently low temperatures. The correlations between different replicas define an order parameter $q^{\alpha \beta}(\varepsilon) \sim \sum_{i} \langle s_{i}^{\alpha} s_{i}^{\beta} \rangle$ for the low-temperature glassy phase \cite{Parisi:2002}. The existence of one or more (coexisting) equilibrium states can be monitored by observing if $q^{\alpha \beta}(\varepsilon)$ is continuous or discontinuous when $\varepsilon$ changes sign \cite{Parisi:1997}, and the presence of a discontinuity is related to the spontaneous breaking of the replica symmetry $q^{\alpha \beta} (0) = q^{\beta \alpha} (0)$ \cite{Bray:1978,Mezard:1983ce,Parisi:1983dx}. In the context of the CTP action the replica symmetry breaking is expressed on the level of the frequency-dependent propagators, i.e., $G_{\alpha \beta}(\omega) \neq G_{\beta \alpha}(\omega)$, $\alpha, \beta \in \{ +, - \}$.}
Clearly, their effect is identical to imposing the boundary constraint $\varphi_{+}^{a}(t = t_{f}) = \varphi_{-}^{a}(t = t_{f})$ at final time $t_{f}$.

The presence of the infinitesimal $i0^{+}$ contributions furthermore ensures the proper causality structure of the propagators and enforces a relation between the positive and negative frequency modes. In the given field-representation the free propagators read:
\begin{equation}
G(p) = \begin{pmatrix} G_{++}(p) & G_{+-}(p) \\ G_{-+}(p) & G_{--}(p) \end{pmatrix},
\label{Eq:FreePropagatorPMStateBasis}
\end{equation}
where
\begin{eqnarray}
G_{++}(p) &=& \frac{-1}{\omega^{2} - \omega^{2}(\boldsymbol{p}) + i 0^{+}} , \\
G_{+-}(p) &=& 2 \pi i\, \theta(\omega) \delta(\omega^{2} - \omega^{2}(\boldsymbol{p})) ,
\end{eqnarray}
and the following symmetry relations hold:
\begin{eqnarray}
G_{++}(p) &=& - \lbrack G_{--}(p) \rbrack^{\dagger} ,
\label{Eq:SymmetryRelationsFreePropagatorA} \\
G_{+-}(p) &=& G_{-+}(-p) .
\label{Eq:SymmetryRelationsFreePropagatorB}
\end{eqnarray}
The diagonal elements of the propagator \eqref{Eq:FreePropagatorPMStateBasis} correspond to the usual time-ordered (and reverse time-ordered) Green functions. However, not all of the components are independent. Apart from the relations \eqref{Eq:SymmetryRelationsFreePropagatorA} and \eqref{Eq:SymmetryRelationsFreePropagatorB} the following algebraic identity holds
\begin{equation}
G_{++}(p) + G_{--}(p) = G_{+-}(p) + G_{-+}(p) ,
\label{Eq:AlgebraicIdentity}
\end{equation}
as one may easily verify. Eqs.\ \eqref{Eq:SymmetryRelationsFreePropagatorA} -- \eqref{Eq:AlgebraicIdentity} reduce the number of independent degrees of freedom and are valid also beyond perturbation theory. Given these relations one might argue that the propagator \eqref{Eq:FreePropagatorPMStateBasis} can be expressed in terms of three independent real-valued components only. However, the presence of a fluctuation theorem \cite{Callen:1951vq, Kubo:1957mj, *Kubo:1966} and the dispersion relation \cite{Toll:1956cya} further reduces the number of independent degrees of freedom. Both of these relations are most easily derived in the retarded/advanced (RA) basis. Therefore, as we have done already in Secs.\ \ref{Sec:Real-time n-point functions and generating functionals} and \ref{Sec:Nonperturbative functional renormalization group in the real-time formalism} we will switch to this basis, where the properties of the action and the consequences of the $i 0^{+}$ insertion become somewhat more transparent. In that basis, the free action takes the following form
\begin{widetext}
\begin{equation}
S_{0}[\varphi,\tilde{\varphi}] = \frac{1}{2} \int_{p} ~ \big( \varphi(-p) , \, \tilde{\varphi}(-p) \big)\, \begin{pmatrix} 0 & ( \omega + i 0^{+} )^{2} - \omega^{2}(\boldsymbol{p}) \\ ( \omega - i 0^{+} )^{2} - \omega^{2}(\boldsymbol{p}) & i 0^{+} \end{pmatrix} \begin{pmatrix} \varphi(p) \\ \tilde{\varphi}(p) \end{pmatrix} .
\label{Eq:FreeFieldTheory} 
\end{equation}
\end{widetext}
Here, we see that the infinitesimal regulating contribution appears as a positive-definite imaginary contribution $\sim \tilde{\varphi}^{2}$ to the action. Not only does this guarantee the convergence of the generating functional, but it also ensures that the dynamic correlations of the system are related to the response with respect to an external perturbation. In the RA-representation, the free propagator is given by
\begin{equation}
G(p) = \begin{pmatrix} i F(p) & G^{\textrm{A}}(p) \\ G^{\textrm{R}}(p) & 0 \end{pmatrix} ,
\label{Eq:FreePropagatorRetardedAdvancedBasis}
\end{equation}
where
\begin{eqnarray}
G^{\textrm{R}}(p) &=& \frac{-1}{\omega^{2} - \omega^{2}(\boldsymbol{p})} + i \pi \sgn(\omega)\, \delta(\omega^{2} - \omega^{2}(\boldsymbol{p})) , 
\label{Eq:RetardedPropagatorLambda}
\\
F(p) &=& \pi \delta(\omega^{2} - \omega^{2}(\boldsymbol{p})) ,
\label{Eq:CorrelationFunctionLambda}
\end{eqnarray}
define the free retarded propagator and statistical correlation function, respectively. By conjugation symmetry \eqref{Eq:ConjugationSymmetryEffectiveAction} the advanced propagator $G^{\textrm{A}}$ is seen to satisfy: 
\begin{equation}
G^{\textrm{A}}(p) = \lbrack G^{\textrm{R}}(p) \rbrack^{\dagger} .
\end{equation}
All components in \eqref{Eq:FreePropagatorRetardedAdvancedBasis} are either purely real or imaginary, i.e., $\lbrack G^{\textrm{R/A}}(p) \rbrack^{\dagger} = G^{\textrm{R/A}}(-p)$ and $\lbrack i F(p) \rbrack^{\dagger} = -i F(-p)$, in contrast to the propagators in the $(\varphi_{+}, \varphi_{-})$ basis. This is one of the advantages of expressing the action in the form \eqref{Eq:FreeFieldTheory} -- it clearly emphasizes the relevant degrees of freedom. Furthermore, from the propagators \eqref{Eq:RetardedPropagatorLambda} and \eqref{Eq:CorrelationFunctionLambda} we may easily read off the following relation
\begin{equation}
F(\omega, \boldsymbol{p}) = \sgn(\omega) \Im G^{\textrm{R}}(\omega, \boldsymbol{p}) = \frac{\rho(\omega, \boldsymbol{p})}{2 |\omega|}  ,
\label{Eq:FluctuationDissipationTheoremRetardedAdvanced}
\end{equation}
where in the last step, we have defined the spectral density $\rho(\omega, \boldsymbol{p})$. In fact, this relation is exact -- it holds for any theory that is prepared in the groundstate ($T=0$). Eq.\ \eqref{Eq:FluctuationDissipationTheoremRetardedAdvanced} is a fluctuation theorem which relates the correlation for zero point fluctuations $F(\omega,\boldsymbol{p})$ to the spectral density (or alternatively the imaginary part of the response function). When the system is in its groundstate, all it can do is to absorb energy and without any coupling to a dissipative environment, the only possibility for non-zero fluctuations is to allow for virtual particle-antiparticle pair creation (and annihilation). This point of view can is supported by the following derivation which starts from the relation
\begin{equation}
\int_{\omega} \left( \alpha + |\omega| \right)^{2} F_{a}(\omega, \boldsymbol{p}) \geq 0 , \label{Eq:IntegratedFRelation}
\end{equation}
where the correlation function is expressed in the mass eigenbasis, $F_{a b}(p) = F_{a}(p) \delta_{a b}$. The components $F_{a}(p)$, $a = 1, \ldots, N$, are real which follows from the reflection symmetry $\lbrack F(p) \rbrack^{\dagger} = F(-p)$, while positivity follows from the stability of the action.\footnote{In the mass eigenbasis, where the propagators are diagonal, the statistical correlation function can be written as:
\begin{displaymath}
F_{a}(p) = \big\vert G^{\textrm{R}}_{a}(p) \big\vert^{2} \frac{(2\pi)^{d+1}}{\vol_{d+1}} \frac{1}{i} \frac{\delta^{2} \Gamma[\phi,\tilde{\phi}]}{\delta \tilde{\phi}_{a}(-p) \delta \tilde{\phi}_{a}(p)} > 0 ,
\end{displaymath}
where $\Gamma$ is the effective action (and corresponds to $S_{0}$ in the free theory).} Exploiting these properties, we may write Eq.\ \eqref{Eq:IntegratedFRelation} which is obviously satisfied for any real-valued parameter $\alpha$. In particular, it holds true at the minimum (with respect to $\alpha$) for which we obtain
\begin{equation}
\left\{ \int_{\omega} F_{a}(\omega, \boldsymbol{p}) \right\} \left\{ \int_{\omega} \omega^{2} F_{a}(\omega, \boldsymbol{p}) \right\}  \geq \left\{ \int_{\omega} |\omega| F_{a}(\omega, \boldsymbol{p}) \right\}^{2}.
\label{Eq:UncertaintyRelation}
\end{equation}
We use the following identities
\begin{eqnarray}
&& \hspace{-30pt} \Delta \varphi_{a}^{2}(\boldsymbol{p}) \equiv \int_{\omega} F_{a}(\omega,\boldsymbol{p}) \nonumber\\ &=& \frac{1}{2} \int_{\boldsymbol{x}} \left. \left\langle \left\{ \hat{\varphi}_{a}(x^{0},\boldsymbol{x}) , \hat{\varphi}_{a}(y^{0},\boldsymbol{0}) \right\} \right\rangle \right\vert_{x^{0} = y^{0}} e^{-i \boldsymbol{p} \cdot \boldsymbol{x}} , \\
&& \hspace{-30pt} \Delta \pi_{a}^{2}(\boldsymbol{p}) \equiv \int_{\omega} \omega^{2} F_{a}(\omega,\boldsymbol{p}) \nonumber\\
&=& \frac{1}{2} \int_{\boldsymbol{x}} \left. \left\langle \left\{ \hat{\pi}_{a}(x^{0},\boldsymbol{x}) , \hat{\pi}_{a}(y^{0},\boldsymbol{0}) \right\} \right\rangle \right\vert_{x^{0} = y^{0}} e^{-i \boldsymbol{p} \cdot \boldsymbol{x}} ,
\end{eqnarray}
where $\hat{\pi}(x^{0},\boldsymbol{x}) = \frac{\partial}{\partial x^{0}} \hat{\varphi}(x^{0},\boldsymbol{x})$ defines the conjugate momentum to $\hat{\varphi}$. Together with the fluctuation theorem \eqref{Eq:FluctuationDissipationTheoremRetardedAdvanced} we write
\begin{equation}
\int_{\omega} |\omega| F_{a}(\omega, \boldsymbol{p}) = \frac{1}{2} \int_{\omega} \rho_{a}(\omega, \boldsymbol{p}) = \frac{1}{2} .
\label{Eq:FieldCommutatorRelation}
\end{equation}
In the last step of Eq.\ \eqref{Eq:FieldCommutatorRelation} we have employed the sum rule for the spectral density (see below), which follows by using the definition of the spectral function
\begin{equation}
\hspace{-3pt} \int_{\omega} \rho_{a}(\omega,\boldsymbol{p}) = -\frac{i}{2} \int_{\boldsymbol{x}} \left. \left\langle \left\lbrack \hat{\varphi}_{a}(x^{0},\boldsymbol{x}) , \hat{\pi}_{a}(y^{0},\boldsymbol{0}) \right\rbrack \right\rangle \right\vert_{x^{0} = y^{0}} e^{-i \boldsymbol{p} \cdot \boldsymbol{x}} ,
\end{equation} 
and the canonical equal-time commutation relations for the field and conjugate momentum operators. This allows us to express Eq.\ \eqref{Eq:UncertaintyRelation} in the well-known form:
\begin{equation}
\Delta \varphi_{a}^{2}(\boldsymbol{p}) \Delta \pi_{a}^{2}(\boldsymbol{p}) \geq \frac{1}{4} .
\label{Eq:HeisenbergUncertaintyRelation}
\end{equation}
Finally, using the above expressions for the free propagators \eqref{Eq:RetardedPropagatorLambda} and \eqref{Eq:CorrelationFunctionLambda} one may easily check that the uncertainty relation is in fact saturated in the groundstate, where $\Delta \varphi_{a}^{2}(\boldsymbol{p}) = 1/(2 \omega_{a}(\boldsymbol{p}))$. Thus, for any finite momentum the groundstate features nonvanishing zero-point fluctuations, which shows that Eq.\ \eqref{Eq:FluctuationDissipationTheoremRetardedAdvanced} is satisfied nontrivially.

For completeness, we also provide the following frequency integrals:
\begin{eqnarray}
\int_{\omega} \frac{\rho_{a}(\omega, \boldsymbol{p})}{2 \omega} = 0 , \qquad \int_{\omega} \rho_{a}(\omega,\boldsymbol{p}) = 1 ,
\label{Eq:SpectralSumRule}
\end{eqnarray}
that follow from the canonical equal-time commutation relations. In general, Eqs.\ \eqref{Eq:FieldCommutatorRelation}, \eqref{Eq:HeisenbergUncertaintyRelation}, and the normalization of the spectral function \eqref{Eq:SpectralSumRule} hold true also in the interacting theory. This applies both to the bare and renormalized fields, for which the kinetic term has coefficient one in the effective action.\footnote{However, note that certain truncations of the effective action (encountered in the framework of the functional RG, cf.\ \mbox{Sec.\ \ref{SubSec:Low-energy effective theory}}) might not retain the full spectrum of the theory, and in these cases the sum rule \eqref{Eq:SpectralSumRule} has to be modified accordingly. In particular, this depends on the quality of the employed approximation for the frequency-dependence of the self-energy.}

Apart from the fluctuation theorem, there is a dispersion relation that relates the real and imaginary parts of response function. It is a direct consequence of analyticity of the retarded (advanced) propagator in the upper (lower) half of the complex plane
\begin{equation}
\frac{1}{2}\Re G^{\textrm{R}} (\omega,\boldsymbol{p}) = - \PV \int_{\omega'} \frac{\Im G^{\textrm{R}}(\omega',\boldsymbol{p})}{\omega - \omega'} ,
\end{equation}
where $\PV$ denotes the Cauchy principal value. The reader may easily verify that this relation is satisfied by the free propagators, but it also holds for the exact propagators. Thus, it is clear that the propagator \eqref{Eq:FreePropagatorRetardedAdvancedBasis} of any system prepared in its ground state can be described in terms of a single real-valued function. In fact, this statement is also true at nonvanishing temperatures (see Sec.\ \ref{SubSec:Low-energy effective theory}). This greatly simplifies the diagrams that one needs to evaluate in perturbative calculations and similarly, in the framework of the nonperturbative functional RG.

Let us move on to the interacting case. The vertices in the CTP action $S = S_{0} + S_{\lambda}$ derive directly from the construction of the theory in the RTF illustrated in Sec.\ \ref{Sec:Real-time n-point functions and generating functionals}. For the $\lambda \varphi^{4}$ theory, we obtain two vertices in the physical RA-representation, parametrized by the same coupling $\lambda$, but different prefactors:
\begin{eqnarray}
S_{\lambda}[\varphi,\tilde{\varphi}] &=& - \frac{\lambda}{3!} \int_{x} \!\bigg\{ \tilde{\varphi}^{a}(x) \varphi_{a}(x) \varphi^{2}(x) \nonumber\\ &&  +\: \frac{1}{4} \varphi^{a}(x) \tilde{\varphi}_{a}(x) \tilde{\varphi}^{2} (x) \bigg\} .
\label{Eq:Interactions}
\end{eqnarray}
Their origin derives from the field degrees of freedom that propagate on the two distinct branches of the time contour $\mathcal{C}$ (cf.\ Sec.\ \ref{Sec:Real-time n-point functions and generating functionals}) which can be seen explicitly, by changing the field representation:
\begin{equation}
S_{\lambda}[\varphi_{+},\varphi_{-}] = - \frac{\lambda}{4!} \int_{x} \left\{ \varphi_{+}^{4}(x) - \varphi_{-}^{4} (x) \right\} .
\end{equation}
The additional minus sign originates from the reverse time-ordering along the $\mathcal{C}_{-}$ path. Thus, the two coupling terms in the RTF are simply inherited from the doubling of the field degrees of freedom on the $\mathcal{C}_{+} \oplus\, \mathcal{C}_{-}$ time integration contour.

In the presence of interactions, both the mass parameter and the coupling of the theory are renormalized. This is due to the coherent interaction of bare particles with excitations from the vacuum and can be compensated by absorbing the associated UV divergences in appropriate counterterms. At nonvanishing temperature, however, we expect an additional IR renormalization which is due to incoherent collisions with renormalized quasiparticles. These two processes lead to two different scales in the problem, which are characterized by the scattering length and mean free path of the system. The purpose of the following two sections is to illustrate the effect of these competing processes on the basis of an appropriate truncation of the CTP effective action. Eqs.\ \eqref{Eq:FreeFieldTheory} -- \eqref{Eq:Interactions} define our microscopic model that we will use as a starting point for the RG investigation.

\subsection{Low-energy effective theory}
\label{SubSec:Low-energy effective theory}

In the framework of the nonperturbative functional RG, we employ an \emph{ansatz} for the scale-dependent CTP effective action, which corresponds to a suitable truncation of the gradient expansion and an expansion in field monomials (see, e.g., Refs.\ \cite{Morris:1994ki,Morris:1997xj}). Projecting the flow \eqref{Eq:FlowEquation} onto those contributions included within our \emph{ansatz}, we obtain a finite set of RG equations for the corresponding parameters and couplings. This necessarily constitutes an approximation of the otherwise exact flow equation. Here, we provide the specific form of the truncated CTP effective action in the $O(N)$ model and discuss the properties of the physical spectrum and dynamical behavior as a result of the finite truncation. Fluctuations in the equilibrium state are reflected in the CTP effective action and it will therefore depend on temperature. We consider different temperature regimes and discuss how they are reflected within our \emph{ansatz}.

In the real-time formalism, the CTP effective action is most conveniently expressed in the frequency-momentum representation. We work in the RA-basis and assume that the effective action takes the following form
\begin{widetext}
\begin{equation}
\Gamma[\phi,\tilde{\phi}] = \frac{1}{2} \int_{p} \big( \phi(-p) , \, \tilde{\phi}(-p) \big) \begin{pmatrix} 0 &  Z^{|\!|} \omega^{2} - Z^{\bot} \boldsymbol{p}^{2} + i \Omega \omega / \beta \\ Z^{|\!|} \omega^{2} - Z^{\bot} \boldsymbol{p}^{2} - i \Omega \omega / \beta  & i ( \Omega \omega / \beta ) \coth (\beta \omega / 2)  \end{pmatrix} \! \begin{pmatrix} \phi (p) \\ \tilde{\phi} (p) \end{pmatrix} - \int_{x} \, \mathcal{U}(\phi,\tilde{\phi}) ,
\label{Eq:1PIGeneratingFunctionalAnsatz}
\end{equation}
\end{widetext}
to second order in the gradient expansion. In contrast to the spatial momenta, an infinite series of frequency terms is taken into account, which is necessary to implement the fluctuation-dissipation theorem \cite{Callen:1951vq, Kubo:1957mj, *Kubo:1966} at nonvanishing temperature exactly at the level of the propagators (cf.\ \eqref{Eq:FDT}). The systematic construction of more elaborate truncations of the equilibrium effective action in the RTF significantly benefits from symmetry considerations \cite{Sieberer:2015hba}. In particular, one may show that fluctuation relations between different $(m,n)$-point functions, $\Gamma^{(m,n)} \equiv \frac{\delta^{m+n} \Gamma}{\delta \phi^{m} \delta\tilde{\phi}^{n}}$, that characterize the properties of the equilibrium state, follow as Ward-Takahashi identities associated with time-reversal symmetry \cite{Altland:2010,Gaspard:2013,Sieberer:2015hba}.

The generalized potential $\mathcal{U} = \mathcal{U}(\phi,\tilde{\phi})$ includes terms that appear to lowest order in the frequency and momentum expansion, while the wavefunction renormalization $Z^{\bot}$ and the renormalization factor $Z^{|\!|}$ parametrize those terms that enter at quadratic order in momentum and frequency, respectively. The coefficient $\Omega$ measures the strength of the imaginary contributions to the 1PI two-point functions $\Gamma^{(1,1)}$ and $\Gamma^{(0,2)}$. All renormalization factors $Z^{\bot}$, $Z^{\vert\!\vert}$, and $\Omega$ are assumed to be field-independent but depend on the RG scale parameter $s$. The consistency of Eq.\ \eqref{Eq:1PIGeneratingFunctionalAnsatz} requires that $\Omega > 0$, $\beta > 0$, a constrained imposed by the stability of the effective action, and $Z^{\vert\!\vert}$, $Z^{\bot} \geq 0$.\footnote{Note, that our definition of the renormalization constants is different from the standard convention, i.e., $Z^{\bot}$ is defined as the inverse of the field renormalization defined in Refs.\ \cite{Itzykson:1980rh,Weinberg:1995mt}.} $\mathcal{U}(\phi,\tilde{\phi})$ is also scale dependent and defines the mass parameters and couplings of the theory. That is, the spectrum of the theory is given in terms of the $N$ eigenvalues of the mass matrix squared:\footnote{Here, we define the two-point functions in the frequency-momentum representation with an appropriate factor for normalization:
\begin{eqnarray}
&& \hspace{-10pt} \Gamma^{(1,1)}(p) = \big( \Gamma^{(1,1)}(-p) \big)^{\dagger} = \frac{(2\pi)^{d+1}}{\vol_{d+1}} 
\frac{\delta^{2} \Gamma[\phi,\tilde{\phi}]}{\delta \tilde{\phi}(-p) \delta \phi(p)} , \\
&& \hspace{-10pt} \Gamma^{(0,2)}(p) = \frac{(2\pi)^{d+1}}{\vol_{d+1}} 
\frac{\delta^{2} \Gamma[\phi,\tilde{\phi}]}{\delta \tilde{\phi}(-p) \delta \tilde{\phi}(p)} .
\end{eqnarray}}
\begin{equation}
m_{a b}^{2} \equiv - \lim_{p \rightarrow 0}  \left. \Gamma^{(1,1)}_{a b}(p) \right\vert_{\min} = \left. \frac{\partial^{2} \mathcal{U}}{\partial \phi^{a} \partial \tilde{\phi}^{b}} \right\vert_{\min},
\label{Eq:MassSpectrum}
\end{equation}
where the second functional derivative is evaluated at the global minimum of the effective potential:
\begin{equation}
\left. \frac{\partial V}{\partial \phi^{a}} \right\vert_{\min} \equiv - \lim_{p \rightarrow 0} \left. \Gamma^{(0,1)}_{a}(p) \right\vert_{\min} = 0 .
\end{equation}
In equilibrium, the field configuration at the minimum of the effective potential is homogeneous and, without loss of generality, we assume that the field expectation value points in the $1$-direction:
\begin{equation}
\phi_{a} = v \delta_{a 1} , \quad \tilde{\phi}_{a} = 0 .
\label{Eq:EffectivePotentialMinimum}
\end{equation}
The propagators are most conveniently expressed in the basis, where the mass matrix is diagonal, i.e., $m^{2}_{a b} = m^{2}_{a} \delta_{a b}$, and the two-point functions are diagonal. In particular, we have
\begin{equation} 
\Gamma^{(1,1)}_{a b}(p) \, \big|_{\min} = \left\lbrack Z^{|\!|} \omega^{2} - Z^{\bot} \omega_{a}^{2}(\boldsymbol{p}) - i \Omega \omega / \beta \right\rbrack \delta_{a b} ,
\label{Eq:2ptFunctionMinimumA}
\end{equation}
with the characteristic field-dependent frequencies $\omega_{a}(\boldsymbol{p}) = \pm \sqrt{\boldsymbol{p}^{2} + m_{R, a}^{2}(v)}$, $a = 1, \ldots , N$, expressed in terms of the renormalized masses $m_{R, a}^{2} = m_{a}^{2}/ Z^{\bot}$. The two-point function $\Gamma^{(2,0)}$ is exactly zero, while
\begin{equation}
\Gamma^{(0,2)}_{a b}(p) \, \big|_{\min} = i ( \Omega \omega / \beta ) \coth (\beta \omega / 2) \delta_{a b} ,
\label{Eq:2ptFunctionMinimumB}
\end{equation}
at the minimum of the effective potential \eqref{Eq:EffectivePotentialMinimum}. Within our truncation there are no field-dependent contributions to $\Gamma^{(0,2)}$. This is enforced by two requirements: (1) that the linear frequency contribution should be field-independent and (2) that a fluctuation-dissipation theorem should hold. Before we comment on the properties of fluctuations and consistent truncations of the CTP effective action, we provide the general form of the propagators within our \emph{ansatz} \eqref{Eq:1PIGeneratingFunctionalAnsatz}. In the basis where the mass-matrix is diagonal, the retarded propagator reads
\begin{equation}
G^{\textrm{R}}_{a b} (p) = \frac{-1}{Z^{\vert\!\vert} \omega^{2} - Z^{\bot} \omega_{a}^{2}(\boldsymbol{p}) - R(\boldsymbol{p}) + i \Omega \omega /\beta} \, \delta_{a b},
\label{Eq:RetardedPropagatorAnsatz}
\end{equation}
and the statistical correlation function takes the form
\begin{equation}
F_{a b} (p) = \frac{(\Omega \omega/\beta) \coth ( \beta \omega / 2 )}{\left\lbrack Z^{\vert\!\vert} \omega^{2} - Z^{\bot} \omega_{a}^{2}(\boldsymbol{p}) - R(\boldsymbol{p}) \right\rbrack^{2} + \left( \Omega \omega /\beta \right)^{2}} \, \delta_{a b} .
\label{Eq:StatisticalPropagatorAnsatz}
\end{equation}
In contrast to the free propagators, they include the momentum-dependent function $R(\boldsymbol{p})$ in the denominator which provides a regularization of massless modes. The propagators corresponding to the 1PI CTP effective action are obtained only in the limit when the RG scale parameter is removed $k = 0$ and all modes have been taken into account, i.e., $R(\boldsymbol{p}) = 0$.

With the given form of the scale-dependent propagators it is easy to check that the following fluctuation-dissipation theorem (FDT) is satisfied:
\begin{equation}
F_{a b} (\omega, \boldsymbol{p}) = \frac{1}{\omega} \left( n(\omega) + \frac{1}{2} \right) \rho_{a b}(\omega,\boldsymbol{p}) ,
\label{Eq:FDT}
\end{equation}
where $n(\omega) = (e^{\beta \omega} - 1)^{-1}$ is the thermal occupation number, and the spectral density is given by 
\begin{equation}
\rho_{a b}(\omega,\boldsymbol{p}) = 2 \omega \Im G^{\textrm R}_{a b}(\omega, \boldsymbol{p}) .
\end{equation}
Let us briefly comment on the properties of Eq.\ \eqref{Eq:FDT} and its relation to the fluctuation theorem stated earlier in Sec.\ \ref{SubSec:Microscopic Theory (T = 0)}. As far as our setup goes, we consider a closed system at a nonvanishing temperature, i.e., it is understood that different macroscopic subsets of the system are mutually in equilibrium. In a thermal state, we have the possibility that a given subsystem might fluctuate and exchange energy or particles with another subsystem. The total energy is conserved in this process and thus, strictly speaking, there is no dissipation. However, when degrees of freedom outside a given subset are integrated out, their combined effect might yield a dissipative coupling to a (classical) random potential in the resulting effective theory (see, e.g., Ref.\ \cite{Feynman:1963fq}). In a similar spirit, the successive integrating out of modes that underlies the Wilsonian RG \cite{Wilson:1973jj} provides a low-energy effective description of the dynamics, where dissipation appears naturally through thermal fluctuations. Our aim is to show how the relativistic $O(N)$ vector model acquires a dissipative coupling and in what way it determines the dynamic universality class when the system is tuned to a second-order phase transition. To understand the transition from unitary to dissipative dynamics Eq.\ \eqref{Eq:1PIGeneratingFunctionalAnsatz} should interpolate between zero-point fluctuations at the smallest scales and macroscopic thermal fluctuations that appear only through the RG averaging procedure. This is achieved by taking into account the exact form of the fluctuation theorem \eqref{Eq:FDT} in the framework of the CTP effective action. Indeed, taking the limit $\beta = 1/T \rightarrow \infty$, we see that the FDT reduces to the fluctuation theorem provided in Sec.\ \ref{SubSec:Microscopic Theory (T = 0)} and by construction, the linear FDT is implemented exactly within our \emph{ansatz}. Similar fluctuation theorems also hold for higher $n$-point functions characterizing the nonlinear response of the system \cite{Wang:1998wg}. We will come back to the nonlinear version of the FDT in Sec.\ \ref{SubSec:Local interaction approximation}, when we construct consistent truncations of the functional RG.

Equation \eqref{Eq:1PIGeneratingFunctionalAnsatz} constitutes the simplest possible \emph{ansatz} that allows us to understand the competition between coherent propagation of particles and dissipation. It is important to understand the limitations of such an approximation. For that purpose let us take a closer look at the limits of the effective CTP action and the associated dynamic behavior. At $T = 0$ the CTP effective action \eqref{Eq:1PIGeneratingFunctionalAnsatz} maps onto the free microscopic action \eqref{Eq:FreeFieldTheory}, up to wavefunction renormalization factors $Z^{\vert\!\vert}$ and $Z^{\bot}$, that satisfy $Z^{\vert\!\vert} = Z^{\bot} = 1$ in the noninteracting case. In the presence of interactions -- when all fluctuations have been integrated out -- the phase structure of the model is captured fully by the parametrization of the effective potential, while the dynamics is governed by $N$ propagating modes
\begin{equation}
\omega = \pm \big( Z^{\bot} / Z^{\vert\!\vert} \big)^{\frac{1}{2}} \,\omega_{a}(\boldsymbol{p}) + \frac{i}{2} \big( \Omega / Z^{\vert\!\vert} \big) \beta^{-1} + \mathcal{O}\left( \beta^{-3}\right) .
\label{Eq:LowTRegime}
\end{equation}
In the low-temperature regime the characteristic frequencies are renormalized, i.e., to leading order: $\omega_{R,a}(\boldsymbol{p}) = \pm \big( Z^{\bot} / Z^{\vert\!\vert} \big)^{\frac{1}{2}} \,\omega_{a}(\boldsymbol{p}) + \mathcal{O}\left( \beta^{-1}\right)$. A nonvanishing imaginary part corresponding to a finite dissipative width appears at subleading order in the low-temperature expansion. If $\Omega/Z^{\vert\!\vert}$ is sufficiently small and $Z^{\bot}$, $Z^{\vert\!\vert} \simeq 1$, one may speak of well-defined single-particle excitations (quasiparticles) with a nonvanishing mass and finite lifetime.

This is in contrast to the high-temperature limit, where we are concerned with the dynamics of collective modes
\begin{equation}
\omega = i \!\left( Z^{\bot} / \Omega \right) \omega_{a}^{2}(\boldsymbol{p}) \beta + \mathcal{O}\left( \beta^{3} \right) .
\label{Eq:HighTRegime}
\end{equation}
Their renormalized characteristic frequencies are given by $\omega_{R,a}(\boldsymbol{p}) = \big( Z^{\bot} / \Omega \big) \,\omega_{a}^{2}(\boldsymbol{p}) \beta$.\footnote{Note, that another set of finite-frequency modes decouples in the limit of small $\beta$ and does not contribute to the low-energy dynamics.} We first consider the hydrodynamic regime $|\boldsymbol{p}| \xi \ll 1$ in the disordered phase, where the correlation length  $\xi = m_{R}^{-1}$ is finite. We observe that the characteristic frequencies do not vanish in this regime -- any long-wavelength excitation will relax locally to the thermal state, with the typical relaxation time $\tau = \omega_{R}^{-1} \sim \xi^{2}$. Thus, the hydrodynamics in the disordered phase is described completely in terms of non-propagating relaxational modes. As we lower the temperature to its critical value, the correlation length $\xi$ diverges and the order parameter experiences critical slowing down $\tau \sim \xi^{z}$ (for $\boldsymbol{p} = 0$), where $z$ is the dynamic critical exponent. The value of this exponent depends critically on the presence of slow (massless) modes and on whether they couple to the order parameter \cite{DeDominicis:1977fw,Hohenberg:1977ym}. While the presence of such massless modes is ruled out in the disordered phase (assuming that there are no conserved quantities that couple to the fields), this is not so in the symmetry-broken phase, where $N - 1$ Nambu-Goldstone modes govern the low-energy dynamics. It is these modes that will -- in principle -- couple to the order parameter and determine the dynamic universality class of the phase transition \cite{Hohenberg:1977ym,Tauber:2014}.

In effective theories for critical dynamics, this mode-coupling appears as a phenomenological input from hydrodynamics \cite{Dzyaloshinskii:1980}. However, within a first-principles approach, that starts from the microscopic action and attempts to connect to the low-energy dynamics, it is clear that they must follow from general symmetry considerations, that is, the fulfillment of conservation laws and Ward identities \cite{Baym:1961zz,Baym:1962sx}. Within the framework of the functional RG these identities are satisfied only with the accuracy of the employed truncation \cite{Katanin:2004,Enss:2005} and in that sense, our \emph{ansatz} \mbox{Eq.\ \eqref{Eq:1PIGeneratingFunctionalAnsatz}} defines a nonconserving approximation to the low-energy dynamics.

We therefore expect that within our approximation the low-energy dynamics in the vicinity of the continuous phase transition should follow that of \mbox{Model A} \cite{Halperin:1972,Hohenberg:1977ym} which is defined in terms of a nonconserved order parameter in the presence of thermal noise. This makes it clear that our truncation will not allow us to distinguish different types of relaxational behavior (as required, e.g., to identify the proper dynamic universality class for the given microscopic model). Nevertheless, Eq.\ \eqref{Eq:1PIGeneratingFunctionalAnsatz} serves as a simple, well-controlled starting point to explore the RG flow in the RTF and to identify how dissipative dynamics emerges in the effective description at large scales. We leave a systematic study of different truncations and their associated possibilities for the dynamic critical behavior for future work.

\subsection{Scaling dimensions and dynamic scaling relations}
\label{SubSec:Scaling dimensions and dynamic scaling relations}

It is instructive to consider the dimensions of the fields and parameters that appear in the microscopic theory and their scaling behavior when fluctuations are taken into account. The scaling dimension $\Delta_{\Phi} \equiv \lbrack \Phi \rbrack$ of the (possibly composite) scaling field $\Phi$ is defined in terms of its behavior under scaling transformations $\boldsymbol{x} \rightarrow e^{-s} \boldsymbol{x}$ and $t \rightarrow e^{-s z} t$:
\begin{equation}
\Phi(x) = e^{-s \Delta_{\Phi}} \Phi( e^{-s z} t, e^{-s} \boldsymbol{x}) , 
\label{Eq:Scaling}
\end{equation}
where $s = \ln k/\Lambda$ is the scaling parameter (as defined in Sec.\ \ref{Sec:Nonperturbative functional renormalization group in the real-time formalism}). $\Phi$ could correspond to the $\varphi_{+}$ and $\varphi_{-}$ fields (or their counterparts in the RA-basis) in the CTP action \eqref{Eq:FreeActionPMStateBasis}, or the corresponding effective degrees of freedom that appear in our \emph{ansatz} for the scale-dependent CTP effective action, Eq.\ \eqref{Eq:1PIGeneratingFunctionalAnsatz}. In general, each of these fields might have different associated scaling dimensions.

The scaling dimensions of the spatial and time derivatives follow from their properties under scaling transformations:
\begin{equation}
\lbrack \boldsymbol{\partial} \rbrack = 1 , \qquad \lbrack \partial_{0} \rbrack = z .
\end{equation} 
In the presence of spacetime symmetries the scaling dimensions are not independent and the dynamic exponent $z$ can be expressed in terms of the scaling dimension of the spatial derivative operator. In particular, at $T = 0$, Lorentz symmetry enforces $z = 1$ exactly for the relativistic $O(N)$ vector model. However, to define a nonvanishing temperature, we need to specify a fixed reference frame. This breaks the Lorentz symmetry and for a thermal ensemble we find $z \neq 1$.

To determine the dynamic scaling exponent $z$, or the scaling dimension $\Delta_{\Phi}$ of the scaling field $\Phi$, is a difficult problem in general: It requires to solve for the spectrum of the dilatation operator, i.e.
\begin{equation}
\left( \Delta_{\Phi} + z t \partial_{0} + \boldsymbol{x} \cdot \boldsymbol{\partial} \right) \Phi(t, \boldsymbol{x}) = 0 .
\end{equation}
In the absence of interactions and at zero temperature however, this is an easy exercise: Assuming for the moment that the fields $\phi_{+}$ and $\phi_{-}$ (or equivalently $\varphi_{+}$ and $\varphi_{-}$) are scaling, we obtain their scaling dimensions:\footnote{The microscopic fields $\varphi_{\alpha}$ and effective fields $\phi_{\alpha, k}$ (in the effective theory at scale $k$) have the same scaling dimensions in a free theory, i.e., $\varphi_{\alpha} = \phi_{\alpha, k}$.}
\begin{equation}
\Delta_{+} = \Delta_{-} = \frac{1}{2} ( D - 2 ) ,
\label{Eq:CanonicalScalingFields}
\end{equation}
where $D = d + z$, and $z = 1$ at zero temperature. The equality $\Delta_{+} = \Delta_{-}$ follows from the constraint imposed by the fluctuation relation at $T = 0$ and does not apply for nonvanishing $T$ (see Eqs.\ \eqref{Eq:ScalingFormRetardedPropagator} -- \eqref{Eq:FluctuationRelationT0} below). In fact, for $T \neq 0$, the fields $\phi_{+}$ and $\phi_{-}$ do not correspond to scaling fields (see Sec.\ \ref{SubSubSec:Classical scaling regime} below) and the corresponding scaling dimensions $\Delta_{\pm}$ are not well-defined.

In the presence of interactions, the fields are renormalized and the canonical dimensions \eqref{Eq:CanonicalScalingFields} are modified. This holds true even at zero temperature, where the difference in the scaling dimensions is captured by a single multiplicative factor $Z_{\alpha}^{\bot}$
\begin{equation}
\sqrt{Z_{\alpha}^{\bot}} \phi_{\alpha} = \phi_{R,\alpha} , 
\label{Eq:MultiplicativeRenormalization}
\end{equation}
$\alpha \in \{ + , - \}$ and $\phi_{R,\alpha}$ define the renormalized degrees of freedom corresponding to the bare fields $\phi_{\alpha}$ (in the limit $k\rightarrow \Lambda$, $Z^{\bot} \rightarrow 1$). Although the renormalization factors appear as a simple rescaling of the fields, it should be kept in mind that they are in fact associated to the renormalization of composite operators, i.e., $\sim Z_{\alpha}^{\bot} \big(\boldsymbol{\partial} \phi_{\alpha}\big)^{2}$. In the RA-representation, we use the following definition of $Z^{\bot}$
\begin{equation}
Z_{+}^{\bot} \big(\boldsymbol{\partial} \phi_{+}\big)^{2} - Z_{-}^{\bot} \big(\boldsymbol{\partial} \phi_{-}\big)^{2} \equiv Z^{\bot} \boldsymbol{\partial} \tilde{\phi}^{a} \cdot \boldsymbol{\partial} \phi_{a} ,
\label{Eq:WavefunctionRenormalizationRA}
\end{equation}
while $Z^{\vert\!\vert}$ is defined as
\begin{equation}
Z_{+}^{\vert\!\vert} \big(\partial_{0} \phi_{+}\big)^{2} - Z_{-}^{\vert\!\vert} \big(\partial_{0} \phi_{-}\big)^{2} \equiv Z^{\vert\!\vert} \partial_{0} \tilde{\phi}^{a} \partial_{0} \phi_{a} .
\label{Eq:WavefunctionRenormalizationRA}
\end{equation}

\subsubsection{Quantum scaling regime (\,$T = 0$)}
\label{SubSubSec:Quantum scaling regime}

At $T = 0$, we have $Z^{\vert\!\vert} = Z^{\bot} \equiv Z$ by Lorentz symmetry. Here, our interest lies in the scaling behavior at the continuous quantum phase transition where the fields exhibit scaling. That is, in the IR limit $0 \leq k / \Lambda \ll 1$, when all fluctuations have been taken into account and the microscopic parameters have been tuned appropriately, the wavefunction renormalization satisfies $Z^{\bot} \sim e^{-s \eta^{\bot}}$. In this quantum critical regime we may define the anomalous scaling exponent
\begin{equation} 
\eta^{\bot} = - \frac{\partial}{\partial s} \ln Z^{\bot} ,
\end{equation}
and it is clear that $Z^{\vert\!\vert} \sim e^{-s \eta^{\vert\!\vert}}$ with $\eta^{\vert\!\vert} = \eta^{\bot} \equiv \eta$. From the multiplicative renormalization of the fields we find that the scaling dimension of the renormalized composite operator $\boldsymbol{\partial} \tilde{\phi}_{R,a} \cdot \boldsymbol{\partial} \phi_{R}^{a} $ is given by
\begin{equation}
\Delta + \widetilde{\Delta} = D - 2 + \eta^{\bot} .
\label{Eq:ScalingDimensionKinetic}
\end{equation}
This result follows from an appropriate rescaling of the fields $\phi_{R}$ and $\tilde{\phi}_{R}$, from which we deduce the scaling dimensions:
\begin{equation}
\Delta = \widetilde{\Delta} = \frac{1}{2} ( D - 2 + \eta^{\bot} ) .
\end{equation}
As in the free massless theory they are degenerate, which follows simply from the properties of the propagators under scaling transformations
\begin{eqnarray}
G^{\textrm{R}}(\omega, \boldsymbol{p}) &=& e^{- s ( \Delta + \widetilde{\Delta} - D)} G^{\textrm{R}}(e^{s z} \omega, e^{s} \boldsymbol{p}) , \label{Eq:ScalingFormRetardedPropagator} \\
F(\omega, \boldsymbol{p}) &=& e^{- s ( 2 \Delta - D )} F(e^{s z} \omega, e^{s} \boldsymbol{p}) ,
\label{Eq:ScalingFormStatisticalCorrelator}
\end{eqnarray}
and the presence of the $T = 0$ fluctuation theorem 
\begin{equation}
F(\omega) = \sgn(\omega) \Im G^{\textrm{R}}(\omega) .
\label{Eq:FluctuationRelationT0}
\end{equation}

\subsubsection{Classical scaling regime (\,$T \neq 0$)}
\label{SubSubSec:Classical scaling regime}

At $T \neq 0$ the characteristic fluctuations in the vicinity of the classical phase transition (CPT) show a different behavior that is dictated by the scaling form of the FDT. In particular, we observe that 
\begin{equation}
F(\omega) = \frac{T}{\omega} \Im G^{\textrm R}(\omega) , 
\end{equation}
in the scaling region $|\omega| \ll T$. We might encounter such a scaling regime if $T$ is much larger than the cutoff scale, i.e., $0 \leq |\omega| \leq \Lambda^{z} \ll T$. Using the scaling form of the propagators Eqs.\ \eqref{Eq:ScalingFormRetardedPropagator} and \eqref{Eq:ScalingFormStatisticalCorrelator}, we derive the scaling dimensions for the fields in the RA-representation:
\begin{equation}
\Delta = \widetilde{\Delta} - z ,
\label{Eq:BrokenScalingDegeneracy}
\end{equation}
Thus, we observe that the presence of a nonvanishing temperature lifts the degeneracy in the scaling spectrum. This is equivalent to the statement that the retarded and statistical propagators acquire different scaling properties at the classical phase transition (and the difference is quantified in terms of the dynamic critical exponent $z$). In contrast, at the QCP, both the correlations and the response are described by the same scaling behavior ($\Delta = \widetilde{\Delta}$). Although, Eq.\ \eqref{Eq:BrokenScalingDegeneracy} indicates that the fields must scale differently, it does not provide the explicit form for the scaling dimensions at nonvanishing temperature. In Sec.\ \ref{Sec:Classical regime}, we show that relation \eqref{Eq:BrokenScalingDegeneracy} is solved by
\begin{equation}
\Delta = \frac{1}{2} ( d - 2 + \eta^{\bot} ) , \quad \widetilde{\Delta} = \frac{1}{2} ( d - 2 + \eta^{\bot} ) + z .
\end{equation}
We may therefore conclude that there is a crossover between the quantum and classical scaling regimes that is characterized by a continuous transition in the scaling dimensions of the fields, parametrized by the effective dimension $d_{\varepsilon}$:
\begin{equation}
\Delta = \frac{1}{2} \left( d_{\varepsilon} - 2 + \eta^{\bot}(d_{\varepsilon}) \right) , 
\end{equation}
i.e., from $d_{\varepsilon} = d + 1$ ($\omega \gg T$) to $d_{\varepsilon} = d$ ($\omega \ll T$). In the literature this is often referred to as dimensional reduction. However, from our discussion it should be clear that this statement applies only to the scaling properties of the theory -- the real dimensionality of the system remains unchanged.

\subsubsection{Dynamic scaling relations}
\label{SubSubSec:Dynamic scaling relations}

Let us consider the dynamic properties at the continuous phase transition: In general two parameters, $Z^{\bot} / Z^{\vert\!\vert}$ and $\Omega / Z^{\vert\!\vert}$, are necessary to specify the dynamics of our model (cf.\ Eqs.\ \eqref{Eq:LowTRegime} and \eqref{Eq:HighTRegime}), while the wavefunction renormalization $Z^{\bot}$ contributes to the renormalized mass spectrum. At zero temperature the finite renormalization of the theory requires only a single independent parameter $Z^{\bot}$ (while $Z^{\bot} / Z^{\vert\!\vert} = 1$ and $\Omega = 0^{+}$) and there is only a single anomalous scaling exponent, i.e., $\eta \equiv \eta^{\bot} = \eta^{\vert\!\vert}$, that features in the scaling spectrum. In contrast, at $T \neq 0$ we have up to three anomalous exponents that contribute to the dynamic scaling in the vicinity of the classical phase transition. They are given by:
\begin{equation}
\eta^{\vert\!\vert, \bot} = - \frac{\partial}{\partial s} \ln Z^{\vert\!\vert, \bot} , \qquad \eta^{\Omega} = - \frac{\partial}{\partial s} \ln \Omega .
\end{equation}
With these we obtain the dynamic critical exponent $z$ from the scaling behavior of the retarded propagator \eqref{Eq:ScalingFormRetardedPropagator} and the statistical correlation function \eqref{Eq:ScalingFormStatisticalCorrelator}. If $Z^{\bot}$, $Z^{\vert\!\vert}$, and $\Omega \neq 0$, then each of the corresponding anomalous dimensions will contribute in the scaling regime. Using the scaling assumption, we may derive the following relations within our truncation
\begin{eqnarray}
Z^{\vert\!\vert} > 0 &:& \hspace{8pt} - \eta^{\vert\!\vert} + 2 z - 2 + \eta^{\bot} = 0 , \label{OmegaZeroScaling} \\
\Omega > 0 &:& \hspace{10.5pt} - \eta^{\Omega} + z - 2 + \eta^{\bot} = 0 . \label{ZvertZeroScaling}
\end{eqnarray}
If both $Z^{\vert\!\vert}$ and $\Omega$ are nonvanishing, we arrive at the following scaling relation:
\begin{equation}
2 + 2 \eta^{\Omega} - \eta^{\bot} - \eta^{\vert\!\vert} = 0 .
\label{Eq:ScalingRelation}
\end{equation}
Thus, in general, it is possible to characterize the dynamic scaling in terms of two independent anomalous exponents only, which we choose to be $\eta^{\bot}$ and $\eta^{\Omega}$. The presence of additional symmetries may further reduce the number of independent exponents that characterize the dynamic correlations. In particular, in the presence of Lorentz symmetry ($\Omega = 0^{+}$) we observe that: $z = 1 + \frac{\eta^{\vert\!\vert} - \eta^{\bot}}{2} \equiv 1$.

To summarize, regarding the dynamic critical behavior, we may distinguish between two scaling regimes:
\begin{eqnarray}
z = \left\{ \begin{array}{l l} 1 & , \qquad T = 0 , \\
2 - \eta^{\bot} + \eta^{\Omega}  & , \qquad T \neq 0 . \\ \end{array} \right.
\end{eqnarray}
At zero temperature, in the absence of a dissipative coupling $\Omega$, the system is invariant under Lorentz transformations, which implies: $\eta^{\vert\!\vert} = \eta^{\bot} \equiv \eta$, and we find that the dynamic critical exponent $z = 1$, independent of the local interactions (provided that they do not break the Lorentz symmetry). However, the presence of a nonvanishing temperature leads to an additional relevant parameter, which leads to a dynamic scaling exponent $z > 1$.

\subsection{Local interaction approximation}
\label{SubSec:Local interaction approximation}

Our \emph{ansatz} for the scale-dependent CTP effective action relies on a truncation of an expansion in gradients, as well as on a truncation of an infinite series of vertices. Here, we address the properties of the finite vertex expansion, which rests on two assumptions:
\begin{itemize}
\item[(I)] We assume that the generalized potential $\mathcal{U} = \mathcal{U}(\phi,\tilde{\phi})$ takes the form of a formal series expansion that is local both in space and time:
\begin{equation}
\hspace{25pt} \mathcal{U} = \tilde{\phi}^{a} \frac{\partial V}{\partial \phi^{a}} + \sum_{n \geq 2} \, \tilde{\phi}^{a_{1}} \tilde{\phi}^{a_{2}} \cdots \tilde{\phi}^{a_{n}} \mathcal{F}_{a_{1} a_{2} \ldots a_{n}} ,
\label{Eq:SeriesExpansionPotential}
\end{equation}
where $V = V(\phi)$ corresponds to the effective potential and the coefficients $\mathcal{F}_{a_{1} a_{2} \ldots a_{n}} = \mathcal{F}_{a_{1} a_{2} \ldots a_{n}} (\phi)$ define the fluctuation amplitudes. 
\item[(II)] We assume that a truncation of the series \eqref{Eq:SeriesExpansionPotential} at 4th order in the fields provides a reasonable approximation of our theory in the low-energy regime. In this case, the only nonvanishing amplitudes $\mathcal{F}_{a_{1} a_{2} \ldots a_{n}}$ that contribute in the generalized potential are given by:
\begin{eqnarray}
\mathcal{F}_{a_{1} a_{2}} &=& \delta_{a_{1} a_{2}} F_{2,0} + \phi_{a_{1}} \phi_{a_{2}} F_{2,2} , \\
\mathcal{F}_{a_{1} a_{2} a_{3}} &=& \delta_{( a_{1} a_{2} } \phi_{a_{3} )}  F_{3,1}  , \\
\mathcal{F}_{a_{1} a_{2} a_{3} a_{4}} &=&  \delta_{( a_{1} a_{2}} \delta_{a_{3} a_{4} )}  F_{4,0} .
\end{eqnarray}
Here, the brackets denote a complete symmetrization in the indices, e.g., \mbox{$\delta_{( a_{1} a_{2} } \phi_{a_{3} )} = \frac{1}{3} \left( \delta_{a_{1} a_{2}} \phi_{a_{3}} + \delta_{a_{1} a_{3} } \phi_{a_{2}} + \delta_{a_{2} a_{3}} \phi_{a_{1}} \right)$}.
\end{itemize}

Typically one expects that such a finite expansion is valid to a good approximation in the vicinity of a continuous phase transition, where higher order contributions in the fields are irrelevant in the RG sense. In the real-time formalism however, this might not be sufficient to resolve the appropriate low-energy dynamics. The vertices might carry a frequency- or momentum-dependence and this information needs to be taken into account to address the question of the relevance or irrelevance of hydrodynamic modes in the low-energy limit. Indeed, using a local series expansion \eqref{Eq:SeriesExpansionPotential} in our \emph{ansatz} for the CTP effective action we may only determine the dynamics of the order parameter, without any coupling to long-wavelength modes. Thus, within our truncation, we expect that the IR dynamics is purely dissipative and the dynamic universality class of the theory is given by that of Model A \cite{Hohenberg:1977ym}, in line with our arguments at the end of Sec.\ \ref{SubSec:Low-energy effective theory}.

\begin{figure*}[!t]
\centering
\includegraphics[width=0.33\textwidth]{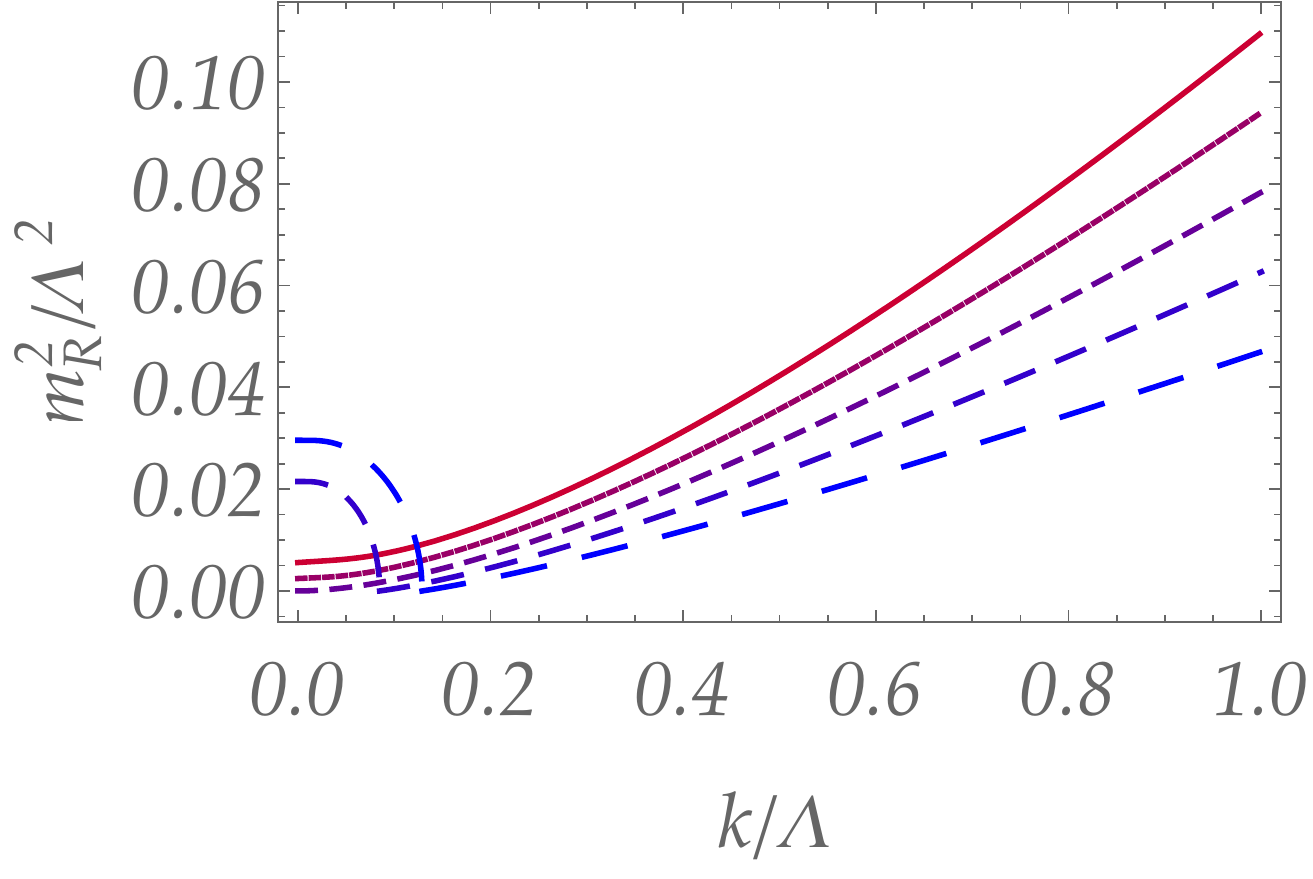} 
\includegraphics[width=0.31\textwidth]{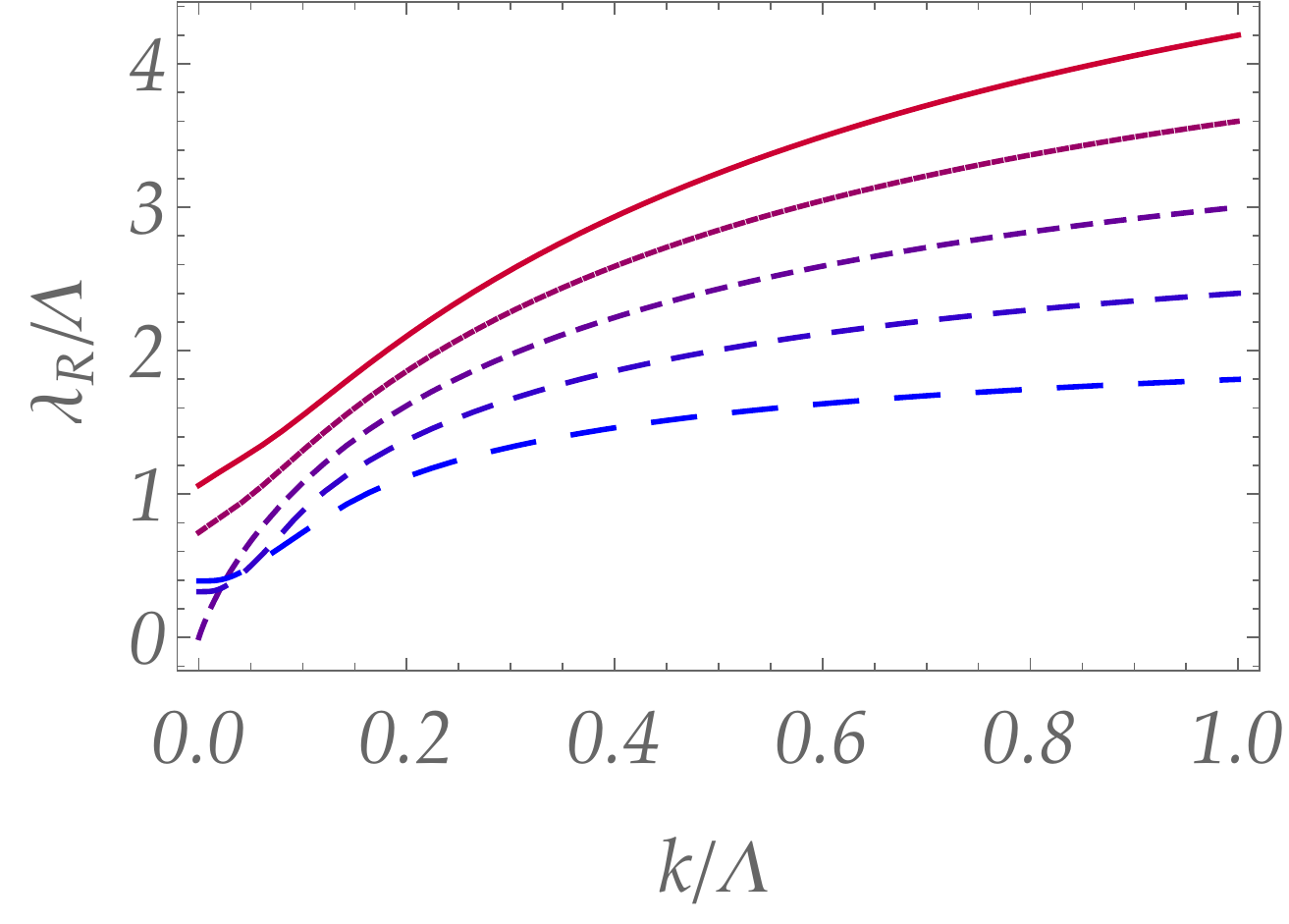}
\includegraphics[width=0.34\textwidth]{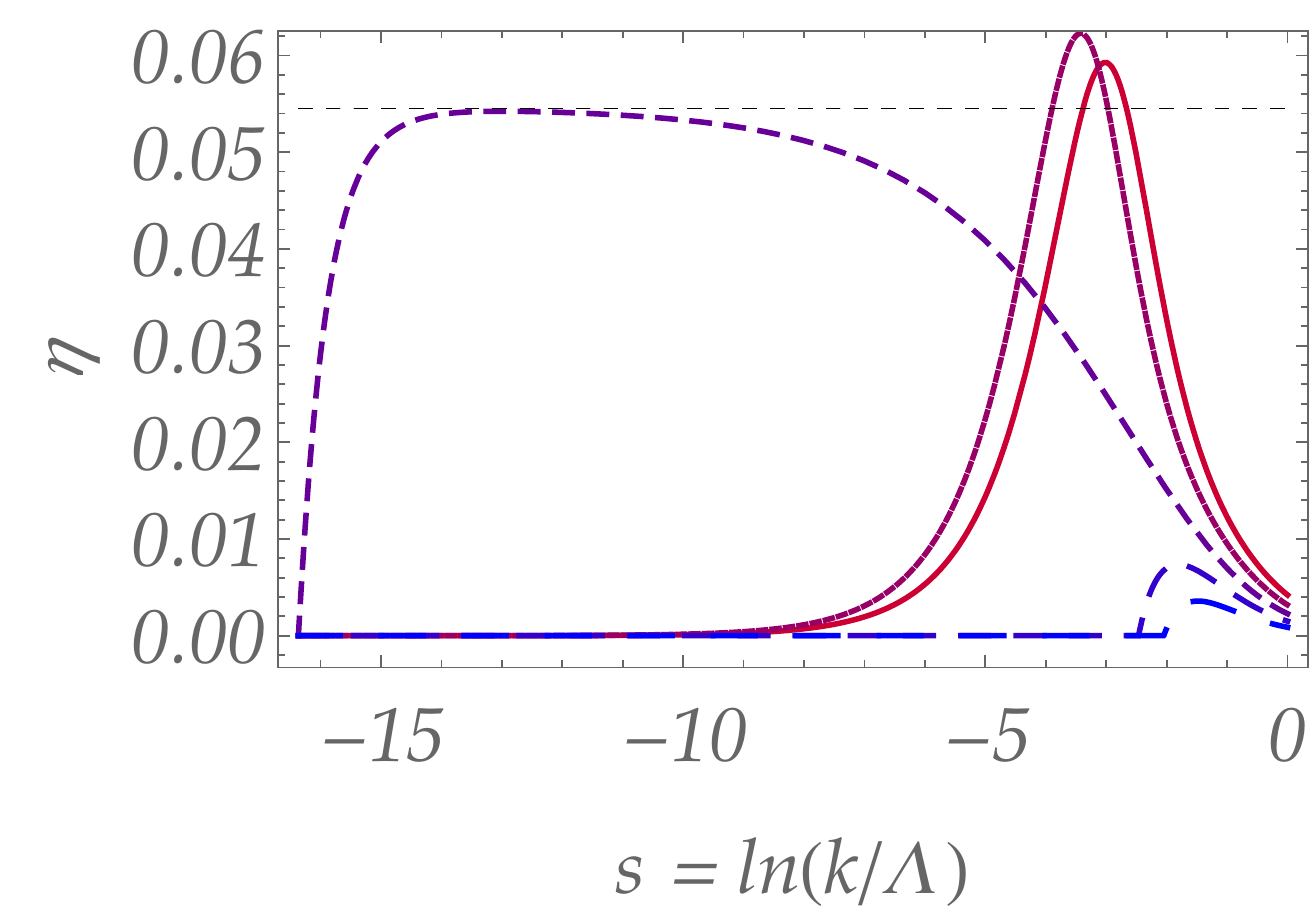}
\caption{\label{Fig:ZeroTemperatureRGFlow}$N = 1$ scalar model ($D = 3$, $T = 0$): As we tune the (bare) quartic coupling $\bar{\lambda}_{\Lambda} = \lambda_{\Lambda} / \Lambda$ at a fixed value of the squared field expectation value $\bar{v}_{\Lambda}^{2} = v^{2}_{\Lambda} / \Lambda = 0.078$, we observe a phase transition, when the renormalized mass $m_{R}^{2}$ vanishes. We define $m_{R}^{2} = \lambda_{R} v_{R}^{2}$ in the symmetry broken regime and $v_{R}^{2} = k \bar{v}^{2}$, $\lambda_{R} = k \bar{\lambda}$. This is shown in the left panel where follow the RG flow of $m_{R}^{2}$ as the bare coupling $\bar{\lambda}_{\Lambda}$ is tuned through its critical value $\bar{\lambda}_{\Lambda, \textrm{cr}}$. In the middle panel we display the scale dependence of the renormalized coupling. At the quantum critical point the correlation length $\xi \equiv m_{R}^{-1}$ diverges and the theory exhibits scaling. This scaling at the QCP is governed by the IR stable Wilson-Fisher fixed point, for which we determine the scaling exponents $\eta \simeq 0.055$ and $z = 1$. The scale dependence of the anomalous dimension is shown in the right panel. If $\bar{\lambda}_{\Lambda} = \bar{\lambda}_{\Lambda, \textrm{cr}}$ the anomalous dimension assumes a constant value in the scaling regime ($k \rightarrow 0$).}
\end{figure*}

The presence of a fluctuation theorem in the groundstate or thermal equilibrium constrains the form of the generalized potential. One finds that in general the vertices are not independent (see, e.g., Ref.\ \cite{Wang:1998wg} where the fluctuation relations are given on the level of the $3$- and $4$-point functions). These nonlinear fluctuation relations take a simple form if one uses a local vertex expansion. Together with the conjugation symmetry $\mathcal{U}(\phi,\tilde{\phi})^{\ast} = - \mathcal{U}(\phi, -\tilde{\phi})$, we find at 4th order in the fields: $\mathcal{F}_{a_{1} a_{2}} = \mathcal{F}_{a_{1} a_{2} a_{3} a_{4}} = 0$. However, we point out that we obtain a nonvanishing contribution $(i \Omega / \beta^{2}) \int_{x} \tilde{\phi}^{2}$ to the CTP effective action from the zero-frequency limit of the quadratic term $\tilde{\phi}^{2}$ in Eq.\ \eqref{Eq:1PIGeneratingFunctionalAnsatz}. The simplification on the level of vertices is essentially a consequence of our assumption of locality and the requirement of consistency that we impose on our \emph{ansatz}. That is, in our construction of the 1PI CTP effective action, we have chosen to incorporate the properties of the equilibrium state exactly by imposing fluctuation relations between $(m+n)$-point functions $\Gamma^{(m,n)}$. Any consistent truncation of the full equilibrium CTP effective action should take into account these relations \cite{Sieberer:2015hba} and this is what we have done here.

In the following it will be useful to consider an alternative representation of the fields. In particular, we introduce the following invariants under $O(N)$ transformations
\begin{equation}
\sigma_{1} = \frac{1}{2} \phi^{a} \phi_{a} , \quad \sigma_{2} = \phi^{a} \tilde{\phi}_{a} , \quad \sigma_{3} = \frac{1}{2} \tilde{\phi}^{a} \tilde{\phi}_{a} ,
\end{equation}
in terms of which $\mathcal{U} = \mathcal{U}( \sigma_{1} , \sigma_{2}, \sigma_{3})$. This choice of parametrization makes the $O(N)$-symmetry of the theory manifest and proves to be especially convenient in the derivation of the RG equations. Using the constraints that follow from conjugation symmetry and nonlinear fluctuation relations, we obtain the following form of the generalized potential:
\begin{equation}
\mathcal{U} = m^{2} \sigma_{2} + \lambda_{1,2} \left( \sigma_{1} - v^{2} / 2 \right) \sigma_{2} + \lambda_{2,3} \sigma_{2} \sigma_{3} ,
\label{Eq:PotentialTruncationInvariants}
\end{equation}
which we expand around the scale-dependent minimum $\left. \sigma_{1} \right\vert_{\min} = v^{2} / 2$. In the disordered phase (symmetric regime), $m^{2} > 0$ and $v^{2} = 0$, while in the ordered phase (symmetry broken regime), $m^{2} = 0$ and $v^{2} > 0$. Since the functional RG provides a regularization of IR divergences associated to massless modes, we can follow the RG flow of the theory, through the phase transition, from the symmetry broken into the symmetric phase. Thus, we will encounter both scenarios when we solve for the RG flow equations.

Note that all parameters and couplings of our model are real-valued. A generic \emph{ansatz} for the generalized potential might also include imaginary couplings, e.g., $i \lambda_{1,3} \sigma_{1} \sigma_{3}$ (see our discussion at the end of \mbox{Sec.\ \ref{Sec:Real-time n-point functions and generating functionals}}). The reason that they are not present here, is a consequence of the fluctuation relations. Furthermore, not all couplings in Eq.\ \eqref{Eq:PotentialTruncationInvariants} are independent. We may establish a relation between $\lambda_{1,2}$ and $\lambda_{2,3}$. We demonstrate this relation explicitly at the example of the flow equations (cf.\ \mbox{Sec.\ \ref{Sec:Quantum regime}}).

\section{Quantum regime ($T = 0$)}
\label{Sec:Quantum regime}

The general form of the flow equations within our truncation is given in Appendix \ref{Sec:Renormalization group equations}. Here, we provide the RG flow equations specifically in the quantum regime. This section summarizes well-known results \cite{Wetterich:1989xg,Wetterich:1991be,Tetradis:1993ts} and serves to provide a complete picture in the context of the quantum-to-classical transition (cf.\ Secs.\ \ref{Sec:Classical regime} and \ref{Sec:Quantum-classical transition}). The flow equations are derived by using a Euclidean regulator function after a Wick rotation to imaginary times and frequencies. We employ the $(d+1)$-dimensional Litim regulator function \cite{Litim:2000ci, *Litim:2001up}. To solve the flow equations it is useful to apply our knowledge of the possible scaling solutions (cf.\ \mbox{Sec.\ \ref{SubSec:Scaling dimensions and dynamic scaling relations}}): We define dimensionless renormalized parameters and couplings in terms of which it is convenient to identify possible fixed point (FP) solutions of the RG flow. Let us consider the scaling dimensions for the parameters and couplings that enter in our model at $T = 0$:
\begin{eqnarray}
\lbrack v^{2} \rbrack &=& D - 2 + \eta^{\bot} , 
\label{Eq:ScalingDimensionFieldExpectationValue} \\
\lbrack \lambda_{m,n} \rbrack &=& 4 - D - 2 \eta^{\bot} .
\label{Eq:ScalingDimensionCouplings}
\end{eqnarray}
We may define the following quantities
\begin{eqnarray}
\bar{v}^{2} &=& Z^{\bot} k^{2 - D} v^{2} , \\
\bar{\lambda}_{m,n} &=& (Z^{\bot})^{-2} k^{D - 4} \lambda_{m,n} ,
\end{eqnarray}
that are scale-independent at the critical point. This is immediately clear, since we have $Z^{\bot} \sim e^{- s \eta^{\bot}}$ in the scaling regime, so that the scaling behavior of the parameters $v^{2}$ and couplings $\lambda_{m,n}$ is exactly canceled. To highlight the fact that we have rescaled the fields in such a way to compensate for the scaling \eqref{Eq:ScalingDimensionFieldExpectationValue} and \eqref{Eq:ScalingDimensionCouplings} in the quantum critical regime, we denote the corresponding parameters and couplings with a bar. We also define $\bar{\Omega} \equiv \Omega$, $\bar{Z}^{\vert\!\vert} \equiv Z^{\vert\!\vert}$, and $\bar{Z}^{\bot} \equiv Z^{\bot}$ even though these parameters have not been rescaled.

In the absence of a temperature scale Lorentz symmetry requires $z = 1$ and the RG equations simplify considerably. In particular, we find that no dissipative coupling is generated by the RG flow:
\begin{equation}
\frac{\partial \bar{\Omega}}{\partial s} = 0 .
\end{equation}
Furthermore, no imaginary couplings are generated and therefore they do not enter in the low-energy effective theory. This holds true independent of the choice of our truncation. In the following we use the relation $\bar{Z}^{\vert\!\vert} = \bar{Z}^{\bot} \equiv \bar{Z}$ and define the anomalous dimension $\eta \equiv - \frac{\partial}{\partial s} \ln \bar{Z}$ also outside the critical region. Thus, $\eta$ will in general be scale-dependent, i.e., $\eta = \eta_{k}$ and takes on its critical value only in the scaling regime.

\begin{figure}[!t]
\centering
\includegraphics[width=0.35\textwidth]{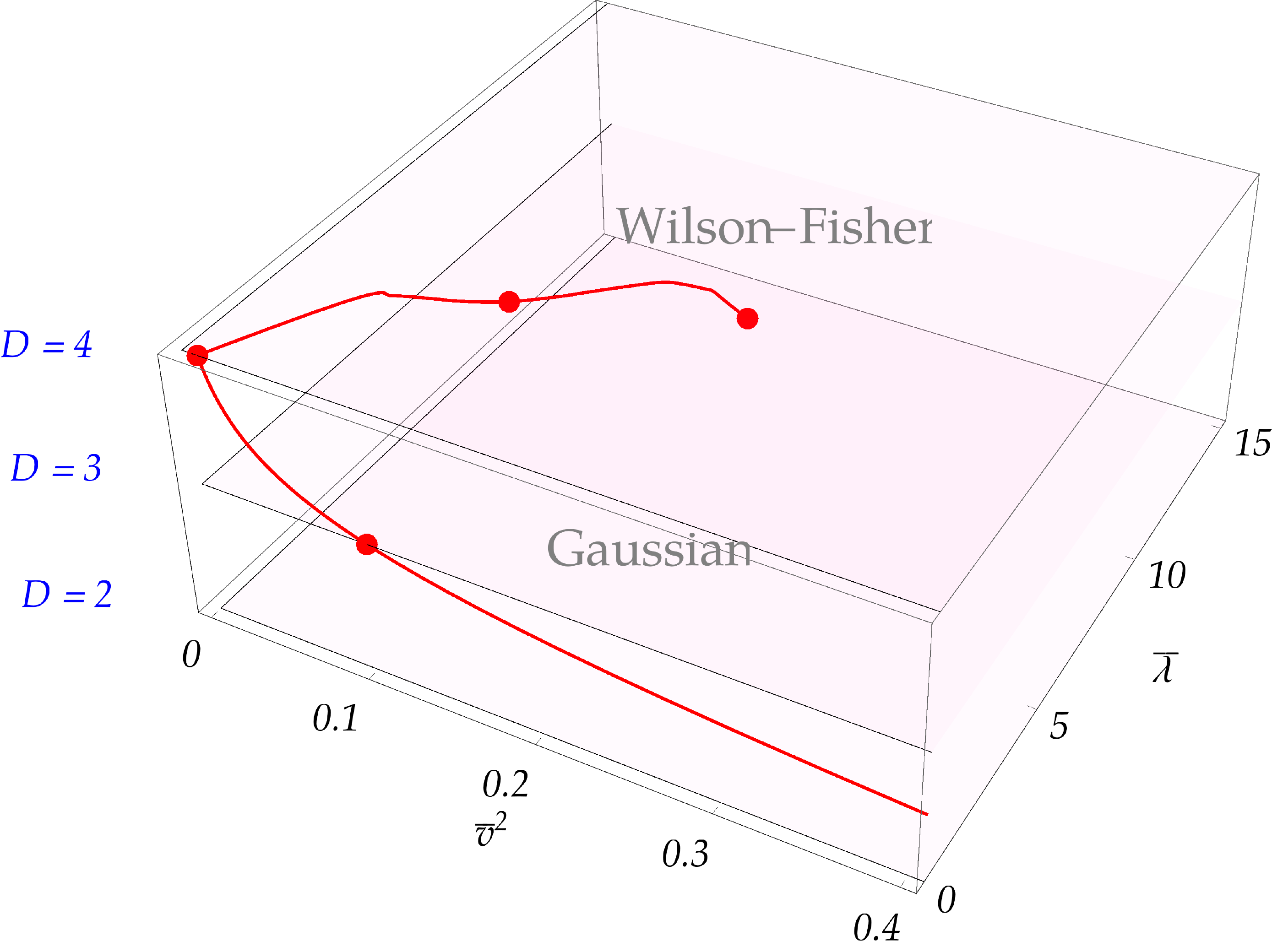}
\caption{\label{Fig:ZeroTemperaturFPs}$N = 1$ scalar model ($T = 0$): Gaussian FP and Wilson-Fisher FP as a function of dimension \mbox{$D = d + 1$}. Both fixed points merge at the upper critical dimension $D_{cr} = 4$.}
\end{figure}

\begin{figure}
\centering
\hskip -22pt \includegraphics[width=0.38\textwidth]{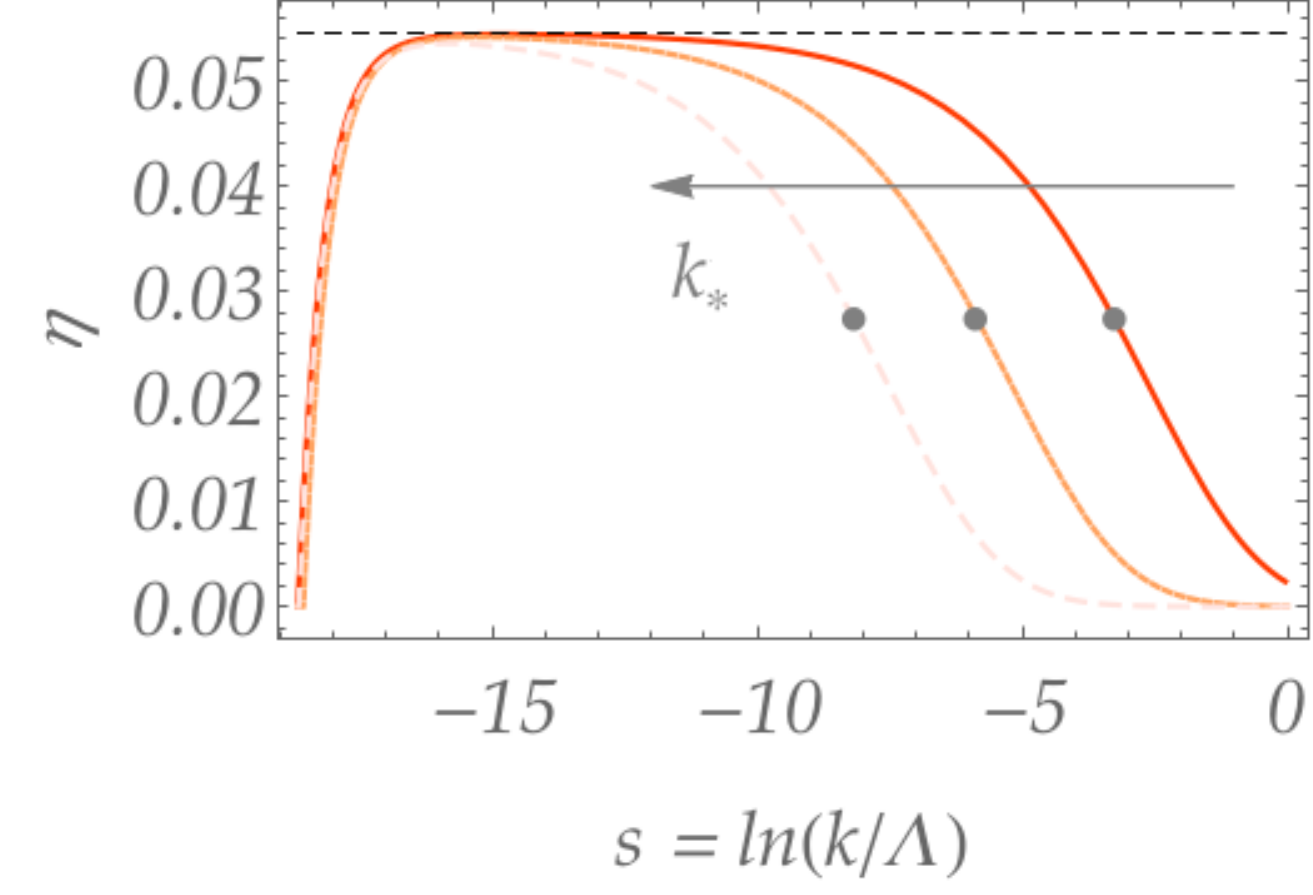} \\[10pt]
\includegraphics[width=0.37\textwidth]{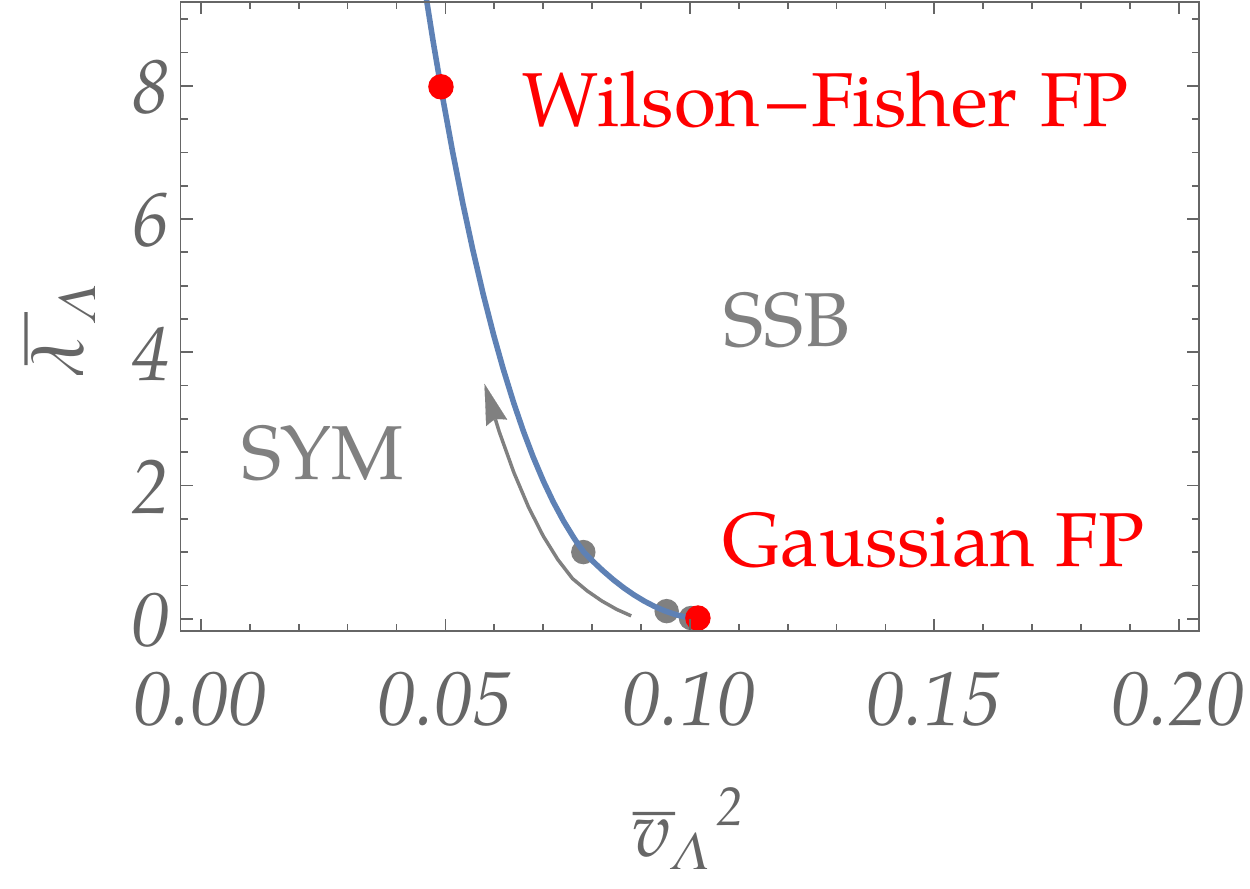}
\caption{\label{Fig:ZeroTemperaturCrossoverScale}$N = 1$ scalar model ($D = 3$, $T = 0$): The crossover scale $k_{\ast}$ that separates the canonical from the nontrivial scaling region is set by the quartic coupling at the cutoff scale $\Lambda$. There is a one-to-one correspondence between the scale $k_{\ast}$ (upper panel) to points that lie on the curve connecting the Gaussian FP with the Wilson-Fisher FP (lower panel). As we move along the phase boundary (indicated by the arrow), separating the spontaneous symmetry broken phase (SSB) from the symmetric phase (SYM), towards the Wilson-Fisher FP the crossover scale is shifted to the left (upper panel). Here, the maximum correlation length $\xi \equiv m_{R}^{-1}$ takes a finite value since both the bare values of the coupling $\bar{\lambda}_{\Lambda}$ and field expectation value $\bar{v}_{\Lambda}^{2}$ have not been tuned perfectly to the phase separation line.}
\end{figure}

The RG flow equations for the squared field expectation value and quartic couplings are given by:
\begin{eqnarray}
\hspace{-20pt} \frac{\partial \bar{v}^{2}}{\partial s} &=& (2 - D - \eta) \bar{v}^{2} \nonumber\\ && +\: \delta_{D} (\eta) \left\{ \frac{3}{(1 + \bar{v}^{2} \lambda_{1,2})^{2}} + N-1 \right\} , \label{Eq:ZeroTemperaturev2Flow} \\
\hspace{-20pt} \frac{\partial \bar{\lambda}_{1,2}}{\partial s} &=& (D - 4 + 2\eta) \bar{\lambda}_{1,2} \nonumber\\ && +\: \delta_{D} (\eta) \bar{\lambda}_{1,2}^{2} \! \left\{ \frac{9}{(1 + \bar{v}^{2} \bar{\lambda}_{1,2})^{3}} + N-1 \right\} , \label{Eq:ZeroTemperatureLambdaFlow} 
\end{eqnarray}
\begin{eqnarray}
\hspace{-20pt} \frac{\partial \bar{\lambda}_{2,3}}{\partial s} &=& (D - 4 + 2\eta) \bar{\lambda}_{2,3} \nonumber\\ && +\: \delta_{D} (\eta) \bar{\lambda}_{1,2} \bar{\lambda}_{2,3} \! \left\{ \frac{9}{(1 + \bar{v}^{2} \bar{\lambda}_{1,2})^{3}} + N-1 \right\} .
\end{eqnarray}
Here, the factor
\begin{equation}
\delta_{D}(\eta) = \frac{1}{2^{D - 1} \pi^{\frac{D}{2}} \Gamma((D+2)/ 2)} \left(1 - \frac{\eta}{D + 2} \right) ,
\end{equation}
originates from the spherical integration in momentum space for the contributing diagrams. Observe that the quartic couplings satisfy the following relation:
\begin{equation}
\bar{\lambda}_{1,2} \frac{\partial \bar{\lambda}_{2,3}}{\partial s} = \bar{\lambda}_{2,3} \frac{\partial \bar{\lambda}_{1,2}}{\partial s} ,
\label{Eq:CouplingsRGFlowRelation}
\end{equation}
which clearly shows that they are not independent. From the microscopic action \eqref{Eq:FreeFieldTheory} we see that: $\lambda_{1,2} = 4 \lambda_{2,3} \equiv \lambda / 3$. By an appropriate rescaling of $\lambda$ and by virtue of Eq.\ \eqref{Eq:CouplingsRGFlowRelation} this relation between the couplings holds true for all RG trajectories in the quantum regime ($T = 0$).

To close the system of RG equations, we need to provide the anomalous dimension:
\begin{equation}
\eta = \delta_{D} (0) \frac{\bar{v}^{2} \bar{\lambda}_{1,2}^{2}}{(1 + \bar{v}^{2} \bar{\lambda}_{1,2})^{2}} .
\label{Eq:ZeroTemperatureAnomalousDimension}
\end{equation}
We find two possible fixed points (imposing the requirement that the squared field expectation value $\bar{v}^{2}$ is positive, and both couplings $\bar{\lambda}_{1,2}$ and $\bar{\lambda}_{2,3}$ are nonnegative): 
\begin{itemize}
\item[(I)] \emph{Gaussian fixed point}
\begin{equation}
\bar{\lambda}_{1,2} = \bar{\lambda}_{2,3} = 0 ,
\end{equation}
and $\eta = 0$.
\item[(II)] \emph{Wilson-Fisher fixed point} ($2 < D < 4$):  
\begin{equation}
\bar{\lambda}_{1,2} = 4 \bar{\lambda}_{2,3} = \frac{\bar{\lambda}}{3} > 0
\end{equation} 
and $\eta$ determined by Eq.\ \eqref{Eq:ZeroTemperatureAnomalousDimension} where  $\bar{\lambda}_{1,2}$ and $\bar{v}^{2}$ are set to their respective fixed point values.
\end{itemize}
The behavior of the RG flow can be illustrated by considering the $D = 3$ scalar theory \mbox{($N = 1$)} as an example, cf.\ Fig.\ \ref{Fig:ZeroTemperatureRGFlow}. In this case both fixed point solutions are present and the theory admits nontrivial IR scaling in the Ising universality class. We obtain $\eta \simeq 0.055$ at the phase transition. This result should be compared with $\eta = 0.03639(15)$ from the high-temperature expansion \cite{Campostrini:2002} which seems to be the most precise determination to date (see, e.g., Ref.\ \cite{Pelissetto:2002} for a compilation of results and corresponding references). The discrepancy is not very surprising, since we have chosen a very crude truncation. Typically higher orders in the gradient expansion, e.g., to 4th order \cite{Morris:1997xj,Canet:2003qd,Litim:2001dt,Litim:2010tt} provide a better numerical estimate of the scaling exponents in the context of the functional RG. In $D = 4$ dimensions the only fixed point is the Gaussian one and the static and dynamic scaling properties are characterized by the respective mean-field values: $z = 1$ and $\eta = 0$. The trajectories of both fixed points in the space of parameters and couplings, as we continuously vary the dimension $D = d + 1$, is shown in \mbox{Fig.\ \ref{Fig:ZeroTemperaturFPs}}.

The RG flow in the $O(N)$ model features a characteristic scale at which fluctuations become important. This is shown in Fig.\ \ref{Fig:ZeroTemperaturCrossoverScale} at the example of the $D = 3$ scalar model ($N = 1$): Starting from the microscopic action, the anomalous scaling exponent crosses over from the canonical scaling behavior with $\eta = 0$ in the vicinity of the Gaussian FP, to the Wilson-Fisher FP with $\eta \neq 0$ at some characteristic scale $k_{\ast}$. This crossover scale depends only on the value of the dimensionless quartic coupling $\bar{\lambda}_{\Lambda} \equiv \lambda_{\Lambda} \Lambda^{D - 4}$ and is to a good approximation independent of $\bar{v}^{2}_{\Lambda}$. When the coupling $\bar{\lambda}_{\Lambda}$ is small, e.g., for $D = 4 - \epsilon$ close to the upper critical dimension, the $\bar{\lambda}_{\Lambda}$ dependence of $k_{\ast}$ is seen to originate from the perturbative one-loop corrections. That is, in an RG treatment these perturbative contributions become large when the RG scale parameter $k$ approaches $k_{\ast}$. In Fig.\ \ref{Fig:ZeroTemperaturCrossoverScale} we see that this perturbative argument remains valid even if $\epsilon = 4 - D$ is of order one.

\section{Classical regime ($T \neq 0$)}
\label{Sec:Classical regime}

In the presence of a nonvanishing temperature we obtain a different set of flow equations. Here, we examine the scenario where the temperature is comparable (or much larger) than the cutoff scale $\Lambda$, i.e., $\beta \Lambda^{z} = 1 / T_{\Lambda} \lesssim 1$, and the thermal occupation of modes $n(\omega) \sim T_{\Lambda} ( \omega / \Lambda^{z} )^{-1} \gg 1$, is strongly enhanced. In this regime, the FDT takes the form:
\begin{equation}
F(\omega, \boldsymbol{p}) = \frac{T_{\Lambda}}{\omega \Lambda^{-z}} \Im\, G^{\textrm R}(\omega, \boldsymbol{p}) .
\label{Eq:HighTemperatureFDT}
\end{equation}

As in the zero temperature case, it is useful to express the flow equations in terms of the dimensionless renormalized couplings and parameters, which allows us to easily identify scale-invariant behavior. However, in contrast to the scaling behavior given in Eqs.\ \eqref{Eq:ScalingDimensionFieldExpectationValue} and \eqref{Eq:ScalingDimensionCouplings}, there is an additional scale dependence in the classical regime which is introduced by the temperature $T \sim \Lambda^{z}$. By an appropriate rescaling of the fields (cf.\ Sec.\ \ref{SubSec:Scaling dimensions and dynamic scaling relations}) we find that the scaling dimensions of the parameters and couplings are modified. In particular, the rescaled field expectation value squared and couplings take the form:
\begin{eqnarray}
\dbar{v}^{2} &=& \bar{v}^{2} (k/\Lambda)^{z} ,
\label{Eq:RescaledFieldExpectationValue} \\
\dbar{\lambda}_{m,n} &=& \bar{\lambda}_{m,n} (k / \Lambda)^{(m + n - 4) z} .
\label{Eq:RescaledCouplings}
\end{eqnarray}
This rescaling is sufficient to eliminate the explicit $k$-dependence in the flow equations. Apart from the parameters and couplings in Eqs.\ \eqref{Eq:RescaledFieldExpectationValue} and \eqref{Eq:RescaledCouplings}, we define $\dbar{Z}^{\vert\!\vert, \bot} = Z^{\vert\!\vert, \bot}$ and $\dbar{\Omega} = \Omega$ in the classical regime. The anomalous dimensions are given by: $\eta^{\vert\!\vert, \bot} \equiv - \frac{\partial}{\partial s} \ln \dbar{Z}^{\vert\!\vert, \bot}$ and $\eta^{\Omega} \equiv - \frac{\partial}{\partial s} \ln \dbar{\Omega}$ also outside the critical region, i.e., in general they can be scale-dependent $\eta^{\vert\!\vert, \bot} = \eta^{\vert\!\vert, \bot}_{k}$ and $\eta^{\Omega} = \eta^{\Omega}_{k}$. Fixed points of the RG equations are identified by identifying constant values for all $(\eta^{\bot}, \eta^{\vert\!\vert}, \eta^{\Omega})$.

In the classical regime the RG equations for the squared field expectation value and the couplings are given by:
\begin{widetext}
\begin{eqnarray}
\frac{\partial \dbar{v}^{2}}{\partial s} &=& (2 - d - \eta^{\bot}) \dbar{v}^{2} + T_{\Lambda} \delta_{d} (\eta^{\bot}) \left\{ \frac{3}{(1 + \dbar{v}^{2} \dbar{\lambda}_{1,2})^{2}} + N-1 \right\} ,
\label{Eq:NonvanishingTemperaturev2Flow} \\
\frac{\partial \dbar{\lambda}_{1,2}}{\partial s} &=& (d - 4 + 2\eta^{\bot} ) \dbar{\lambda}_{1,2} + \dbar{\lambda}_{1,2}^{2} T_{\Lambda} \delta_{d} (\eta^{\bot}) \left\{ \frac{9}{(1 + \dbar{v}^{2} \dbar{\lambda}_{1,2})^{3}} + N-1 \right\} , 
\label{Eq:NonvanishingTemperatureLambda12Flow} \\
\frac{\partial \dbar{\lambda}_{2,3}}{\partial s} &=& (d + 2 z - 4 + 2\eta^{\bot}) \dbar{\lambda}_{2,3} + \dbar{\lambda}_{1,2} \dbar{\lambda}_{2,3} T_{\Lambda} \delta_{d}(\eta^{\bot}) \left\{ \frac{9}{(1 + \dbar{v}^{2} \dbar{\lambda}_{1,2})^{3}} + N-1 \right\} .
\label{Eq:NonvanishingTemperatureLambda23Flow}
\end{eqnarray}
The anomalous dimensions associated to the renormalization factors read
\begin{eqnarray}
\eta^{\bot} &=& T_{\Lambda} \delta_{d}(0) \frac{\dbar{v}^{2} \dbar{\lambda}_{1,2}^{2}}{(1 + \dbar{v}^{2} \dbar{\lambda}_{1,2})^{2}} , \\
\eta^{\Omega} &=& \delta_{d}(\eta^{\bot}) \frac{T_{\Lambda}}{\dbar{v}^{2}} \left\{ 1 + \frac{1}{(1 + \dbar{v}^{2} \dbar{\lambda}_{1,2})^{2}} - \frac{32}{\left\lbrack 4 + \dbar{v}^{2} \dbar{\lambda}_{1,2} (2 + \dbar{\kappa} \dbar{v}^{2} \dbar{\lambda}_{1,2})\right\rbrack^{2}} \right\} , \label{Eq:AnomalousDimensionOmega} \\
\eta^{\vert\!\vert} &=& \frac{\eta^{\Omega}}{2} - \frac{\delta_{d}(\eta^{\bot})}{2} \frac{T_{\Lambda}}{\dbar{v}^{2}} \frac{1}{\dbar{\kappa} \dbar{v}^{2} \dbar{\lambda}_{1,2}} \left\{ 1  - \frac{1}{(1 + \dbar{v}^{2} \dbar{\lambda}_{1,2})^{2}} - \frac{128 \, \dbar{v}^{2} \dbar{\lambda}_{1,2} \left\lbrack 1 - ( \dbar{\kappa} \dbar{v}^{2} \dbar{\lambda}_{1,2} )^{2} \right\rbrack }{\left\lbrack 4 + \dbar{v}^{2} \dbar{\lambda}_{1,2} (2 + \dbar{\kappa} \dbar{v}^{2} \dbar{\lambda}_{1,2}) \right\rbrack^{3}} \right\} ,
\end{eqnarray}
and are derived for $\dbar{\kappa} > 0$ and $\dbar{\Omega} > 0$. Apart from a possibly nonvanishing anomalous dimension $\eta^{\Omega}$, and the lifting
\end{widetext}
of the degeneracy of $\eta^{\vert\!\vert} = \eta^{\bot}$, we observe two important differences to the RG flow equations at $T = 0$, cf.\ Eqs.\ \eqref{Eq:ZeroTemperaturev2Flow} -- \eqref{Eq:ZeroTemperatureAnomalousDimension}: (1) Instead of $D = d + 1$ we see that the spatial dimension $d$ appears in the canonical scaling contributions to the RG flow. (2) The loop contributions are not only controlled by the couplings $\lambda_{1,2}$ and $\lambda_{2,3}$ but also by the temperature.

\begin{figure}[!t]
\centering
\includegraphics[width=0.40\textwidth]{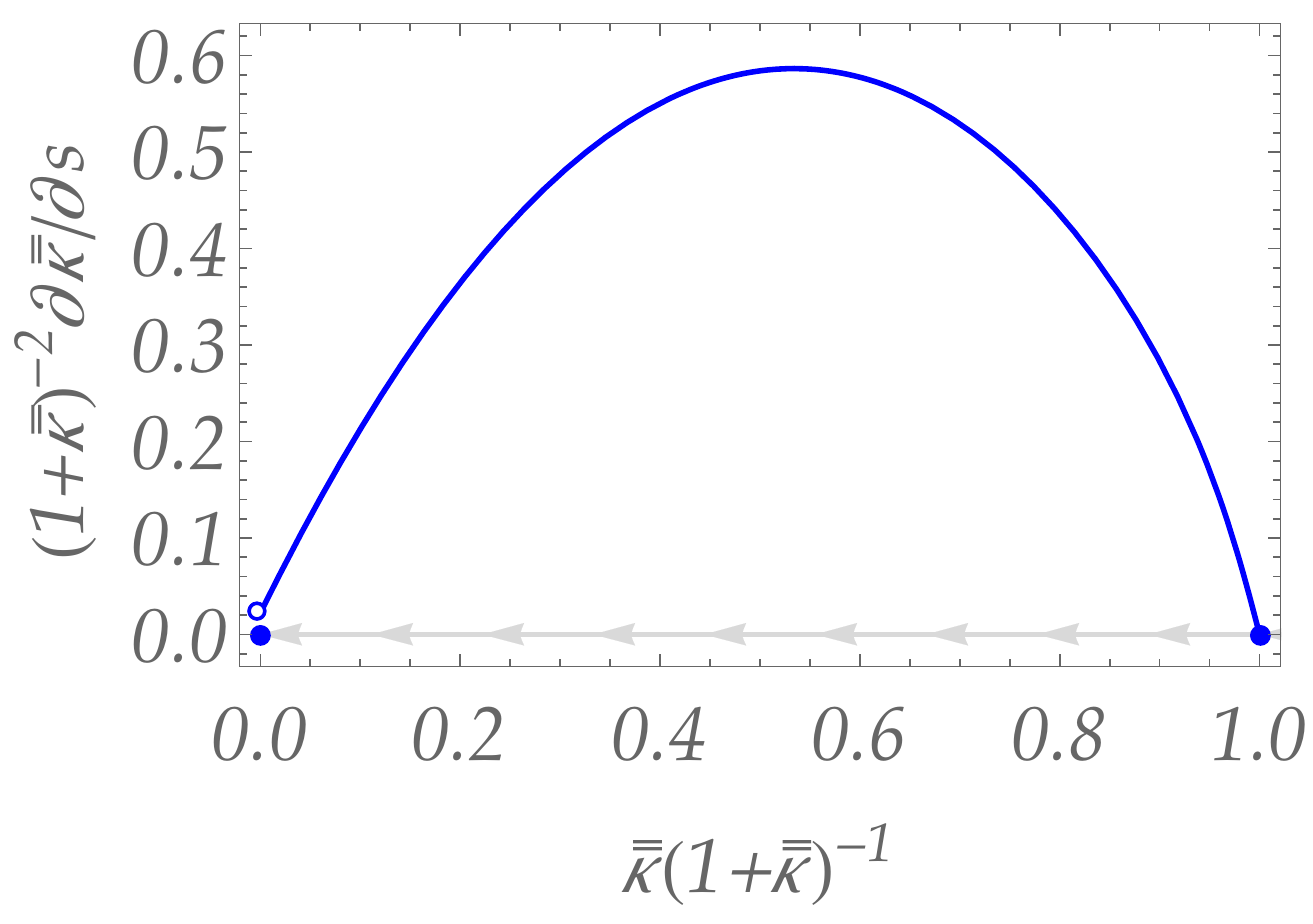} \\[5pt]
\hskip 2pt \includegraphics[width=0.38\textwidth]{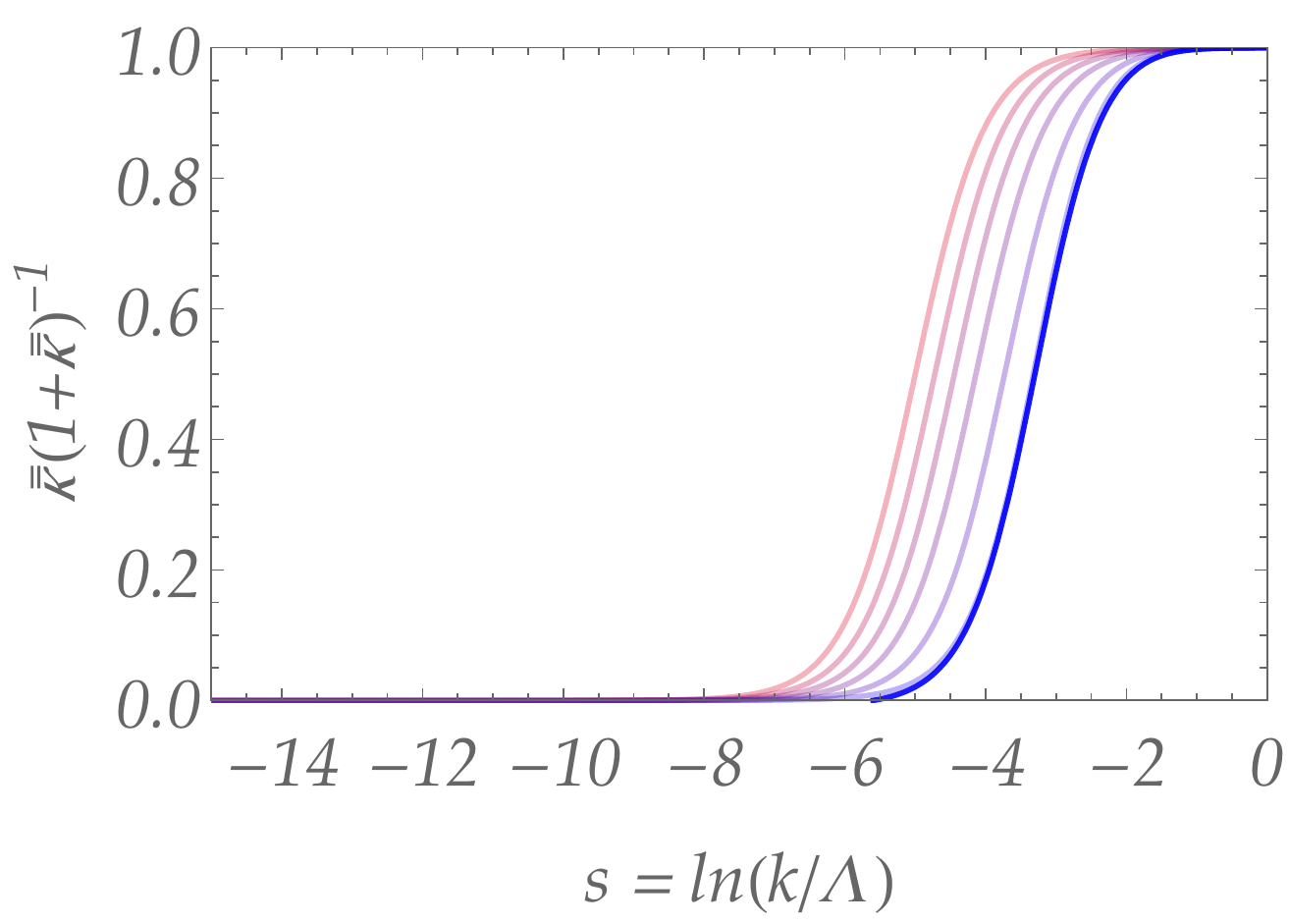} 
\caption{\label{Fig:RenormalizationGroupFlowKineticParameter}Upper panel: Renormalization group flow for the parameter $\dbar{\kappa}$ evaluated at the Wilson-Fisher FP ($T_{\Lambda} = T_{\Lambda,cr}$) for the $N = 1$ scalar model. We map the infinite range of values $\dbar{\kappa} \in \lbrack 0, \infty)$ to a compact interval $\dbar{\kappa} \mapsto \dbar{\kappa} (1+\dbar{\kappa})^{-1}$ in order to illustrate the $\dbar{\kappa}\rightarrow \infty$ limit. The solution at $\dbar{\kappa} = 0$ (dissipative dynamics) is IR attractive as indicated by the arrows on the horizontal axis. The limit $\dbar{\kappa} \rightarrow \infty$ corresponds to the relativistic fixed point and is IR repulsive. Lower panel: Dynamic crossover between the two scaling solutions $\dbar{\kappa} = 0$ and $\dbar{\kappa} \rightarrow \infty$ shown for different temperatures, $T_{\Lambda} \leq T_{\Lambda,cr}$ ($T_{\Lambda,cr}$: solid, blue curve). As the temperature increases the crossover scale is shifted towards the UV.}
\end{figure}

\begin{figure}[!t]
\centering
\includegraphics[width=0.41\textwidth]{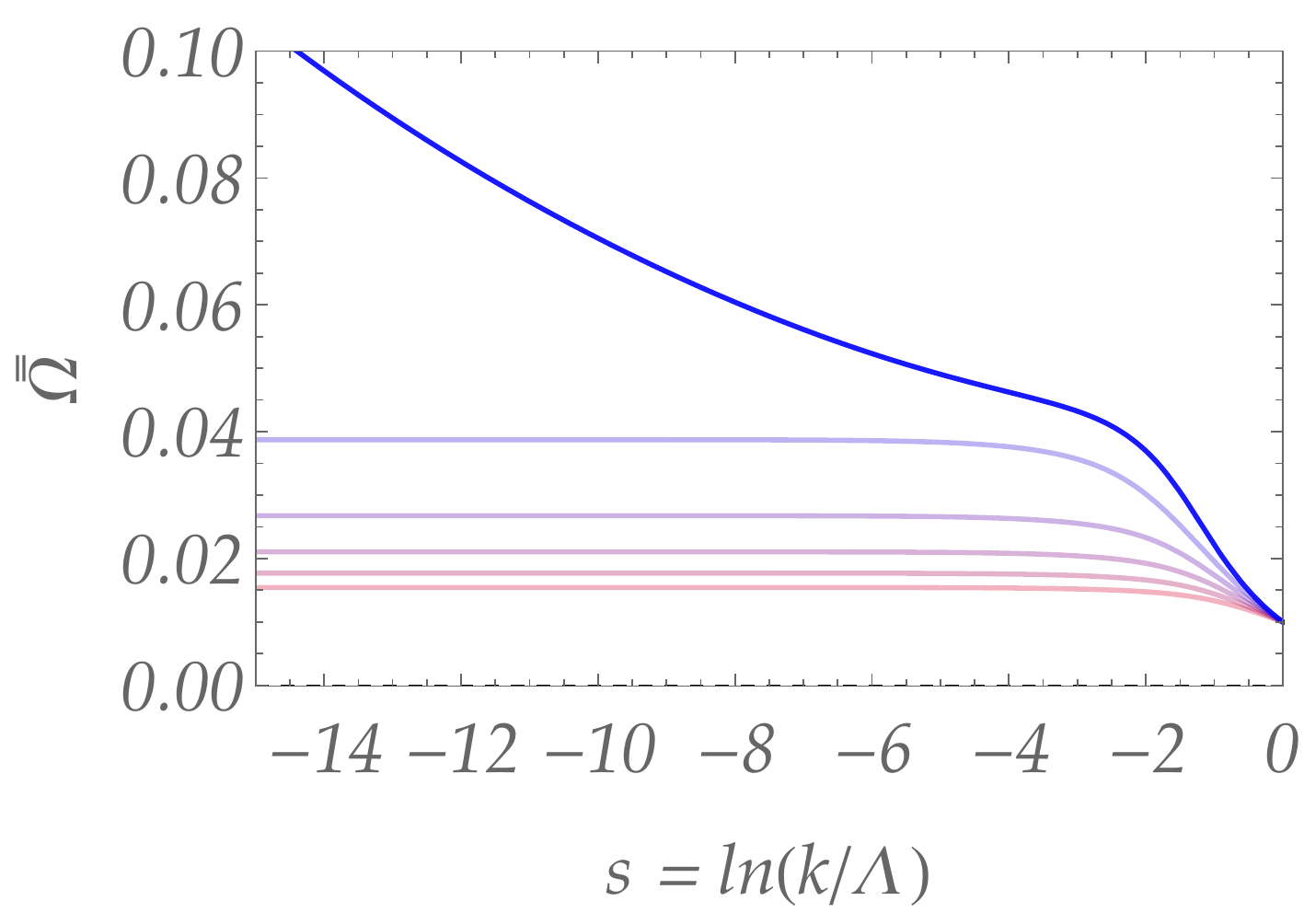} \\
\hskip 5pt \includegraphics[width=0.395\textwidth]{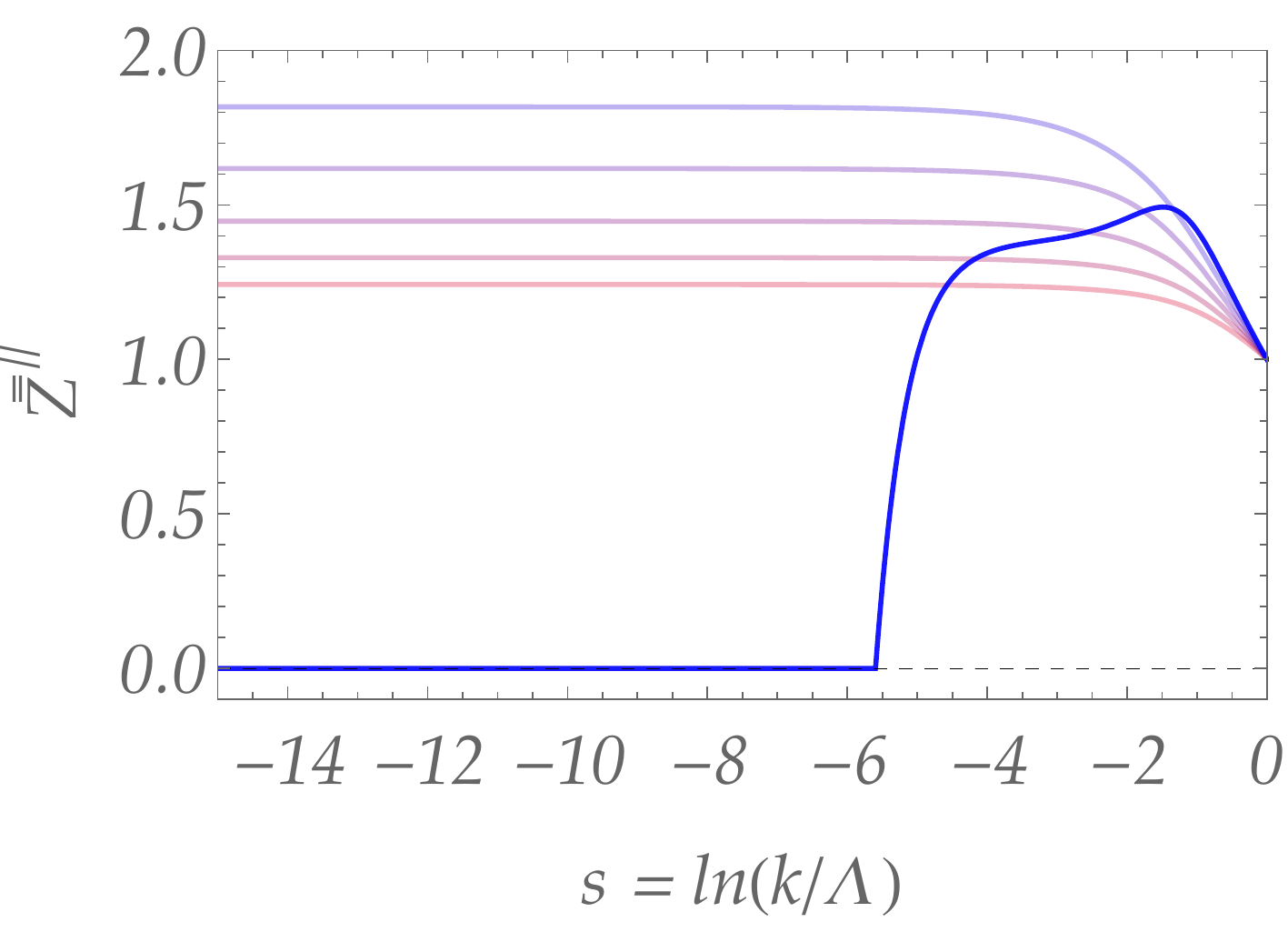}
\caption{\label{Fig:DynamicCrossover}The scale dependence of the renormalization parameters $\dbar{\Omega}$ and $\dbar{Z}^{\vert\!\vert}$ for the $N = 1$ scalar model is shown for different values of the temperature: $T_{\Lambda} \leq T_{\Lambda, cr}$ ($T_{\Lambda, cr}$: solid, blue curve). As the temperature is increased towards its critical value, the parameter $\dbar{\Omega}$ displays scaling behavior, i.e., $\dbar{\Omega} \sim e^{-s \eta^{\Omega}}$ and $\eta^{\Omega} = \textrm{const.}$, while $Z^{\vert\!\vert}$ vanishes at some finite scale before the asymptotic scaling regime is reached.}
\end{figure}

Note, that both anomalous dimensions $\eta^{\Omega}$ and $\eta^{\vert\!\vert}$ depend on the parameter 
\begin{equation}
\dbar{\kappa} = e^{2 s} \dbar{Z}^{\bot} \dbar{Z}^{\vert\!\vert} / ( T_{\Lambda} \dbar{\Omega} )^{2} . 
\end{equation}
Its significance is clear from the discussion at the end of Sec.\ \ref{SubSec:Low-energy effective theory}: It defines a ratio between the parameters, $\dbar{Z}^{\bot} / \dbar{Z}^{\vert\!\vert}$ and $\dbar{\Omega} /\dbar{Z}^{\vert\!\vert}$, that determine the low-energy dynamics of our model. In particular, we note the two interesting limiting scenarios: 1) $\dbar{\kappa} \rightarrow \infty$: $\dbar{Z}^{\vert\!\vert} = \dbar{Z}^{\bot} = 1$, $T_{\Lambda} \dbar{\Omega} \rightarrow 0^{+}$ corresponding to the microscopic model with relativistic dynamics and vanishing $\dbar{\Omega}$ ($T_{\Lambda} \neq 0$), and 2) $\dbar{\kappa} = 0$: $\dbar{Z}^{\vert\!\vert} = 0$, $\dbar{Z}^{\bot} \geq 1$, $T_{\Lambda} \dbar{\Omega} > 0$ in the classical regime.

To close the set of flow equations \eqref{Eq:NonvanishingTemperaturev2Flow} -- \eqref{Eq:FlowEquationKappa} we provide the flow equation for the coefficient $\dbar{\kappa}$
\begin{equation}
\frac{\partial \dbar{\kappa}}{\partial s} = \left\lbrack 2 ( 1 + \eta^{\Omega} ) - \eta^{\vert\!\vert} - \eta^{\bot} \right\rbrack \dbar{\kappa} = \big( z + \eta^{\Omega} - \eta^{\vert\!\vert} \big) \dbar{\kappa} ,
\label{Eq:FlowEquationKappa}
\end{equation}
and use the relation $z = 2 - \eta^{\bot} + \eta^{\Omega}$ that defines the dynamic critical exponent.

Even without solving the full set of equations, we may infer the dynamics of this model and determine the associated fixed points. However, this discussion is somewhat subtle -- substituting the given expressions for the anomalous dimensions into Eq.\ \eqref{Eq:FlowEquationKappa} one might conclude that there is only one solution for nonnegative $\dbar{\kappa}$, which is reached in the $\dbar{\kappa}\rightarrow \infty$ limit (see Fig.\ \ref{Fig:RenormalizationGroupFlowKineticParameter}). Further fixed points exist, but all of them lie in the region of $\dbar{\kappa} < 0$.\footnote{The presence of such solutions within our truncation might be related to the neglect of higher order $\mathcal{O}(\omega^{3})$ terms in the frequency expansion of $\Gamma^{(1,1)}$ that may stabilize the large-frequency behavior of the renormalization group flow.} Such negative solutions are clearly unphysical and are not allowed within our truncation (which requires that $\dbar{Z}^{\vert\!\vert}, \dbar{Z}^{\bot} \geq 0$, and $\dbar{\Omega} > 0$). Hence, when the renormalization group flow reaches $\dbar{\kappa} = 0$ we stop the evolution of $\dbar{Z}^{\vert\!\vert}$ which is then set to zero for the remaining flow. Thus, it appears that within our truncation only two independent scaling solutions exist, corresponding to $\dbar{\kappa} = 0$ ($Z^{\vert\!\vert} \equiv 0$) and $\dbar{\kappa} \rightarrow \infty$ ($Z^{\vert\!\vert} > 0$). Thermal fluctuations drive the system towards the IR stable FP at $\dbar{\kappa} = 0$ and we are able to identify a dynamic crossover (cf.\ \mbox{Figs.\ \ref{Fig:RenormalizationGroupFlowKineticParameter}, \ref{Fig:DynamicCrossover}}). This scenario is of course in line with our expectation that the dynamics is governed by nonrelativistic relaxational modes well below the thermal scale. We provide further interpretation of these results in the concluding section (Sec.\ \ref{Sec:Conclusions}) where we also comment on how different truncations might affect this picture of the dynamic crossover between the $\dbar{\kappa} = 0$ and $\dbar{\kappa} \rightarrow \infty$ scaling solutions.

\begin{figure}[!t]
\centering
\includegraphics[width=0.42\textwidth]{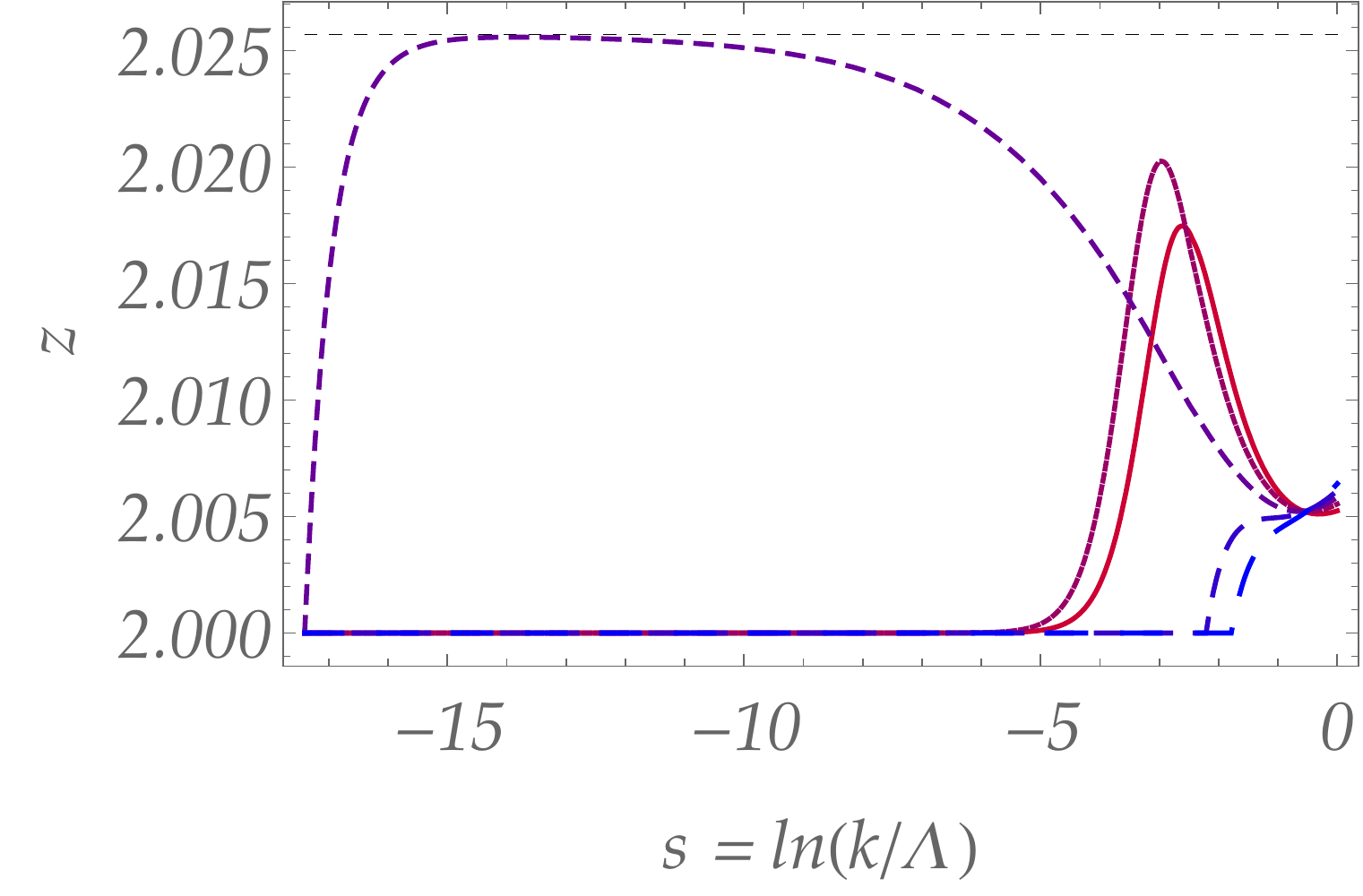}
\caption{\label{Fig:NonvanishingTemperatureRGFlow}$N = 1$ scalar model ($d = 3$, $T_{\Lambda} > 0$): As we tune the temperature $T_{\Lambda}$ through its critical value $T_{\Lambda, \textrm{cr}}$ at fixed values of $\dbar{v}_{\Lambda}^{2} = v^{2}_{\Lambda} / \Lambda = 0.078$ and $\dbar{\lambda}_{\Lambda} = \lambda_{\Lambda} / \Lambda = 1$, we observe a phase transition, where the theory exhibits scaling. We show the scale dependence of the dynamic scaling exponent $z = 2 - \eta^{\bot} + \eta^{\Omega}$ and find $z \simeq 2.025$ in the scaling regime.}
\end{figure}

Let us focus on this asymptotic scaling regime where the $\dbar{\kappa} = 0$ solution takes over ($Z^{\vert\!\vert} = 0$). We observe that in this regime, Eq.\ \eqref{Eq:AnomalousDimensionOmega} is identical to the anomalous dimension $\eta^{\Omega}$ of the relaxation coefficient derived in the context of the functional RG for \mbox{Model A} \cite{Canet:2006xu,Mesterhazy:2013naa}. We identify the following scaling solutions in this regime:
\begin{itemize}
\item[(I)] \emph{Gaussian fixed point}:
\begin{equation}
\dbar{\lambda}_{1,2} = \dbar{\lambda}_{2,3} = 0
\end{equation} 
and the anomalous dimensions vanish, i.e., $\eta^{\bot} = \eta^{\Omega} = 0$. The dynamic critical exponent is $z = 2$.
\item[(II)] \emph{Wilson-Fisher fixed point}  ($2 < d < 4$): The squared field expectation value $\dbar{v}^{2}$ and the quartic coupling $\dbar{\lambda}_{1,2}$ show a temperature dependence:
\begin{eqnarray}
\dbar{v}^{2}(T_{\Lambda}) &=& T_{\Lambda} \dbar{v}^{2}(1) , \label{Eq:TemperatureScalingFPFieldExpectationVal} \\
\dbar{\lambda}_{1,2}(T_{\Lambda}) &=& T_{\Lambda}^{-1} \dbar{\lambda}_{1,2}(1) , \label{Eq:TemperatureScalingFPLambda}
\end{eqnarray}
while
\begin{equation}
\dbar{\lambda}_{2,3}(T_{\Lambda}) = 0 ,
\end{equation}
at the fixed point. The anomalous dimensions are nonvanishing: $\eta^{\bot}, \eta^{\Omega} \neq 0$. 
\end{itemize} 
Although both FPs depend on the temperature, the scaling exponents do not. This is easily checked, if we substitute the scaling form \eqref{Eq:TemperatureScalingFPFieldExpectationVal} and \eqref{Eq:TemperatureScalingFPLambda} in our result for the anomalous dimension
\begin{equation}
\eta^{\bot} = \delta_{d}(0) \frac{\dbar{v}^{2}(1) \dbar{\lambda}_{1,2}^{2}(1)}{\left\lbrack 1 + \dbar{v}^{2}(1) \dbar{\lambda}_{1,2}(1) \right\rbrack^{2}} , 
\end{equation}
and the dynamic critical exponent
\begin{widetext}
\begin{eqnarray}
 z = 2 - \eta^{\bot} + \frac{\delta_{d}(\eta^{\bot})}{\dbar{v}^{2}(1)} \left\{ 1 + \frac{1}{\left\lbrack 1 + \dbar{v}^{2}(1) \dbar{\lambda}_{1,2}(1) \right\rbrack^{2}} - \frac{8}{\left\lbrack 2 + \dbar{v}^{2}(1) \dbar{\lambda}_{1,2}(1) \right\rbrack^{2}} \right\} .
\end{eqnarray}
\end{widetext}
We therefore conclude, that as a function of temperature both FPs describe a continuous line of phase transitions which lie in the same universality class. Depending on spatial dimension either one of the two FPs is IR stable and characterizes the scaling properties at the transition. In particular, it is the Wilson-Fisher FP that is stable below the upper critical dimension $d_{cr} = 4$ and we find that the dynamic scaling is determined by the universality class of Model A (within the limits of our truncation).

We proceed to discuss the behavior of the theory as we vary the temperature $T_{\Lambda}$ through its critical value $T_{\Lambda,\textrm{cr}}$ (in units of the cutoff scale $\Lambda$). Our results are illustrated at the example of the $N = 1$ scalar model in $d = 3$ dimensions, cf.\ \mbox{Figs.\ \ref{Fig:RenormalizationGroupFlowKineticParameter}, \ref{Fig:DynamicCrossover}, and \ref{Fig:NonvanishingTemperatureRGFlow}}. As we have already discussed the static properties of the three-dimensional Ising universality class in the previous section, we focus only the dynamic properties. As we have already argued, we expect a dynamic crossover well below the thermal scale. This is verified by examining Fig.\ \ref{Fig:DynamicCrossover} where we see that the renormalization coefficient $Z^{\vert\!\vert}$ runs to zero before the system enters the scaling region. The scaling behavior in this asymptotic regime $k \ll \Lambda$ is fully characterized by the Wilson-Fisher FP, for which we obtain the dynamic critical exponent $z \simeq 2.025$, in the dynamic universality class of Model A. Our result agrees well with Monte Carlo estimates: $z = 2.032(4)$ \cite{Grassberger:1995}, $z = 2.055(10)$ \cite{Ito:2000}, field-theory methods $z = 2.0237(55)$ \cite{Krinitsyn:2006}, and is consistent with the functional RG results from Ref.\ \cite{Canet:2006xu} to the given order of our truncation.

The following picture emerges: In the classical regime, we obtain a continuous line of phase transitions that lie in the $d$-dimensional static universality class of the $O(N)$ model. The critical exponents at the phase transition depend on nature of the IR stable fixed point. For $N \geq 2$ and $2 < d < 4$ the Wilson-Fisher FP is the stable one and we observe nontrivial scaling, i.e., $\eta \neq 0$, $z > 2$. For $d \geq 4$ the Gaussian FP takes over and we obtain mean-field exponents, $\eta = 0$, $z = 2$.  For $N = 1$ (Ising universality class), the lower critical dimension is \mbox{$d = 1$} and we observe nontrivial scaling for any dimension between $1 < d < 4$. The dynamic critical exponent is $z > 2$ for \mbox{$1 < d < 4$} (which is consistent with the results from Ref.\ \cite{Bausch:1981zz} and those compiled in Ref.\ \cite{Hohenberg:1977ym}).

\section{Limiting behavior of renormalization group equations in the classical regime}
\label{Sec:Limiting behavior of renormalization group equations in the classical regime}

Here, we examine the properties of the $O(N)$ Wilson-Fisher scaling solution in the classical regime ($T \neq 0$), by considering an expansion around the upper critical dimension of the model $d_{cr} = 4$ and the behavior in the large-$N$ limit. Although we do not expect to reproduce the known results from the $\epsilon$- or the $1/N$-expansion (which is due to threshold effects in the nonperturbative functional RG), this analysis nevertheless provides important insights into the quality of our truncation in these limiting cases.

\subsection{Expansion around the upper critical dimension}
\label{SubSec:Expansion around the upper critical dimension}

\begin{table*}[!t]
\renewcommand{\arraystretch}{1.6}
\renewcommand{\tabcolsep}{12pt}
\begin{tabular}{lccc}
& Regulator function & $a_{r} = X_{r} Y_{r;1} / Y_{r;2}^{2}$ & $c_{r} = Y_{r;3} / X_{r} - 1$ \\
\hline\hline
Exponential cutoff \cite{Wetterich:1991be,Tetradis:1992qt} & $r_{\textrm{exp}} = ( e^{y} - 1 )^{-1}$ & $1/2$ & $6 \ln \left(4 / 3\right) - 1$\\
Litim cutoff \cite{Litim:2000ci,*Litim:2001up} & $r_{\textrm{opt}} = \left( 1/y - 1 \right) \theta(1-y) $ & $1/2$ & $1/2$ \\
Sharp cutoff \cite{Wegner:1972ih,Liao:1992fm,*Liao:1999sh} & $r_{\textrm{sharp}} = 1/\theta(y-1) - 1$ & $\infty$ & $-1$ \\
\hline
$\epsilon$-expansion  \cite{Wilson:1972, Brezin:1972, Halperin:1972} && $1/2$ & $6 \ln \left(4 / 3\right) - 1$ \\
\hline\hline
\end{tabular}
\caption{\label{Tab:RegulatorFunctionsAnomalousDimension}Numerical coefficients $a_{r}$ and $c_{r}$ that determine the quality of a given cutoff in the local interaction approximation are compared with the $\mathcal{O}(\epsilon^{2})$ result from the $\epsilon$-expansion, i.e., $\eta^{\bot} = a_{r} \frac{N+2}{(N+8)^{2}} \epsilon^{2}$ and $z = 2 + c_{r} \eta^{\bot}$. Overall, the exponential regulator provides an $\mathcal{O}(\epsilon^{2})$ improvement for both $\eta^{\bot}$ and $\eta^{\Omega}$. This is in contrast to the optimized Litim and sharp cutoff functions that fail to reproduce the known results. Note, that the sharp cutoff limit and its shortcomings are well-documented in the literature (see, e.g., \cite{Morris:1994ki,Aoki:1996fn,Litim:2000ci,Berges:2000ew}).}
\end{table*}

As we have already noted in the previous section, the coupling $\dbar{\lambda}_{2,3}$ vanishes exactly at the Wilson-Fisher fixed point. Thus, the only relevant parameters that characterize this scaling solution are the coupling $\dbar{\lambda}_{1,2}$ and the field expectation value $\dbar{v}$. For the following discussion it will prove useful to introduce the parametrization: $x = \dbar{v}^2\dbar{\lambda}_{1,2}$ and $y = \delta_d(0)T_{\Lambda}\dbar{\lambda}_{1,2}$, for which fixed point equations take a somewhat simpler form
\begin{eqnarray}
0 &=& (2 - d - \eta^{\bot}) x + \frac{\delta_{d}(\eta^{\bot})}{\delta_{d}(0)} \!\left\{ \frac{3}{(1 + x)^{2}} + N-1 \right\} y , \nonumber\\ && \label{Eq:FiniteTFixedPointCondition01} \\
0 &=& (d - 4 + 2\eta^{\bot} ) + \frac{\delta_{d}(\eta^{\bot})}{\delta_{d}(0)} \!\left\{ \frac{9}{(1 + x)^{3}} + N-1 \right\} y . \nonumber\\ && \label{Eq:FiniteTFixedPointCondition02}
\end{eqnarray}
The anomalous dimension is given by
\begin{equation}
\eta^{\bot} = \frac{x y}{(1 + x)^{2}} ,
\end{equation}
and dynamic critical exponent takes the form
\begin{equation}
z = 2 + \frac{\delta_{d}(\eta^{\bot})}{\delta_{d}(0)} \frac{2 (x+1)}{(x+2)^2} \eta^{\bot} . 
\end{equation}
Equations \eqref{Eq:FiniteTFixedPointCondition01} and \eqref{Eq:FiniteTFixedPointCondition02} imply that the leading contribution to the parameters $x$ and $y$ in an expansion around the upper critical dimension $\epsilon = 4 - d$ is of order $\epsilon$. Up to second order, we find
\begin{widetext}
\begin{eqnarray}
&& \hspace{-10pt} x = \frac{N+2}{2 (N+8)} \epsilon + \frac{(N+2)^{2}(N+33)}{4(N+8)^3} \epsilon^{2} + \mathcal{O}(\epsilon^{3}) , \\
&& \hspace{-10pt} y = \frac{1}{N+8} \epsilon + \frac{25 (N+2)}{2 (N+8)^3} \epsilon^{2} + \mathcal{O}(\epsilon^{3}) .
\end{eqnarray}
Thus, within our truncation the anomalous dimension is given by
\begin{equation}
\eta^{\bot} = \frac{1}{2} \frac{N+2}{(N+8)^2} \epsilon^{2} + \frac{N+2}{(N+8)^{2}} \left\lbrack \frac{14 N + 37}{(N+8)^2} - \frac{1}{4} \right\rbrack \epsilon^{3} + \mathcal{O}(\epsilon^{4}) ,
\label{Eq:Epsilon_Expansion_Static_Critical_Exponent}
\end{equation}
to third order in $\epsilon$, while the dynamic critical exponent reads
\begin{equation}
z = 2 + \left\lbrack \frac{1}{2} + \mathcal{O}(\epsilon^{2}) \right\rbrack \eta^{\bot} ,
\label{Eq:Epsilon_Expansion_Dynamic_Critical_Exponent}
\end{equation}
to the same order as the anomalous dimension. Let us comment to what extent the limiting behavior \eqref{Eq:Epsilon_Expansion_Static_Critical_Exponent} and \eqref{Eq:Epsilon_Expansion_Dynamic_Critical_Exponent} reproduces the known $\epsilon$-expansion result. For the anomalous scaling exponent $\eta^{\bot}$ we observe that the $\mathcal{O}(\epsilon^{2})$ contribution is exactly reproduced, while higher-order terms appear to be inconsistent \cite{Wilson:1972, Brezin:1972}. This is in contrast to other critical exponents. In particular, from the eigenvalues of the stability matrix at the Wilson-Fisher FP, we determine the correlation length exponent
\begin{eqnarray}
\nu &=& \frac{1}{2} + \frac{(N+2)}{4(N+8)} \epsilon + \frac{(N+2)}{8(N+8)^{3}} \left\lbrack N^{2} + \frac{1}{3} (152 N + 574) \right\rbrack \epsilon^{2} + \mathcal{O}(\epsilon^{3}) ,
\end{eqnarray}
\end{widetext}
and Wegner's exponent $\omega$
\begin{equation}
\omega = \epsilon - \frac{25(N+2)}{2(N+8)^{2}} \epsilon^{2} + \mathcal{O}(\epsilon^{3}) ,
\end{equation}
which governs corrections to asymptotic scaling. Since there are only two independent exponents at the FP, we may determine the remaining exponents by use of scaling relations below the upper critical dimension \cite{Fisher:1967}, e.g., we may apply Fisher's scaling relation $\gamma = \left( 2 - \eta^{\bot} \right)\! \nu$ to determine the susceptibility exponent
\begin{eqnarray}
\hspace{-35pt}\gamma &=& 1 + \frac{(N+2)}{2(N+8)} \epsilon + \frac{(N+2)}{4(N+8)^{3}} \nonumber\\
&& \hspace{10pt} \times\: \left\lbrack N^{2} + \frac{1}{3} (149 N + 550) \right\rbrack \epsilon^{2} + \mathcal{O}(\epsilon^{3}) .
\end{eqnarray}
We observe that the given exponents and scaling relation match those of the $\epsilon$-expansion \cite{ZinnJustin:2002ru} perfectly to linear order however, discrepancies appear for contributions of order $\epsilon^{2}$. In fact, linear contributions in $\epsilon$ are in general independent of the chosen regulator function \cite{Ball:1994ji,Aoki:1996fn}, while this does not hold true for higher order contributions (at least within the employed truncation of the effective action).

We demonstrate the regulator dependence of the two-loop result explicitly for the anomalous dimension which is seemingly consistent at $\mathcal{O}(\epsilon^{2})$. Consider a momentum-dependent regulator $R(\boldsymbol{p}) = Z^{\bot} \boldsymbol{p}^{2} r(\boldsymbol{p}^{2}/k^{2})$ that is general arbitrary up to the specification of $r = r(y)$, $y = \boldsymbol{p}^{2}/k^{2}$. This function that implements the IR cutoff, should be nonnegative $r(y) \geq 0$ and we impose the following limiting properties: $\lim_{y \rightarrow 0} r(y) \simeq 1 /y \rightarrow \infty$ and decay sufficiently fast at infinity, i.e., $\lim_{y\rightarrow \infty} r(y) \rightarrow 0$ and $\lim_{y\rightarrow \infty} r'(y) \rightarrow 0$ (cf.\ Sec.\ \ref{Sec:Nonperturbative functional renormalization group in the real-time formalism}). Using a regulator function that satisfies these conditions but is otherwise arbitrary, we derive the general expression for the anomalous dimension to leading order $\mathcal{O}(\epsilon^{2})$ in the local interaction approximation of the $O(N)$ model:
\begin{equation}
\eta^{\bot} = \frac{(N+2) \epsilon^{2}}{(N+8)^{2}} \frac{X_{r} Y_{r;1}}{Y_{r;2}^{2}} + \mathcal{O}(\epsilon^{3}) ,
\label{Eq:AnomalousDimensionLeadingOrderRegulatorDependence}
\end{equation}
where the coefficients $X_{r}$ and $Y_{r;n}$ are given by
\begin{eqnarray}
X_{r} &=& \int_{0}^{\infty} dy \left(\frac{d}{dy} \frac{1}{1 + r(y)}\right)^{2} , \label{Eq:CoefficientXr} \\
Y_{r;n} &=& \int_{0}^{\infty} dy \, y^{2-n} \frac{d}{dy} \frac{1}{\left( 1 + r(y) \right)^{n}} . \label{Eq:CoefficientYr}
\end{eqnarray}
Their numeric values depend on the specific choice of the regulator function. Let us consider a few examples: We start with the optimized Litim regulator
\begin{equation}
r_{\textrm{opt}}(y) = \left( 1/y - 1 \right) \theta(1-y) ,
\label{Eq:LitimRegulatorShapeFunction}
\end{equation} 
employed in this work, cf.\ Secs.\ \ref{Sec:Quantum regime} -- \ref{Sec:Classical regime}. The corresponding coefficients read $X_{\textrm{opt}} = 1$ and $Y_{\textrm{opt};n} = n/2$. Thus we see that the ratio $X_{\textrm{opt}} Y_{\textrm{opt};1} \left(Y_{\textrm{opt};2}\right)^{-2} = 1/2$, and we confirm our result Eq.\ \eqref{Eq:Epsilon_Expansion_Static_Critical_Exponent} for the anomalous dimension. To check how \eqref{Eq:LitimRegulatorShapeFunction} compares to other regulators, we consider the exponential cutoff
\begin{equation}
r_{\textrm{exp}}(y) = ( e^{y} - 1 )^{-1} ,
\end{equation}
which is also frequently used and serves as a useful benchmark. This function yields $X_{\textrm{exp}} = 1/2$ and the lowest coefficients $Y_{\textrm{exp};n}$ in the series $n = 1, 2, 3, \ldots$ are $1$, $1$, $3 \ln(4/3)$, \ldots, etc. Thus, we obtain the ratio $X_{\textrm{exp}} Y_{\textrm{exp};1} \left(Y_{\textrm{exp};2}\right)^{-2} = 1/2$ and the same leading-order result in the $\epsilon$-expansion for $\eta^{\bot}$ as for the Litim regulator. From these examples we conclude that although the result to two-loop order, Eq.\ \eqref{Eq:AnomalousDimensionLeadingOrderRegulatorDependence} is regulator dependent in general, that it is nevertheless possible to find a class of regulator functions with improved behavior at $\mathcal{O}(\epsilon^{2})$. A counterexample that fails to produce the known result for $\eta^{\bot}$ to leading order is the sharp cutoff \cite{Wegner:1972ih,Liao:1992fm,*Liao:1999sh}, which is known to introduce strong regulator artifacts \cite{Liao:1992fm,*Liao:1999sh,Litim:2000ci,Berges:2000ew}.

For the dynamic critical exponent $z$, the result from the $\epsilon$-expansion is expressed in the form $z = 2 + c \eta^{\bot}$ \cite{Hohenberg:1977ym}, with the coefficient $c = 0.7261 \left( 1 - 1.69 \epsilon + \mathcal{O}(\epsilon^{2}) \right)$ \cite{Halperin:1972}. Here, we observe a clear difference to order $\epsilon^{2}$, where we find $c = 1/2 + \mathcal{O}(\epsilon^2)$ within our truncation, and employing the Litim cutoff. While the anomalous dimension $\eta^{\bot}$ is correctly reproduced to order $\epsilon^{2}$ for this choice of the regulator function, this must not necessarily hold for $\eta^{\Omega}$. To parametrize the regulator dependence of our results, we write $\eta^{\Omega}$ as
\begin{equation}
\eta^{\Omega} = \frac{Y_{r;3}}{X_{r}} \eta^{\bot} + \mathcal{O}(\epsilon^{3}) ,
\end{equation}
where the coefficients $X_{r}$ and $Y_{r;n}$ are defined in Eqs.\ \eqref{Eq:CoefficientXr} -- \eqref{Eq:CoefficientYr}. From the definition of the dynamic critical exponent $z = 2 - \eta^{\bot} + \eta^{\Omega} = 2 + c \eta^{\bot}$, we may therefore express $c$ in the form:
\begin{equation}
c = \frac{Y_{r;3}}{X_{r}} - 1 .
\end{equation}
Using the previously derived values, we obtain to leading order in $\epsilon$: $c_{\textrm{opt}} = 1/2$ for the optimized Litim cutoff, while for the sharp cutoff we find: $c_{\textrm{sharp}} = -1$. Considering this discrepancy, it is quite surprising to note that the same coefficient takes a different value for the exponential cutoff: $c_{\textrm{exp}} = 6 \ln \left(\frac{4}{3}\right) - 1 \simeq 0.72609$, in perfect agreement with the $\epsilon$-expansion.

We may conclude that in general the scaling exponents, both static and dynamic, are regulator dependent at $\mathcal{O}(\epsilon^{2})$, while the result to linear order in $\epsilon$ is correctly reproduced independent of the choice of cutoff (see also \mbox{Refs.\ \cite{Ball:1994ji,Aoki:1996fn}}). Nevertheless, certain regulators might show improved behavior at $\mathcal{O}(\epsilon^{2})$ (at least within the scope of the employed truncation of the scale-dependent effective action). A summary of these results is shown in Tab.\ \ref{Tab:RegulatorFunctionsAnomalousDimension}.

\subsection{Large-$N$ expansion}
\label{SubSec:Large-N expansion}

In the same representation of the parameters \eqref{Eq:FiniteTFixedPointCondition01} and \eqref{Eq:FiniteTFixedPointCondition02}, we may consider the large-$N$ behavior as an alternative approach to understand the properties of our RG equations. In this limit, we find that 
\begin{eqnarray}
x &=& \frac{4-d}{d-2} + \mathcal{O}\left(\! \frac{1}{N^{2}} \!\right) , \\
y &=& (4-d) \frac{1}{N} + \mathcal{O}\left(\! \frac{1}{N^{2}} \!\right) ,
\end{eqnarray}
up to order $1/N$. The result for the anomalous dimension reads
\begin{equation}
\eta^{\bot} = \frac{(d-2)(4-d)^{2}}{4} \frac{1}{N} + \mathcal{O}\left(\! \frac{1}{N^{2}} \!\right) ,
\label{Eq:EtaLargeNResult}
\end{equation}
while the dynamic critical exponent is given by
\begin{equation}
z = 2 + \left\lbrack \frac{4(d-2)}{d^2} + \mathcal{O}\!\left(\frac{1}{N}\right) \right\rbrack \eta^{\bot} ,
\end{equation}
In $d = 3$ dimensions results are readily available in the literature: In the large-$N$ limit the value for $\eta^{\bot}$ is given by $\eta^{\bot} = \frac{8}{3\pi^{2}} \frac{1}{N}$ \cite{Ma:1973}. Clearly, this result does not match the limiting behavior in Eq.\ \eqref{Eq:EtaLargeNResult} which is known to exhibit a regulator dependence within our finite truncation of the vertex expansion \cite{Tetradis:1993ts}. The difference can be explained by the infinite resummation of diagrams in the large-$N$ result, but which cannot be accounted for in the context of our \emph{ansatz} for the CTP effective action. Our result for the dynamic critical exponent should be compared with predictions for the dynamic critical behavior of Model A, for which the dynamic exponent reads $z = 2 + c \eta^{\bot}$, with $c = 0.5$ in the large-$N$ limit \cite{Hohenberg:1977ym, Suzuki:1975}.

\section{Quantum-classical transition}
\label{Sec:Quantum-classical transition}

\begin{figure*}[!t]
\centering
\includegraphics[width=0.345\textwidth]{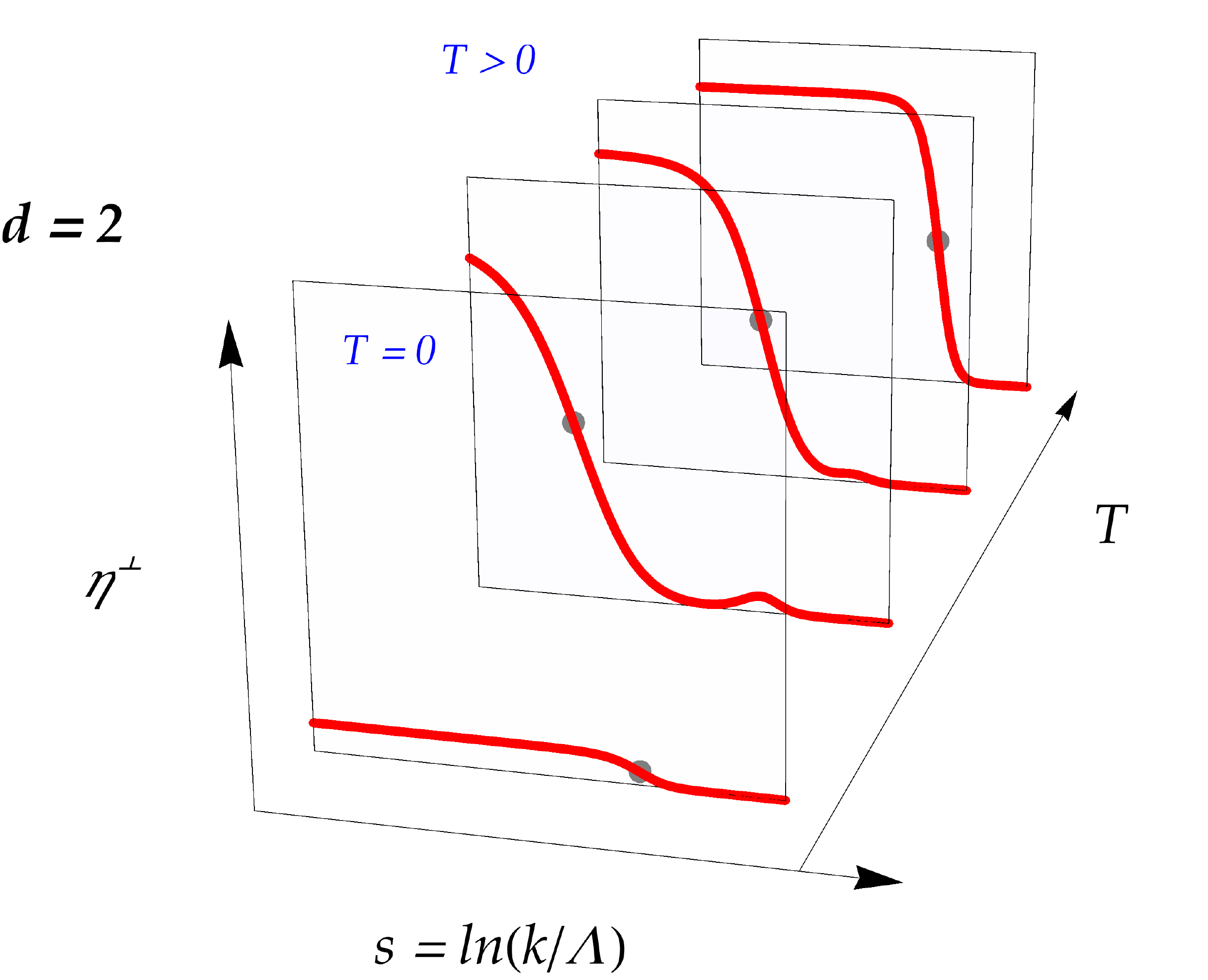} \hskip 20pt
\includegraphics[width=0.345\textwidth]{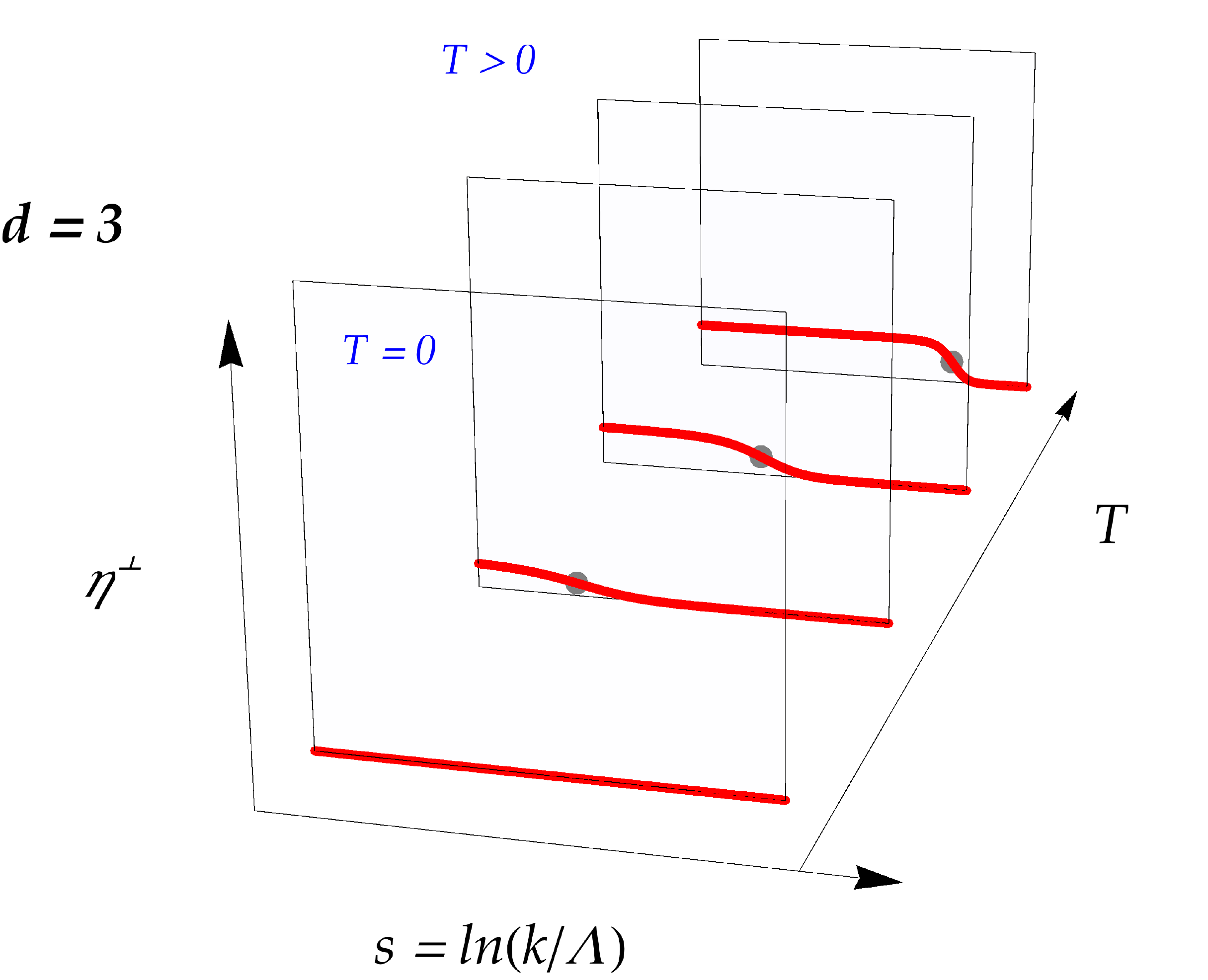} 
\caption{\label{QuantumClassicalCrossoverSchematic}$N = 1$ scalar model for spatial dimensions $d = 2$ and $d = 3$: Schematic scale dependence of the anomalous dimension $\eta^{\bot}$. We tune the bare field expectation value $v_{\Lambda}^{2}$ as a function of the temperature, so that the theory exhibits scaling in the $k\rightarrow 0$ limit. We assume that the bare coupling $\lambda_{\Lambda}$ is fixed at a given value. In $d = 2$ dimensions, where the theory admits a QCP characterized by nontrivial scaling the system features a crossover scale to the IR scaling region (left panel). This scale is set by the quartic coupling. As we turn on temperature, quantum critical fluctuations die out and thermal fluctuations take over. In $d = 3$ dimensions (right panel) the QCP lies in the $(3+1)$-dimensional (Euclidean) universality class which is characterized by mean-field scaling -- fluctuations become important only at nonvanishing temperatures. The curves are shown to scale with $\eta^{\bot} = 1/4$ $(d = 2)$ and $\eta^{\bot} = 0.03639(15)$ $(d = 3)$ for the Ising model.}
\end{figure*}

Although in principle the flow equations can be derived also for the intermediate temperature region (\mbox{$0 < T_{\Lambda} \lesssim 1$}) we find that they take a rather complicated form and therefore require a more elaborate numerical treatment. We postpone this to future work. Here, we examine possible scenarios of the quantum-to-classical crossover based on our understanding of the low and high-temperature limits, developed in Secs.\ \ref{Sec:Quantum regime} and \ref{Sec:Classical regime}. The general picture of the phase structure is the following: The $d$-dimensional $O(N)$ theory is characterized by a continuous line of classical phase transitions that terminates at $T = 0$. Along this line the dynamic scaling exponent is $z \geq 2$. At the end point, the system lies in a different universality class, namely that of the $(d+1)$-dimensional Euclidean theory with $z = 1$.

Essential to the picture of the dynamic crossover is the observation that both in the quantum and classical case we may find two fixed points: the Gaussian FP ($\eta = 0$) and Wilson-Fisher FP ($\eta \neq 0$). The existence of a nontrivial IR stable FP depends subtly on the dimensionality and temperature of the system. In contrast, the dynamic scaling properties turn out to be much more simple -- they completely decouple from the statics for any $T > 0$. This is immediately clear from the inspection of the RG flow Eqs.\ \eqref{Eq:NonvanishingTemperaturev2Flow} -- \eqref{Eq:FlowEquationKappa} where we see that the parameter $\dbar{\kappa}$ and the anomalous dimensions $\eta^{\Omega}$ and $\eta^{\vert\!\vert}$ do not feed back into the RG flow of the static quantities. In fact, we have used this property explicitly to determine the dynamic FPs of the RG flow, i.e., $\dbar{\kappa} = 0$ and $\dbar{\kappa} = \infty$ for any $T > 0$. Thus, the problem of determining the dynamics at nonvanishing temperature reduces to the problem of understanding how the Wilson-Fisher FP and in particular how $\eta^{\bot}$ depends on the effective dimension $d_{\varepsilon}$ (cf.\ Sec.\ \ref{SubSubSec:Classical scaling regime}). Between the quantum and classical regime, $\eta^{\bot}$ varies continuously with $d_{\varepsilon}$, from $d + 1$ to $d$. This fact can be understood from the perspective of Euclidean finite size systems where one observes a continuous dimensional crossover \cite{Uzunov:1995}.

Possible scenarios for the RG flow are illustrated in Fig.\ \ref{QuantumClassicalCrossoverSchematic} at the example of the $N = 1$ scalar theory. We examine the scale-dependent anomalous dimension $\eta^{\bot} = - (\partial / \partial s) \ln Z^{\bot}$ which provides information on relevant fluctuations in different regions of the RG flow. The maximum achievable correlation length $\xi = m_{R}^{-1}$ is determined by how close the microscopic parameters have been tuned to their respective critical values, and we assume that $\xi \rightarrow \infty$ in the limit $k\rightarrow 0$. For small temperatures ($0 \lesssim T_{\Lambda} \ll 1$), the RG flow enters a regime where quantum fluctuations are present but do not play out their full effect. Instead, close to the classical phase transition the characteristic scale at which fluctuations become important are fully determined by the temperature. We point out that the possible situations illustrated in Fig.\ \ref{QuantumClassicalCrossoverSchematic} have to be modified in the case of a system with continuous $O(N)$ symmetry, where for $d = 2$ and $N = 2$ the system experiences Kosterlitz-Thouless transition, while a thermal phase transition is absent for $N \geq 3$.

\section*{Conclusions}
\label{Sec:Conclusions}

In this work, we have applied the nonperturbative functional RG in the real-time formalism to the problem of the low-energy dynamics of a relativistic $O(N)$ theory. Our approach relies on a truncation of the scale-dependent CTP effective action and we have shown that our \emph{ansatz} admits two possible fixed points at $T > 0$ that determine the low-energy dynamics of the order parameter. Only one of them is IR stable which characterizes the relaxational dynamics of a dissipative mode. We find that the dynamic universality class associated to this scaling solution is identical to that of Model A, for which $z = 2 + c \eta^{\bot}$ \cite{Halperin:1972,Hohenberg:1977ym}. This observation is based on an exact equivalence of the flow equations derived in this work in the IR scaling regime, with those obtained in the framework of the effective model given in Refs.\ \cite{Canet:2006xu} and \cite{Mesterhazy:2013naa}. At $T = 0$, we find only one fixed point, with unitary dynamics $z = 1$ protected by Lorentz symmetry. We show that there is a continuous crossover that smoothly connects the microscopic theory ($z = 1$) with a nonrelativistic effective theory ($z \geq 2$) that corresponds to Model A, if the theory is tuned to criticality.

We emphasize that the existence of the \mbox{Model A} fixed point in the high-temperature limit (within our truncation of the scale-dependent effective action) is essentially required by consistency of our \emph{ansatz} (cf.\ Sec.\ \ref{Sec:Classical regime}). The stability of the theory (i.e., $\dbar{Z}^{\vert\!\vert}, \dbar{Z}^{\bot} \geq 0$, $\dbar{\Omega} > 0$) had to be imposed to rule out a renormalization group flow to negative values of $\dbar{Z}^{\vert\!\vert}$ that are clearly unphysical. Although the relativistic UV fixed point seems to be unstable, which is certainly expected at nonvanishing temperatures, the apparent lack of a solution where $\dbar{Z}^{\vert\!\vert} = 0$ is reached asymptotically, is certainly surprising. We suspect that this is an artifact of our truncation that includes only a linear contribution in the frequency-expansion of the self-energy and is therefore insufficient to capture the large-frequency behavior of correlation or response functions. To reach a conclusion on this issue other types of approximations will need to be employed that take into account the full frequency-dependence of the self-energy and vertex functions \cite{Jakobs:2010,Husemann:2012,Giering:2012,Uebelacker:2012}.

Within our truncation, we were able to observe a dynamic crossover induced by thermal fluctuations. However, as far as the low-energy limit is concerned, one might also also imagine a different scenario, where the system features a crossover to the nonrelativistic limit well above the thermal scale. This would have dramatic consequences, since the corresponding nonrelativistic action conserves particle number \cite{Evans:1995yz,ZinnJustin:2002ru}. Thus, the question that remains is: What is the nature of the nonrelativistic limit? We see that an answer to this question is intimately tied to that of the dynamic universality class for the given microscopic theory. In either one of the above scenarios the nonrelativistic action will feature different symmetries, that determine the presence or absence of further slow modes at the phase transition. To capture these possibilities different truncations of the scale-dependent effective action will have to be considered, that are able to take into account conservation laws and Ward identities consistently.

Thus, within our current \emph{ansatz} for the CTP effective action Eq.\ \eqref{Eq:1PIGeneratingFunctionalAnsatz} it therefore not possible to provide a definitive answer to the question of the dynamic universality class of the $O(N)$ theory. It is only possible to follow the transition from the microscopic coherent dynamics (without dissipation) to the situation where the dynamics is dissipative nonrelativistic, without additional conserved quantities -- which corresponds to \mbox{Model A} dynamics in the critical region. The simplest possible extension of this work would be to allow for a diffusive relaxation of the order parameter in addition to a purely dissipative behavior. This might allow one to distinguish between two dynamic fixed points associated to the universality classes of \mbox{Model A} and \mbox{Model B} in the Halperin-Hohenberg classification. A corresponding \emph{ansatz} for the CTP effective action should include a momentum-dependent as well as a field-dependent kinetic coefficient, i.e., $\Omega_{a b} = \Omega_{a b}(\phi) + \Omega'_{a b}(\phi) \boldsymbol{p}^{2} + \mathcal{O}(\boldsymbol{p}^{4})$. We do not expect this to be sufficient however, to address the dynamic properties of the $O(N)$ model in the low-energy limit. Certainly, the interplay of massless Nambu-Goldstone modes, and the possible presence of particle-number and energy-conservation allows for very complicated dynamics with competing mode couplings. To distinguish between generic dynamic universality classes, we expect that a nonlocal expansion in the vertices (both in space and in time) is necessary (see, e.g., Refs.\ \cite{Jakobs:2010,Husemann:2012,Giering:2012,Uebelacker:2012,Tanizaki:2013yba}). Such a treatment is in particular required to address the issue of the nature of the nonrelativistic limit. We refer to future work which is currently in progress.

\begin{acknowledgments}
We thank J.~Berges, L.~F.~Palhares, M.~A.~Stephanov, and H.-U.~Yee for fruitful discussions. This research is supported by DOE grant number No.\ DE-FG0201ER41195 and the European Research Council under the European Union’s Seventh Framework Programme (FP7/2007-2013) / ERC grant agreement 339220. The work of Y.~T. is supported by the JSPS Research Fellowships for Young Scientists (No.25-6615), by the RIKEN iTHES project, the JSPS Strategic Young Researcher Overseas Visits Program for Accelerating Brain Circulation, and by the Program for the Leading Graduate Schools, MEXT, Japan. 
\end{acknowledgments}

\begin{appendix}

\begin{widetext}

\section{Renormalization group equations}
\label{Sec:Renormalization group equations}

The flow equations for the $(m+n)$-point functions $\Gamma^{(m,n)}$ are derived from Eq.\ \eqref{Eq:FlowEquation} by taking functional derivatives with respect to the fields and evaluating these expressions in the homogeneous background field configuration \mbox{$\phi_{a} = v \delta_{a 1}$}, $v \geq 0$, and \mbox{$\tilde{\phi} = 0$}. Within our truncation, we assume that all $(m+n)$-point functions $\Gamma^{(m,n)}$, with $m + n > 4$ are zero. Furthermore, $4$-point functions are assumed to be local in both space and time, while this not true in general for $2$- or $3$-point functions. The presence of a nonlinear fluctuation theorem implies that only two of the $4$-point vertices $\Gamma^{(1,3)}$ and $\Gamma^{(3,1)}$ are nonzero (cf.\ Sec.\ \ref{SubSec:Local interaction approximation}). With these approximations we obtain the flow equations for the (amputated) 1PI two-point function $\Gamma^{(1,1)}$
\begin{equation}
\frac{\partial}{\partial s} \Gamma^{(1,1)} (p) = -\frac{1}{2} \int_{q} \frac{\partial R}{\partial s}\frac{\partial}{\partial R} \Tr  \left\{ F(q) \Gamma^{(3,1)} + 2 F(q) \Gamma^{(2,1)} (-p, p + q) \Re G^{\textrm R}(p+q) \,\Gamma^{(2,1)}(p, q)  \right\} , 
\label{Eq:Gamma11}
\end{equation}
and the (nonvanishing) 1PI $3$- and $4$-point functions
\begin{eqnarray}
\frac{\partial}{\partial s} \Gamma^{(2,1)} (p,p') &=& - \int_{q} \frac{\partial R}{\partial s} \frac{\partial}{\partial R} \Tr \left\{ F(q) \Gamma^{(2,1)}(-q, p + p' + q) \Re G^{\textrm R}(q) \Gamma^{(3,1)} + \ldots ~\right\} , \label{Eq:Gamma21} \\
\frac{\partial}{\partial s} \Gamma^{(3,1)} &=& - \int_{q} \frac{\partial R}{\partial s} \frac{\partial}{\partial R} \Tr \left\{ F(q) \Gamma^{(3,1)} \Re G^{\textrm R}(q) \Gamma^{(3,1)} + \ldots ~\right\} , \label{Eq:Gamma31} \\
\frac{\partial}{\partial s} \Gamma^{(1,3)} &=& - \int_{q} \frac{\partial R}{\partial s} \frac{\partial}{\partial R} \Tr \left\{ F(q) \Gamma^{(3,1)} \Re G^{\textrm R}(q) \Gamma^{(1,3)} + \ldots ~\right\} . \label{Eq:Gamma13} 
\end{eqnarray}
We define the four-momenta $p = (p^{0}, \boldsymbol{p})$, $q = (q^{0}, \boldsymbol{q})$, etc.\ and the frequency-momentum integration $\int_{q} \cdots \equiv \frac{1}{(2\pi)^{d+1}}\int dq^{0} d^{d}\boldsymbol{q}$, while the trace $\Tr \,\{ \cdots \}$ runs over internal field indices. In Eqs.\ \eqref{Eq:Gamma11} -- \eqref{Eq:Gamma13} we have pulled out the derivative operator, $(\partial R / \partial s) ( \partial / \partial R )$ which acts on the regulator function. This greatly simplifies the evaluation of the terms on the RHS. In particular, using a frequency-independent regulator (at nonvanishing temperature), we may calculate the frequency integral on the RHS before evaluating the $R$-derivative. Finally, those terms that are omitted in Eqs.\ \eqref{Eq:Gamma21} -- \eqref{Eq:Gamma13} (denoted by the ellipsis) correspond to all possible contributions (allowed by the diagrammatic rules) where the $4$-point vertices of the first term on the RHS are substituted by the dumbbell diagram as illustrated in Fig.\ \ref{Fig:Diagram}. This is done explicitly in Eq.\ \eqref{Eq:Gamma11}.

\begin{figure*}[!h]
\centering
\includegraphics[width=0.75\textwidth]{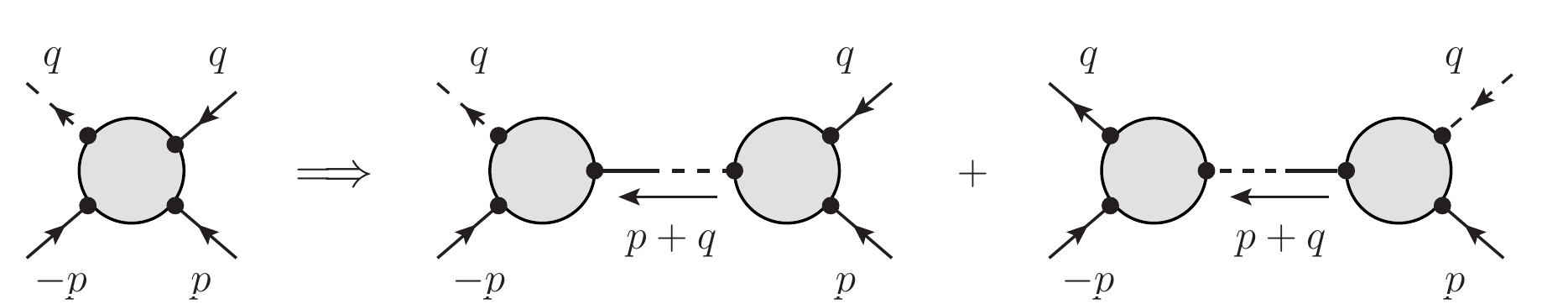}
\caption{\label{Fig:Diagram}Substitution rule illustrated at the example of the amputated $4$-point function $\Gamma^{(3,1)}$ in terms of two $3$-point vertices $\Gamma^{(2,1)}$ (cf.\ Eq.\ \eqref{Eq:Gamma11}). External lines indicate the inflowing momenta and the inserted fields in the RA-representation (full lines correspond to $\phi$ field insertions, while dashed lines to insertions of the $\tilde{\phi}$ field). The notation $\Gamma^{(2,1)}(p,q)$ refers only the incoming momenta. The full-dashed lines which connect the two vertices on the RHS corresponds to (nonperturbative) retarded/advanced propagators, i.e., $G^{\textrm{R}/\textrm{A}}$.}
\end{figure*}

Equations \eqref{Eq:Gamma21} -- \eqref{Eq:Gamma13} provide the RG flow of the generalized potential in the limit of zero external momenta, i.e., $p \rightarrow 0$ and $p' \rightarrow 0$. Similarly, we obtain the flow equations for the wavefunction renormalization $Z^{\bot}$ and the dynamic parameters $\Omega$ and $Z^{\vert\!\vert}$, of our model by a suitable projection onto the flow equation for $\Gamma^{(1,1)}$:
\begin{eqnarray}
\frac{\partial Z^{\bot}}{\partial s} &=& \lim_{p\rightarrow 0}\frac{\partial}{\partial \boldsymbol{p}^{2}} \int_{q} \frac{\partial R}{\partial s} \frac{\partial}{\partial R} \Tr \left\{ F(q) \Gamma^{(2,1)} (-p, p + q) \Re G^{\textrm R}(p + q) \Gamma^{(2,1)}(p, q) \right\} , \label{Eq:ZperbFlow} \\
\frac{\partial Z^{\vert\!\vert}}{\partial s} &=& - \frac{1}{2} \lim_{p\rightarrow 0}\frac{\partial^{2}}{\partial \omega^{2}} \int_{q} \frac{\partial R}{\partial s} \frac{\partial}{\partial R} \Tr \left\{ F(q) \Gamma^{(2,1)} (-p, p + q) \Re G^{\textrm R}(p + q) \Gamma^{(2,1)}(p, q) \right\} , \label{Eq:ZvertFlow} \\
\frac{\partial \Omega}{\partial s} &=& - i \beta \lim_{p\rightarrow 0}\frac{\partial}{\partial \omega} \int_{q} \frac{\partial R}{\partial s} \frac{\partial}{\partial R} \Tr \left\{ F(q) \Gamma^{(2,1)} (-p, p + q) \Re G^{\textrm R}(p + q) \Gamma^{(2,1)}(p, q) \right\} . \label{Eq:OmegaFlow}
\end{eqnarray}
In practice, one usually considers only the contribution from the Goldstone modes on the RHS, since it is understood that they give the dominant contribution to the wavefunction renormalization $Z^{\bot}$ \cite{Tetradis:1993ts,Berges:2000ew} (and similarly to the dynamic coefficients $Z^{\vert\!\vert}$ and $\Omega$). Here, we follow the same strategy and therefore, instead of evaluating the trace we carry out a projection onto the Goldstone modes in Eqs.\ \eqref{Eq:ZperbFlow} -- \eqref{Eq:OmegaFlow}. The diagrams that we evaluate to obtain the RG flow of the parameters are shown in Fig.\ \ref{Fig:Diagram-2}.

\begin{figure*}[!h]
\centering
\includegraphics[width=0.655\textwidth]{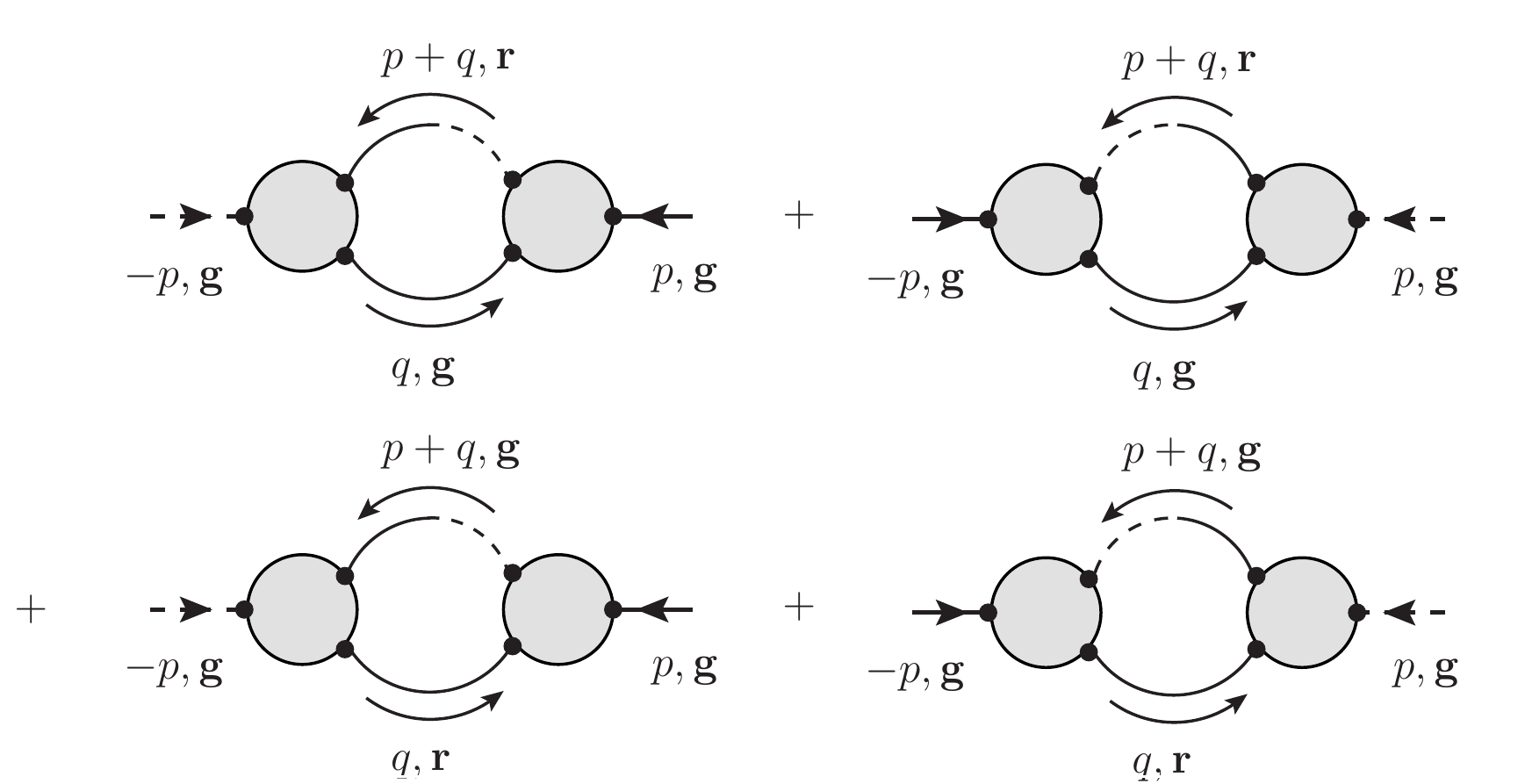}
\caption{\label{Fig:Diagram-2}Diagrams that contribute on the RHS of Eqs.\ \eqref{Eq:ZperbFlow} -- \eqref{Eq:OmegaFlow}. Full-dashed lines corresponds to (nonperturbative) retarded/advanced propagators $G^{\textrm{R}/\textrm{A}}$, while the full lines denote statistical propagators $F$. Full/dashed external legs denote $\phi$/$\tilde{\phi}$-field insertions. Next to the momenta we indicate the relevant modes indicated (\textbf{r}: radial mode, \textbf{g}: Goldstone mode).}
\end{figure*}

\end{widetext}

\end{appendix}

\bibliography{references}

\end{document}